\documentclass[sigconf]{acmart}
	\usepackage{graphicx}
        \usepackage{epstopdf}
	\usepackage{balance}
	\usepackage{amsmath}
	\usepackage{amsthm}
	\usepackage{amssymb}
	\usepackage{enumitem}
	\usepackage{mathtools}
        \usepackage{xspace}
        \usepackage{comment}
	
	\usepackage{multirow, rotating, wasysym, url}
	\usepackage{subfigure}
	\usepackage[ruled,linesnumbered,noend]{algorithm2e}
	\usepackage{hyperref} 
	\usepackage{color}
	\usepackage{soul}
	\usepackage{bbm}
	\usepackage{caption}
	\usepackage{enumerate}
	\usepackage{framed}
	\usepackage{lipsum}
	\usepackage[normalem]{ulem}
	\usepackage{booktabs}
	\usepackage{wrapfig}
        \usepackage{xcolor}
	
	\newcommand{\ie}{\textit{i.e.}}

	\newcommand{\bbb}{\noindent\textbf}

\newtheorem{theorem}{Theorem}

\newtheorem{definition}{Definition}

\newcommand{\sketch}{Carbonyl$_4$\xspace}
\newcommand{\bucket}{Balance Bucket\xspace}
\newcommand{\overflow}{Cascading Overflow\xspace}

\begin{document}

\title{\sketch: A Sketch for Set-Increment Mixed Updates}

\author{Yikai Zhao}
\affiliation{%
  \institution{Peking University}
  \city{Beijing}
  \state{China}
}
\email{zyk@pku.edu.cn}

\author{Yuhan Wu}
\affiliation{%
  \institution{Peking University}
  \city{Beijing}
  \state{China}
}
\email{yuhan.wu@pku.edu.cn}

\author{Tong Yang}
\affiliation{%
  \institution{Peking University}
  \city{Beijing}
  \state{China}
}
\email{yangtong@pku.edu.cn}

 	\begin{abstract}
In the realm of data stream processing, the advent of \textsc{Set-Increment} Mixed (SIM) data streams necessitates algorithms that efficiently handle both \textsc{Set} and \textsc{Increment} operations. We present \sketch, an innovative algorithm designed specifically for SIM data streams, ensuring accuracy, unbiasedness, and adaptability. \sketch introduces two pioneering techniques: the \bucket for refined variance optimization, and the \overflow for maintaining precision amidst overflow scenarios. Our experiments across four diverse datasets establish \sketch’s supremacy over existing algorithms, particularly in terms of accuracy for item-level information retrieval and adaptability to fluctuating memory requirements. The versatility of \sketch is further demonstrated through its dynamic memory shrinking capability, achieved via a re-sampling and a heuristic approach. The source codes of \sketch are available at GitHub \cite{github}.
%
%
\end{abstract}
 	\maketitle


        
        \section{\textbf{Introduction}}

Sketches are algorithms that succinctly approximate per-key information in data streams, enabling functions like point queries \cite{cmsketch, cusketch, csketch}, subset queries \cite{coco, uss, hyperuss}, and Top-$K$ queries \cite{spacesaving, elastic, univmon}. They're applied in diverse areas, including databases \cite{psketch, amssketch, kllpm}, data mining \cite{maxloghash, oph, doph}, networking \cite{lightguardian, sketchlib, flowradar}, and web services \cite{oddsketch}. Unlike hash tables \cite{cuckoo, hopscotch}, which record complete key-value pair information, sketches construct a highly accurate approximate representation of such pairs using sublinear space, offering a configurable balance between space efficiency and accuracy. For platforms with memory constraints, such as handheld devices, IoT networks, and high-performance ASICs, sketches are increasingly preferred due to their ability to deliver near-accurate measurements with tolerable error margins, thus providing satisfactory performance.

However, a notable shortcoming of sketches, when compared to their advantages in space efficiency and accuracy, is their limited support for certain types of updates in data streams. Unlike hash tables, sketches are not inherently designed to handle \textsc{Set} updates. Typically, each item in a data stream corresponds to a modification of a key's value.
Both sketches and hash tables support \textsc{Increment} updates $\langle e, ``+=", v\rangle$, which increment the value associated with a key $e$, akin to the SQL command:
\begin{equation}
\mathtt{UPDATE}\ \mathtt{table}\ \mathtt{SET}\ \mathtt{value} = \mathtt{value} + v\ \mathtt{WHERE}\ \mathtt{key} = e.
\end{equation}
Yet, only hash tables facilitate \textsc{Set} updates $\langle e, ``:=", v\rangle$, which completely replace the value for a key $e$, corresponding to the SQL statement:
\begin{equation}
\mathtt{UPDATE}\ \mathtt{table}\ \mathtt{SET}\ \mathtt{value} = v\ \mathtt{WHERE}\ \mathtt{key} = e.
\end{equation}
To illustrate, we present examples where data streams incorporate both \textsc{Increment} and \textsc{Set} updates, a scenario we refer to as \textsc{Set-Increment} Mixed (SIM) updates.

\bbb{Sensor Data Collection:}
Sensors gather time-series data, with each sensor's ID acting as the key and its readings as the value. To optimize bandwidth, sensors switch between transmitting complete readings upon request (\textsc{Set} updates) and sending incremental changes at other times (\textsc{Increment} updates), resulting in SIM updates \cite{inc-abs-2, inc-abs-1}.

\bbb{Batch Size Statistics:}
In data streams, a batch, or in networking terms, a flowlet, is defined by items with the same key $e$ occurring within a certain time threshold $T$ \cite{batchy, clocksketch, flowlet}. Tracking the size of these batches involves marking the first item as a reset point (\textsc{Set} updates) and subsequently counting additional items as they arrive (\textsc{Increment} updates), leading to SIM updates.

\bbb{Real-time Memory Monitoring:}
Monitoring memory usage in real-time for objects in live programs is crucial for developers. Tools like LLDB and JProfiler track memory by recognizing new or resized objects as \textsc{Set} updates and incremental memory additions as \textsc{Increment} updates \cite{lldb, jprofiler}. This process, analogous to tracking keys in a data stream, also results in SIM updates.

\begin{table}[h]
\centering
\caption{Comparison of update forms supported by different algorithms.}
\begin{tabular}{|l|l|l|}
\hline
                             & \textbf{Exact}                & \textbf{Approximate}        \\ \hline
\textbf{\textsc{Increment}-Only} & Hash Tables                  & Existing Sketches                    \\ 
\textbf{\textsc{Set-Increment} Mixed}            & Hash Tables                   & \textbf{This Study}      \\ \hline
\end{tabular}
\label{tab:update}
\end{table}

Table \ref{tab:update} delineates the capability of hash tables and sketches in managing distinct update forms. Hash tables can accurately process both \textsc{Increment}-Only and \textsc{Set-Increment} Mixed updates, while existing sketches are constrained to \textsc{Increment}-Only updates with approximate results. Our contribution lies in the introduction of an innovative sketch tailored for \textsc{Set-Increment} Mixed updates, which facilitates a more effective approximation of data streams with SIM updates. This development significantly enhances the memory efficiency and adaptability of applications, including the three exemplified cases, optimizing them for more streamlined and versatile deployment.



Before we introduce our solution, it is essential to understand why existing sketches have difficulty with \textsc{Set} updates. Sketches typically come in two varieties: counting sketches and key-value sketches. Counting sketches do not keep track of individual item keys, utilizing shared counters that permit hash collisions. Altering a counter to accommodate a \textsc{Set} value can inadvertently affect other items that collide in that counter, potentially breaching the error limits established by $L_1$ or $L_2$ norms, thus leading to significant errors. 
\textit{
Simply put, consider a task: recording the sum of values for all keys. If there are only \textsc{Increment} updates, we could trivially use a single counter; however, if there are \textsc{Set} updates, we must record the value of each key.}
Key-value sketches maintain keys and values for a subset of items within a bucket array, where items hashed to the same bucket vie for limited space using intricate replacement strategies. These strategies, however, are optimized for \textsc{Increment} updates with minor and restricted increments, falling short in accuracy for \textsc{Set} updates that introduce a larger range of values.

In this work, we introduce \sketch, a sophisticated key-value sketch algorithm for data streams with Set-Increment mixed (SIM) updates, optimizing for limited memory. \sketch adapts to the value distributions of \textsc{Set} and \textsc{Increment} updates, supporting standard queries and offering enhanced accuracy across datasets. Its unbiased estimates ensure controlled errors, and its flexible design allows for dynamic memory management.
We address two main challenges to achieve our objectives. The \textit{Feasibility Challenge} involves managing \textsc{Set} updates to maintain algorithmic accuracy and function within $L_1$ or $L_2$ norm constraints. The \textit{Dynamic Challenge} seeks to adapt to a wide range of \textsc{Set} update values, minimizing error propagation due to hash collisions.

In addressing the feasibility challenge, we extend the Probability Proportional to Size (PPS) sampling-based item competition approach, fundamental to existing unbiased sketches \cite{uss, coco}—termed \textit{unbiased merging} in Section \ref{sec:back}—from the positive integer domain $\mathbb{N}^+$ to the entire real number field $\mathbb{R}$. This extension lays the groundwork for the \textit{\bucket}, tailored to optimize local variance while maintaining unbiased estimates. Traditional replacement strategies in sketches \cite{coco, elastic}, which consistently merge any new item with the smallest existing value in the bucket, may be near-optimal for increment updates with fixed values but are not ideal for set updates where values can be substantially larger. Our \textit{\bucket} introduces a more nuanced approach, considering the size of the update value to determine the most advantageous merging strategy within the bucket, thereby providing an optimal solution for handling set updates.

To tackle the dynamic challenge, we implement an overflow mechanism that connects different \textit{\bucket} within a bucket array, assigning each item to a pair of buckets rather than just one. When processing an update, if the item values in the associated bucket are comparatively large, leading to potential significant errors during competition or replacement, we trigger an overflow. This process moves the item with the smallest value to another bucket that is also associated with that item. Termed as \textit{\overflow}, this strategy unfolds in a cascading fashion, halting only upon activation of an in-built automatic adaptation protocol. Effectively, \textit{\overflow} transitions the focus from minimizing local variance within individual buckets to achieving a broader, near-global variance reduction by iterating through multiple buckets. This provides an expanded scope for accommodating updates with larger values. Unlike the feasibility challenge, which relies on theoretical analysis, the dynamic challenge is predominantly addressed through intricate algorithmic innovation.

\bbb{Key Contributions:}
1) \sketch is introduced as the first algorithm designed for \textsc{Set-Increment} mixed data streams, ensuring accuracy, unbiasedness, and flexibility.
2) Comprehensive experiments on various datasets confirm \sketch's effectiveness for point, subset, and Top-$K$ queries, outperforming existing approaches.
3) Advanced in-place shrinking algorithms are crafted to significantly boost \sketch's flexibility without compromising on performance.
        \section{\textbf{Background}}
\label{sec:back}


\subsection{\textbf{Problem Definition}}
\label{sec:back:def}

We simplify our terminology by referring to data streams with Set-Increment mixed updates as Set-Increment mixed data streams (SIM streams). The formal definition is as follows:

\begin{definition}
    (\textsc{Set-Increment Mixed Data Stream}) A Set-Increment mixed (SIM) data stream $\mathcal{S}=\{op_1, op_2, \cdots, op_n\}$ consists of a sequence of updates $op_i$, each being either a set update $\langle e_i, ``:=", v_i\rangle$ or an increment update $\langle e_i, ``+=", v_i\rangle$. Here, $e_i$ denotes an item from the universal set $\mathcal{U}$, and $v_i$ is a value from the set of real numbers $\mathbb{R}$. The true value of item $e\in\mathcal{U}$ after $n$ updates is denoted as $R(e)$.
\end{definition}

\begin{definition}
    (\textsc{Point Query}) A point query over a SIM data stream $\mathcal{S}$, given an algorithm $\mathcal{A}$, requires $\mathcal{A}$ to estimate the true value $R(e)$ of any item $e\in\mathcal{U}$, providing an approximation $Q(e)$.

    (\textsc{Subset Query}) A subset query for a SIM data stream $\mathcal{S}$, under algorithm $\mathcal{A}$, requires an estimation of the cumulative true value $R(\mathcal{U}')=\sum_{e\in\mathcal{U}'}R(e)$ for any subset $\mathcal{U}'\subseteq\mathcal{U}$, denoted as $Q(\mathcal{U}')$.

    (\textsc{Top-$K$ Query}) For a Top-$K$ query on SIM data stream $\mathcal{S}$, algorithm $\mathcal{A}$ is expected to identify $K$ items with the highest true absolute values $|R(e)|$ and provide their corresponding estimates $Q(e)$.
\end{definition}

\begin{definition}
    (\textsc{$L_1$ Norm}) The $L_1$ norm of a Set-Increment mixed data stream $\mathcal{S}=\{op_1, op_2, \cdots, op_n\}$ is defined as the sum of the absolute values of the updates:
    $$
    \lVert\mathcal{S}\rVert_1=\sum_{i=1}^{n} |v_i|.
    $$
\end{definition}

\subsection{\textbf{Related Work}}
\label{sec:back:work}

We categorize the pertinent algorithms into three types: counting sketches, key-value sketches, and hash tables, assessing their adaptability to handle Set-Increment mixed (SIM) data streams with memory efficiency. Other related works also include \cite{loglog, couper, spear, rfs, rosetta, kllpm, univmon, prsketch, compass}, among others.

\bbb{Counting Sketches.}
Counting sketches are tailored for increment-only data streams and share a common structure. Typically, they utilize a matrix of counters with $d$ rows and $w$ columns to log item values, associating each item with a counter in every row via $d$ distinct hash functions.
Popular counting sketches such as count-min (CM) sketch \cite{cmsketch}, Count sketch \cite{csketch}, conservative-update (CU) sketch \cite{cusketch}, CMM sketch \cite{cmm}, CSM sketch \cite{csm}, SALSA \cite{salsa}, mong others \cite{psketch, stair, adasketch}, are differentiated by their counter update mechanisms.
For instance, CM sketch increments the hashed counters for each item directly across all rows, Count sketch adjusts the increment probabilistically based on the item's key, and CU sketch selectively increases only the smallest hashed counter, functioning exclusively with positive increments, denoted by $v_i>0$.
However, adapting counting sketches to accommodate Set-Increment mixed (SIM) data streams is infeasible, as they lack the mechanism to record individual item keys, rendering the identification of previous item values for set updates $\langle e,``:=",v\rangle$ unattainable.

\bbb{Key-Value Sketches.}
Key-value sketches, primarily designed for increment-only data streams, function similarly to approximate hash tables by logging the keys and estimated values of a selection of items from the universal set $\mathcal{U}$.
For instance, SpaceSaving \cite{spacesaving} (SS) and Unbiased SpaceSaving \cite{uss} (USS) employ a modified heap structure of size $K$ to keep track of the $K$ items with the largest estimated values, handling increment updates where $v_i=1$ to maintain $O(1)$ time complexity. The SpaceSaving$^\pm$ \cite{sspm} algorithm extends this to manage $v_i=-1$. To accommodate set updates, SS, USS, and SS$^\pm$ would need to accept an update complexity of $O(\log K)$.
Other key-value sketches, such as RAP \cite{rap}, Elastic Sketch \cite{elastic} (Elastic) and CocoSketch \cite{coco} (Coco), along with several others \cite{hyperuss,clocksketch,waving,switchsketch,dhs}, replace the counters in a matrix with key-value pairs. Coco, an advanced unbiased sketch, provides fair estimates for each item but is limited to increment updates with non-negative values $v_i\geqslant 0$. Upon receiving an increment update $\langle e, ``+=", v\rangle$, Coco identifies the entry $\langle e', v'\rangle$ with the smallest value among the $d$ hashed counterparts and merges them using the \textit{unbiased merging} technique defined within the positive integer realm $\mathbb{N}^+$ as shown:
$$
\mathtt{MERGE}(\langle e, v\rangle, \langle e', v'\rangle) = \begin{cases}
    \langle e, v+v'\rangle & \text{with probability}\ \frac{v}{v+v'}\\
    \langle e', v+v'\rangle & \text{with probability}\ \frac{v'}{v+v'}
\end{cases}.
$$
Altering Elastic or Coco to facilitate set updates seems straightforward—if an item $e$ is pre-recorded, its estimated value is directly updated; otherwise, the set update is processed as an increment update. However, this approach can lead to substantial inaccuracies, as demonstrated in Section \ref{sec:exp}.

\bbb{Hash Tables.}
A hash table requires $O((1+\epsilon)n)$ slots to store the keys and corresponding values for $n$ items, mapping each item to a slot or series of slots. Widely utilized hash tables, such as the Cuckoo hash table \cite{cuckoo} and Hopscotch hashing \cite{hopscotch}, are designed to provide constant query time. In particular, the Cuckoo hash table uses two hash functions, $h_1$ and $h_2$, to associate each item $e$ with two slots $C[h_1(e)]$ and $C[h_2(e)]$. When inserting an item, if both slots are occupied, the table displaces one of the existing entries to an alternative slot, continuing this `kicking' process until an empty slot is found or a preset kick count is exceeded. Hash tables are inherently capable of handling SIM data streams provided they have sufficient memory. Even with a load factor beyond 100\%, the table can still process updates by discarding entries when the kick limit is reached, although this introduces significant errors, which we demonstrate in Section \ref{sec:exp}.

\bbb{Association with \sketch: }
The \bucket design draws from CocoSketch's data structure, and the \overflow concept takes cues from the Cuckoo hash table's collision strategy. Our \sketch algorithm, however, is uniquely tailored to the numerical dynamics of SIM data streams, detailing operations that extend beyond these initial inspirations.

        \section{\textbf{\sketch Algorithm}}

In this section, we delineate the utilization of a \textit{\bucket} for managing Set-Increment mixed data streams. The approach is crafted to ensure unbiased estimations and to minimize variance on a local scale. 
Subsequently, we present the \textit{\overflow} technique, our novel contribution, aimed at extending variance minimization from a local to a near-global scope.

%

\subsection{\textbf{\bucket}}
\label{sec:algo:bucket}

\begin{figure*}[t]
    \centering
    \includegraphics[width=0.8\textwidth]{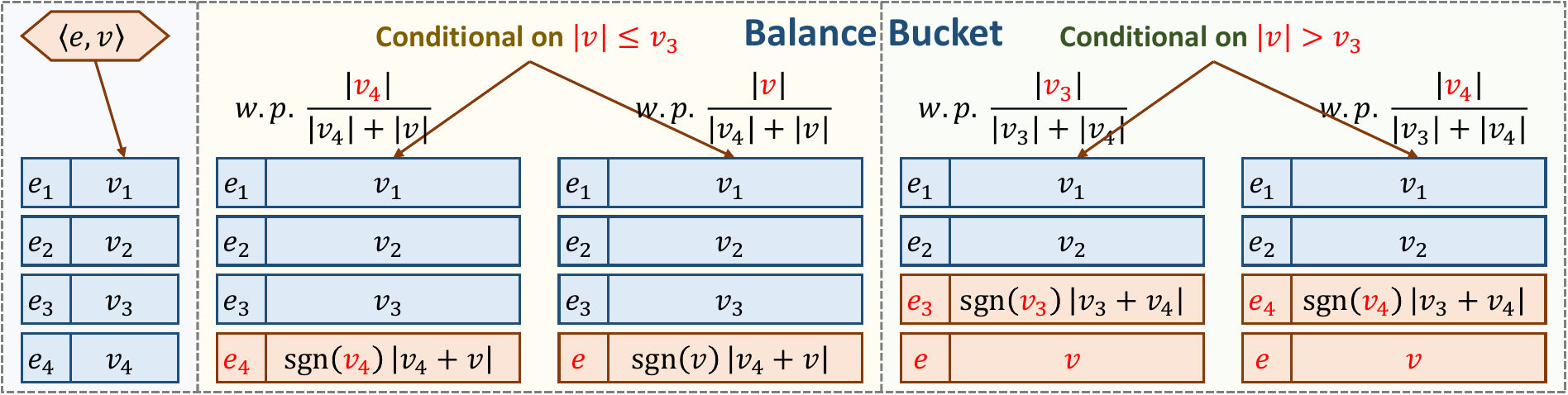}
    \caption{\bucket. Bucket initial containing $d=4$ entries $\langle e_1,v_1\rangle,\cdots,\langle e_4,v_4\rangle$; the right side demonstrates four possible update scenarios when a new update $\langle e,v\rangle$ arrives: there are two possibilities when $|v|\leqslant|v_3|$, and two possibilities when $|v|>|v_3|$.}
    \label{fig:carb:bucket}
\end{figure*}

\bbb{\texttt{MERGE}$^\pm$: Extending to Real Numbers.}
To accommodate both set and increment updates for negative and non-integer values, we extend the concept of unbiased merging, $\texttt{MERGE}(\cdot, \cdot)$, from the domain of positive integers $\mathbb{N}^+$ to a more comprehensive operation, $\texttt{MERGE}^{\pm}(\cdot, \cdot)$, applicable over the real number field $\mathbb{R}$. This is accomplished by considering two entries $\langle e_1, v_1\rangle$ and $\langle e_2, v_2\rangle$ and merging them with a probability $p$ calculated as:
$$
p=\frac{|v_1|}{|v_1|+|v_2|}.
$$
With probability $p$, $\texttt{MERGE}^{\pm}$ returns the item $e_1$ with an updated value $\langle e_1, \texttt{sgn}(v_1)(|v_1|+|v_2|)\rangle$; with the complementary probability $1-p$, it returns item $e_2$ with the updated value $\langle e_2, \texttt{sgn}(v_2)(|v_1|+|v_2|)\rangle$. The function $\texttt{sgn}(v)$ denotes the sign of a real number $v$. We introduce the concept of \textit{merge cost} and assert its minimality for $\texttt{MERGE}^{\pm}$ through the following theorem.

\begin{theorem}
\label{theo:merge}
The operator $\texttt{MERGE}^{\pm}(\langle e_1, v_1\rangle, \langle e_2, v_2\rangle)$ offers an unbiased estimate for the entries $\langle e_1, v_1\rangle$ and $\langle e_2, v_2\rangle$. The variance of this operator, termed the \textbf{merge cost}, is $2|v_1||v_2|$, which is proven to be optimal among all unbiased estimation operators.
\end{theorem}

\begin{proof}
The derivation of Theorem \ref{theo:merge} follows directly from the arguments presented in Theorems 1 and 2 of reference \cite{coco}.
\end{proof}

\bbb{Data Structure:}
The \bucket $B$ consists of $d$ entries, structured as depicted in Figure \ref{fig:carb:bucket}. Each entry, $B[i] = \langle B[i].K, B[i].V \rangle$, stores an item's key, $B[i].K$, and its corresponding estimated value, $B[i].V$, for the $i$-th item in the bucket.

\bbb{Set Update:}
Handling an incoming set update $\langle e, ``\texttt{:=}", v\rangle$ in bucket $B$ unfolds in one of four scenarios:

\noindent\textit{Case 1}:
If item $e$ is present in bucket $B$, with $B[i].K = e$, we directly overwrite the existing value with the new one, i.e., $B[i].V = v$.

\noindent\textit{Case 2}:
If item $e$ is not in bucket $B$ and there is at least one vacant entry, denoted by $B[i]$, we record $e$ in this empty entry as $B[i] = \langle e, v\rangle$.

\noindent\textit{Case 3} (illustrated on the left side of Figure \ref{fig:carb:bucket}):
If item $e$ is not already in bucket $B$ and all entries are occupied, we assume without loss of generality that $$|B[1].V| \geqslant \cdots \geqslant |B[d-1].V| \geqslant |B[d].V|,$$ and $|v| < |B[d-1].V|$. Under this assumption, we merge $\langle e, v\rangle$ with the entry possessing the smallest absolute value, $B[d]$, as:
$$
B[d] = \texttt{MERGE}^\pm(\langle e, v\rangle, B[d])
$$
with a merge cost of $2\cdot|v|\cdot|B[d].V|$.

\noindent\textit{Case 4} (illustrated on the right side of Figure \ref{fig:carb:bucket}):
If item $e$ is not already in bucket $B$, all entries are filled, and $|v| \geqslant |B[d-1].V|$, we first merge the two entries with the smallest absolute values, $B[d-1]$ and $B[d]$, as:
$$
B[d-1] = \texttt{MERGE}^\pm(B[d-1], B[d])
$$
incurring a merge cost of $2\cdot|B[d-1].V|\cdot|B[d].V|$. We then record the new update in $B[d]$ as $B[d] = \langle e, v\rangle$.

\bbb{Increment Update:}
For incoming increment updates $\langle e, ``\texttt{+=}", v\rangle$, the handling within bucket $B$ can be categorized into the following cases:

\noindent\textit{Case 1}:
If item $e$ is already recorded in bucket $B$, identified by $B[i].K = e$, we incrementally update the value: $B[i].V += v$.

\noindent\textit{Case 2, 3, and 4}:
Should item $e$ be unrecorded in bucket $B$, the update is processed as $\langle e, ``\texttt{:=}", v\rangle$, adhering to the \textbf{Set Update} procedures outlined in Cases 2, 3, and 4. 

\bbb{The Rationale of \bucket}:
In the context of Cocosketch, the value $v$ associated with an increment update typically represents the size of a network packet, constrained by a maximum value (e.g., $v \leqslant 1500$), and is often substantially less than the least recorded value in the bucket, $B[d].V$, due to its monotonic increment. 
Conversely, in SIM data streams, both set and increment updates have the potential to decrease recorded values, resulting in scenarios where $v$ could surpass the absolute value of the smallest recorded value, $|B[d].V|$, or even the second smallest, $|B[d-1].V|$. Consequently, discerning the optimal pairing for merging—by comparing $|v|$ with $|B[d-1].V|$—is essential to achieve minimal merge cost and maintain the integrity of the \bucket's data structure.





\bbb{Point Query:}
To ascertain the estimated value of an item $e$, we scan each non-empty entry in the \bucket $B$. If we find an entry corresponding to $e$ (\ie, $B[i].K = e$), we return the recorded estimate $Q(e) = B[i].V$. If $e$ is not found within the bucket, we conclude that $Q(e) = 0$.

\begin{theorem}
\label{theo:bucket:unbias}
    \textsc{(Unbiasedness)}
    For any item $e$ within a universal set $\mathcal{U}$ and subjected to a Set-Increment mixed (SIM) data stream $\mathcal{S}$, let $R(e)$ denote the true value and $Q(e)$ the estimated value by the \bucket. It holds that the expected value of the estimate is equal to the true value:
    $$
    \mathbb{E}[Q(e)] = R(e).
    $$
\end{theorem}

\begin{theorem}
\label{theo:bucket:var}
    \textsc{(Variance Bound)}
    Under the same setting as above, the expected sum of squared deviations between the estimated and true values across all items in $\mathcal{U}$ is bounded by:
    $$
    \mathbb{E}\left[\sum_{e\in \mathcal{U}}(Q(e) - R(e))^2\right] \leqslant \frac{2\lVert\mathcal{S}\rVert_1^2}{d}.
    $$
\end{theorem}

\begin{proof}
    Proofs for the above theorems are presented in detail in Section \ref{sec:math:bucket}.
\end{proof}




\subsection{\textbf{\sketch with \overflow}}

While the \bucket is effective, its capacity must be judiciously constrained to guarantee rapid and consistent update times. To address this, we introduce the fully-fledged \sketch algorithm, which orchestrates multiple small \bucket in tandem, each serving as a fundamental unit within the system.


\begin{figure*}[t]
    \centering
    \includegraphics[width=0.85\textwidth]{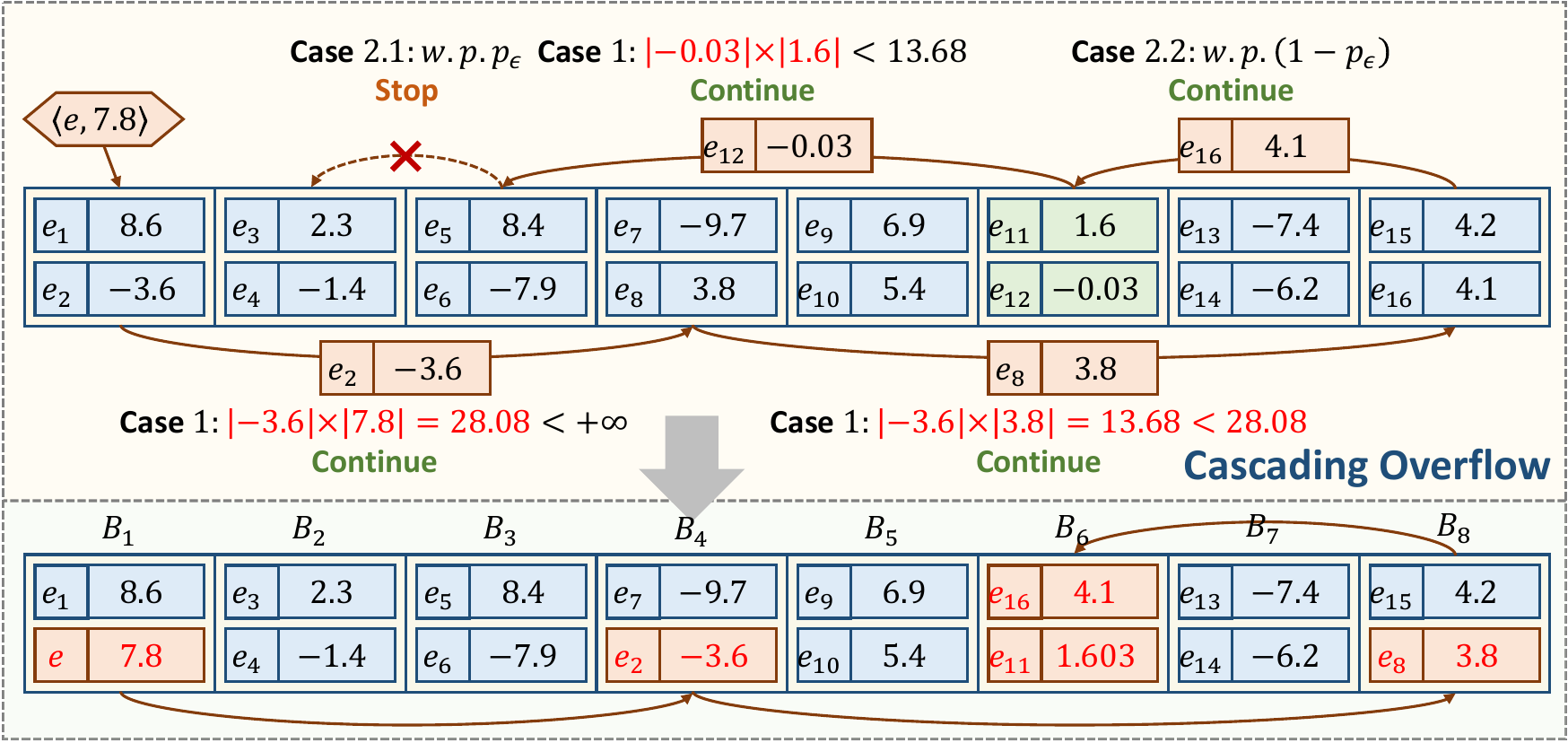}
    \caption{In a \overflow example:
\textbf{(a) Searching Stage:}
The process starts with the update $\langle e, 7.8\rangle$ at bucket $B_1$, with $min_{global}$ initially infinite.
\textit{Step 1:} $min_{local} = 28.08$ prompts an update to $min_{global}$, and the search moves to $B_4$ with $\langle e_2, -3.6\rangle$.
\textit{Step 2:} $min_{global}$ becomes $13.68$, and the search transitions to $B_8$ with $\langle e_8, 3.8\rangle$.
\textit{Step 3:} With $min_{local} \geqslant min_{global}$, the search proceeds to $B_6$ with a chance of stopping.
\textit{Step 4:} A new low for $min_{global}$ at $0.123$ leads to $B_3$ with $\langle e_{12}, -0.03\rangle$.
\textit{Step 5:} The search may stop, with $B_{opt}$ identified as $B_{6}$.
\textbf{(b) Kicking Stage:}
Initiating at $B_1$, the entry $\langle e, 7.8\rangle$ causes a series of displacements across the buckets, ending with a merge in $B_6$.
%
}
    \label{fig:carb:searching}
\end{figure*}

\bbb{Data Structure:}
The architecture of \sketch, as depicted in Figure \ref{fig:carb:searching}, comprises $w$ \bucket  $B_1, \cdots, B_w$. Each \bucket $B_i$ houses $d$ entries. The system employs two hash functions, $h_1(\cdot)$ and $h_2(\cdot)$, which uniformly map an item $e \in \mathcal{U}$ to one of the $w$ buckets, with the stipulation that $h_1(e) \neq h_2(e)$ for any item $e$.

\bbb{Set Update:}
When a set update $\langle e, ``\texttt{:=}", v\rangle$ arrives, \sketch utilizes the hash functions to identify two corresponding \bucket, $B_{h_1(e)}$ and $B_{h_2(e)}$, which we refer to as the hashed buckets for item $e$. If $e$ is already present in either $B_{h_1(e)}$ or $B_{h_2(e)}$, we apply the overwriting strategy from Case 1 outlined in Section \ref{sec:algo:bucket}. Conversely, if $e$ has not been recorded but an empty entry exists in either hashed bucket, we proceed as described in Case 2 from Section \ref{sec:algo:bucket} to record the update. If neither condition is met, we select one hashed bucket at random and initiate a \textbf{\overflow} process, which seeks a near-globally optimal merge candidate within the system.



\overflow consists of two distinct phases: the searching stage and the kicking stage.

\bbb{Searching Stage} (illustrated on the upper portion of Figure \ref{fig:carb:searching}):
The searching stage initiates with the entry $\langle e, v\rangle$ and one of its hashed buckets, referenced as $B_{idx}$. We commence by setting a variable $min_{global}$ to $+\infty$, representing the smallest merge cost encountered thus far. Under the assumption that
$$
|B_{idx}[1].V| \geqslant \cdots \geqslant |B_{idx}[d-1].V| \geqslant |B_{idx}[d].V|,
$$
we compare the incoming value $v$ with the penultimate smallest value $|B_{idx}[d-1].V|$ in the bucket. The objective is to identify a locally optimal merge within $B_{idx}$ and to compute the corresponding minimal merge cost, as detailed in lines 3-7 of the pseudo-code \ref{algo:searching}:
\begin{align*}
min_{local} = \\
\begin{cases}
    2 \cdot |v| \cdot |B_{idx}[d].V| & \text{if } |v| \leqslant |B_{idx}[d-1].V|, \\
    2 \cdot |B_{idx}[d].V| \cdot |B_{idx}[d-1].V| & \text{if } |v| > |B_{idx}[d-1].V|.
\end{cases}
\end{align*}

\begin{itemize}[leftmargin=*]
    \item \textit{Case 1}: If $min_{local}$ is less than the global minimum $min_{global}$, we update $min_{global} = min_{local}$ as per lines 8-10 in pseudo-code \ref{algo:searching}. The smallest entry in $B_{idx}$, $B_{idx}[d]$, is then treated as the insertion candidate, and the search proceeds to the alternate hashed bucket of $B_{idx}[d].K$, following lines 14-17 in the pseudo-code.
    
    \item \textit{Case 2}: If $min_{local} \geqslant min_{global}$, we have two subcases:
        \begin{itemize}[leftmargin=*]
            \item \textit{Case 2.1}: The searching stage halts with probability $p_\epsilon$, detailed in lines 11-13 of the pseudo-code.
            \item \textit{Case 2.2}: The search continues with the remaining probability, echoing the procedure in Case 1, as outlined in lines 14-17 of the pseudo-code.
        \end{itemize}
\end{itemize}
The number of search steps may be capped at a maximum value $M$ to constrain the process.

\bbb{Kicking Stage} (illustrated on the lower part of Figure \ref{fig:carb:searching}):
Upon concluding the searching stage, the kicking stage commences with the entry $\langle e, v\rangle$ and its associated buckets $B_{idx}$. We operate under the assumption that the absolute values within $B_{idx}$ are ordered as 
$$
|B_{idx}[1].V| \geqslant \cdots \geqslant |B_{idx}[d-1].V| \geqslant |B_{idx}[d].V|.
$$
The entry $B_{idx}[d]$ is replaced by $\langle e, v\rangle$ (\ie, $B_{idx}[d] = \langle e, v\rangle$), and the displaced $B_{idx}[d]$ becomes the new candidate for insertion, perpetuating the kick through its alternate hashed bucket.
This process iterates until a bucket $B_{opt}$ is located, where the local minimum merge cost aligns with the near-global minimum, $min_{global}$, identified in the searching stage. The set update mechanics of the \bucket, as delineated in Section \ref{sec:algo:bucket}, are then executed within $B_{opt}$, concluding the kicking phase.
Notably, should an empty entry be encountered during the search, $min_{global}$ becomes zero, and the kicking phase will cease at that juncture, negating further action.

\bbb{The Rationale of \overflow}:
In a SIM data stream, when the incoming value $|v|$ and the penultimate smallest value $|B[d-1].V|$ within a hashed bucket $B$ substantially exceed the mean of the least absolute values recorded across buckets, merging $B[d]$ with either $\langle e, v\rangle$ or $B[d-1]$ becomes suboptimal. 
\overflow's strategy is to identify a near-globally optimal merge within the constraints of the predetermined time complexity. This approach is adaptive, with extensive searching when the initial merge cost is high, and abbreviated steps when the initial cost is already minimal.

\begin{algorithm}[t]
{
\renewcommand\baselinestretch{0.9}\selectfont
\caption{Routine for the searching stage.}
\label{algo:searching}

\SetKwFunction{FMain}{Searching}
\SetKwProg{Fn}{Func}{:}{}
\Fn{\FMain{$idx, \langle e, v\rangle, min_{global}$}} {
    \textcolor{blue}{$\triangleright$ Assume $B_{idx}$ entries are sorted.}\\
    \uIf{$|v|\leqslant |B_{idx}[d-1].V|$} {
        $min_{local} = |v|\times|B_{idx}[d].V|$\;
    }
    \Else{
        $min_{local}=|B_{idx}[d].V|\times|B_{idx}[d-1].V|$\;
    }
    \textcolor{blue}{$\triangleright$ Compute local optimal merge.}\\
    \uIf{$min_{local}<min_{global}$}{
        $min_{global}\gets min_{local}$\;
        \textcolor{blue}{$\triangleright$ Update for better merge.}\\
    }
    \ElseIf{$r\in Uniform(0,1)\leqslant p_{\epsilon}$}{
        \textcolor{blue}{$\triangleright$ Stop with chance $p_\epsilon$.}\\
        \Return;
    }
    $idx\gets \mathtt{Another\_Index}(B[d].K, idx)$\;
    $\langle e, v\rangle\gets B[d]$\;
    \textcolor{blue}{$\triangleright$ Continue with next bucket.}\\
    $\mathtt{Searching}(idx, \langle e, v\rangle, min_{global})$\;
}
}
\end{algorithm}

\begin{algorithm}[t]
{
\renewcommand\baselinestretch{0.9}\selectfont
\caption{Routine for the kicking stage.}
\label{algo:kicking}

\SetKwFunction{FMain}{Kicking}
\SetKwProg{Fn}{Func}{:}{}
\Fn{\FMain{$idx, \langle e, v\rangle, min_{global}$}} {
    \textcolor{blue}{$\triangleright$ Assume $B_{idx}$ entries are sorted.}\\
    \uIf{$|v|\cdot|B_{idx}[d].V|=min_{global}$} {
        $B_{idx}[d]\gets\mathtt{MERGE}^\pm\left(
        \begin{aligned}
            \langle e, v\rangle, B_{idx}[d]
        \end{aligned}\right)$\;
        \textcolor{blue}{$\triangleright$ Merge and end if local equals global.}\\
    }
    \uElseIf{$|B_{idx}[d].V|\cdot|B_{idx}[d-1].V|=min_{global}$} {
        $B_{idx}[d-1]\gets\mathtt{MERGE}^\pm\left(
        \begin{aligned}
            B_{idx}[d], B_{idx}[d-1]
        \end{aligned}\right)$\;
        $B_{idx}[d]\gets\langle e, v\rangle$\;
        \textcolor{blue}{$\triangleright$ Merge and end if local equals global.}\\
    }
    \Else{
        $idx\gets \mathtt{Another\_Index}(B[d].K, idx)$\;
        $\mathtt{Swap}(\langle e, v\rangle, \langle B[d].K, B[d].V\rangle)$\;
        \textcolor{blue}{$\triangleright$ Kick and continue with next bucket.}\\
        $\mathtt{Kicking}(idx, \langle e, v\rangle, min_{global})$\;
    }
}
}
\end{algorithm}

\begin{theorem}
\label{theo:search}
    \textsc{(Expected Number of Searching Steps)}
    Consider a Set-Increment mixed (SIM) data stream $\mathcal{S}$ and \sketch parameterized with $p_\epsilon = \frac{\epsilon}{(1-\sqrt{\epsilon})^2}$. The expected number of steps in the searching stage of the \overflow process is bounded by $O\left(\frac{1}{\epsilon}\right)$.
\end{theorem}

\begin{proof}
    The detailed proof is elaborated in Section \ref{sec:math:search}.
\end{proof}

\bbb{Increment Update}:
When processing an increment update $\langle e, ``\texttt{+=}", v\rangle$, the two hash functions determine the corresponding \bucket $B[h_1(e)]$ and $B[h_2(e)]$. If $e$ is already present in either bucket, we increment the stored value as per the method described in Case 1 of Section \ref{sec:algo:bucket}. If $e$ is not recorded in either bucket, the update is handled as a set update $\langle e, ``\texttt{:=}", v\rangle$, following the complete \textbf{Set Update} procedure.

\bbb{Point Query}:
To retrieve the estimated value for an item $e$, we inspect all entries in the buckets $B_{h_1(e)}$ and $B_{h_2(e)}$. If $e$ is found (\ie, $B_{h_i(e)}[j].K = e$), the estimated value is given by $B_{h_i(e)}[j].V$; if not, the estimated value defaults to zero.

\bbb{Subset Query}:
The estimated sum for a subset $\mathcal{U}'$ within the universal set $\mathcal{U}$ is computed by summing the estimated values of the subset's items across all buckets in \sketch:
$
\sum_{i=1}^{w}\sum_{B_i[j].K \in \mathcal{U}'} B_i[j].V
$.

\bbb{Top-$K$ Query}:
To identify the $K$ items with the highest absolute estimated values, we traverse every entry in all $w$ buckets of \sketch and return the $K$ entries with the largest absolute values.

\begin{theorem}
\label{theo:subset}
    \textsc{(Unbiasedness)}
    Given a SIM data stream $\mathcal{S}$ and a universal set $\mathcal{U}$, for any subset $\mathcal{U}' \subseteq \mathcal{U}$, the \sketch provides an unbiased estimation of the sum. That is, if $R(e)$ is the true value and $Q(e)$ is the estimated value provided by \sketch for any item $e \in \mathcal{U}$, then:
    $$
    \mathbb{E}\left[\sum_{e \in \mathcal{U}'}Q(e)\right] = \sum_{e \in \mathcal{U}'}R(e).
    $$
\end{theorem}

\begin{proof}
    The comprehensive proof is detailed in Section \ref{sec:math:subset}.
\end{proof}

        \section{\textbf{Flexibility of \sketch}}

Flexibility is crucial for approximate algorithms, allowing for dynamic memory allocation to balance between precision and resource availability. This section contrasts the conventional re-build approach for memory adjustments with two novel in-place shrinking algorithms designed for \sketch. These algorithms significantly expedite the shrinking process, improving speed by more than 20-fold. One algorithm aims to surpass the average accuracy of re-build with a near-optimal re-sampling strategy, while the other is specifically tailored to enhance performance for Top-$K$ queries.



\subsection{\textbf{Re-Build based Expanding and Shrinking}}

Adjusting the size of \sketch involves transforming an existing sketch \( C \) with \( w \) buckets into a revised sketch \( C' \) with \( w' \) buckets. When the ratio \( \frac{w}{w'} \) is less than a certain threshold \( \alpha \), which is below the standard load factor for cuckoo hash tables, the new sketch \( C' \) is populated by treating it as a cuckoo hash table with a stopping probability \( p_\epsilon \) set to zero. This means we can directly insert entries from \( C \) into \( C' \) without interruption. Conversely, if the ratio \( \frac{w}{w'} \) meets or exceeds \( \alpha \), the stopping probability \( p_\epsilon \) must be determined based on a complexity level that is deemed acceptable for the operation, before transferring entries from \( C \) to \( C' \).



\begin{figure}[t]
    \centering
    \includegraphics[width=0.4\textwidth]{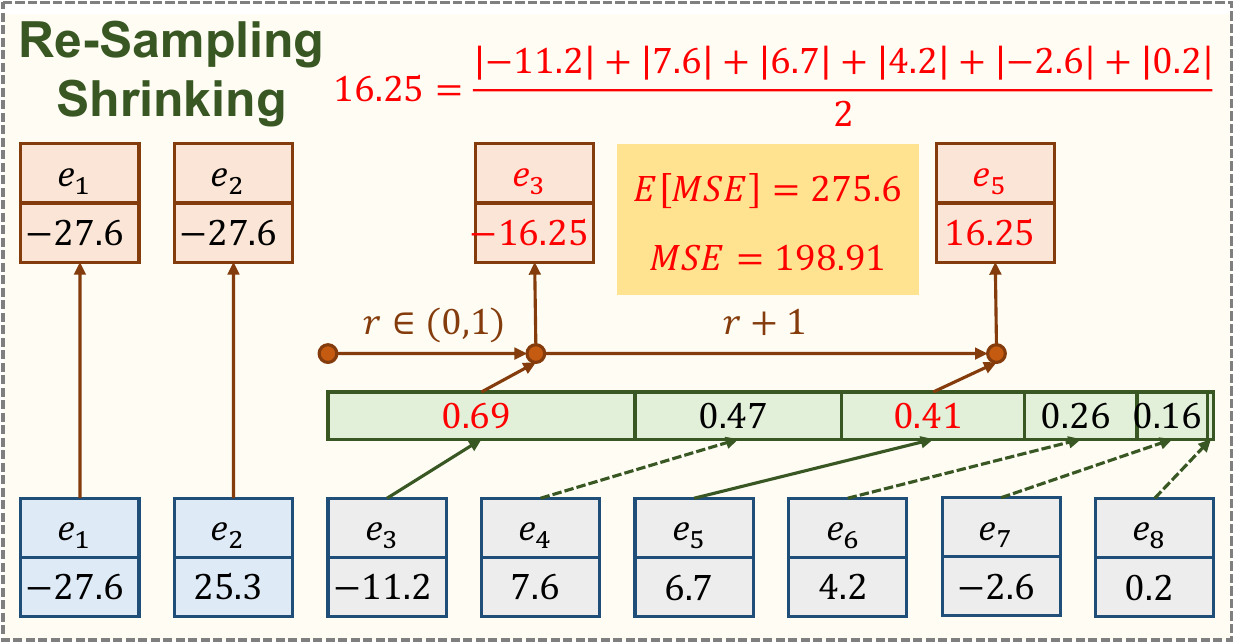}
    \caption{Re-Sampling shrinking. Initially, there are a total of 8 entries (in blue) from two buckets, and after shrinking, there remain 4 entries (in orange) placed in one bucket.}
    \label{fig:shrink:resampling}
\end{figure}

\subsection{\textbf{Re-Sampling based In-place Shrinking}}

While re-build based resizing offers high flexibility and accuracy for adjusting \sketch's memory footprint, it can be time-consuming. This is especially problematic for shrinking, which is often necessitated by the immediate need to reallocate resources to more critical tasks.
For scenarios requiring a reduction of \sketch's memory usage to half, an expedient method is to merge corresponding pairs of buckets, \( B_k \) and \( B_{k+\frac{w}{2}} \), directly into \( B_k \). This in-place operation not only executes swiftly but also requires only a simple adjustment to the hash functions, setting \( h'_i(e) = h_i(e) \% w' \).
This section introduces an optimal re-sampling strategy specifically designed for the efficient merging of bucket pairs when reducing the \sketch size.



\bbb{Re-Sampling Shrinking}:
As depicted in Figure \ref{fig:shrink:resampling}, given two full buckets $B_k$ and $B_{k+\frac{w}{2}}$, our task is to downsize by selecting $d$ entries from the combined $2d$. The entries are sorted by their absolute values in descending order, yielding the sequence $\langle e_1, v_1\rangle, \cdots, \langle e_{2d}, v_{2d}\rangle$. We then assess each entry beginning with the largest, against the criterion:
$$
(d-i+1)\times\frac{|v_{i}|}{\sum_{j=i}^{2d}|v_j|}\geqslant 1.
$$
Entries meeting this condition are placed directly into $B_k$. If not, it implies for all $j \geqslant i$, the probability $p_j=\frac{(d-i+1)|v_{j}|}{\sum_{l=i}^{2d}|v_l|}$ is less than one. Therefore, we need to sample $(d-i+1)$ entries based on their probabilities from the remaining set.

Independent sampling might not yield precisely $(d-i+1)$ entries, necessitating a non-independent approach. We set auxiliary variables $r_{i-1}=0$ and $r_j=\sum_{l=i}^{j} p_l$ for all $j \geqslant i$, then draw a random number $r$ uniformly between $0$ and $1$. For integers $z$ from $0$ to $(d-i)$, we identify $j$ such that $r_{j-1}\leqslant r+z < r_{j}$ and select $\langle e_j, v_j\rangle$ for unbiased re-sampling, placing $\langle e_j, \mathtt{sgn}(v_j)\frac{\sum_{l=i}^{2d}|v_{l}|}{d-i+1}\rangle$ into $B_{k}$.
The time complexity for re-sampling each pair of buckets is $O(d\log(d))$, summing to $O(wd\log(d))$ for the entire \sketch.

\begin{theorem}
\label{theo:resamp}
\textsc{(Optimality)}
The re-sampling method described guarantees unbiased selection from two combined buckets $B_k$ and $B_{k+\frac{w}{2}}$, and achieves the least total variance among all unbiased sampling algorithms.
\end{theorem}

\begin{proof}
Section 1.4 of reference \cite{sampling_1} introduces IPPS (Inclusion Probabilities Proportional to Size) sampling as the optimal variance-minimizing algorithm. Our re-sampling technique is an application of IPPS with a fixed sample size, following the methodology in Section 2.1 of reference \cite{sampling_2}.
\end{proof}

\subsection{\textbf{Heuristic based In-place Shrinking}}

Minimizing total variance through re-sampling is effective for merging buckets $B_k$ and $B_{k+\frac{w}{2}}$, but certain applications, particularly those prioritizing Top-$K$ query accuracy, may require preserving the precision of entries with larger absolute values while maintaining unbiasedness. Addressing this, we introduce a heuristic bucket merging approach tailored to optimize Top-$K$ query performance post-shrinking.

\begin{figure}[t]
    \centering
    \includegraphics[width=0.4\textwidth]{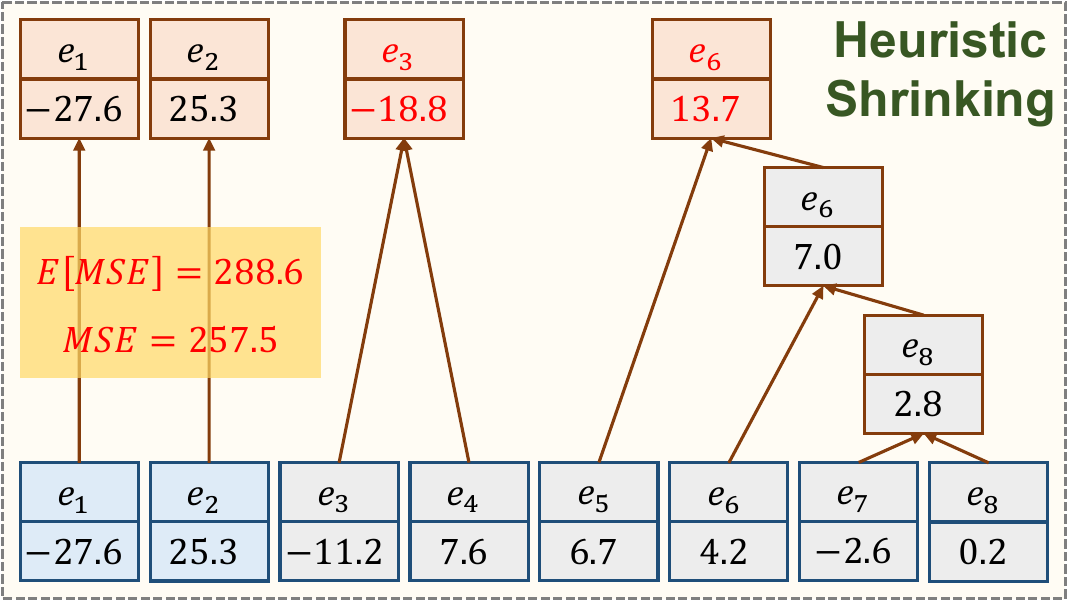}
    \caption{Heuristic shrinking. Initially, there are a total of 8 entries (in blue) from two buckets, and after shrinking, there remain 4 entries (in orange) placed in one bucket.}
    \label{fig:shrink:huerisitc}
\end{figure}

\bbb{Heuristic Shrinking}:
In situations where buckets $B_k$ and $B_{k+\frac{w}{2}}$ lack empty entries, merging $2d$ entries into $d$ necessitates $d$ merge operations, as depicted in Figure \ref{fig:shrink:huerisitc}. The heuristic method prioritizes accuracy for larger absolute values by consistently merging the two smallest entries in each step. Employing a min-heap to facilitate this process, the complexity of each merge is $O(\log(d))$, culminating in a total complexity of $O(wd\log(d))$ for the entire \sketch.




        \section{\textbf{Mathematical Analysis}}
\label{sec:math}

This section is dedicated to substantiating key properties of \sketch through rigorous proofs. We will focus on establishing the unbiased nature and variance constraints of the \bucket, as articulated in Theorem \ref{theo:bucket:unbias} and Theorem \ref{theo:bucket:var} in Subsection \ref{sec:math:bucket}. Additionally, we delve into the computational efficiency of the \overflow process as detailed in Theorem \ref{theo:search} within Subsection \ref{sec:math:search}, and confirm the unbiasedness of subset queries in \sketch, outlined in Theorem \ref{theo:subset}, which is further explored in Subsection \ref{sec:math:subset}.



\subsection{\textbf{Analysis of \bucket}}
\label{sec:math:bucket}

\begin{definition}
Given a Set-Increment mixed (SIM) data stream $\mathcal{S}=\{op_1, \cdots, op_n\}$ and a \bucket $B$, let $Q(e, i)$ be the estimated value of item $e$ by the \bucket after the $i$-th update, and $R(e, i)$ be the true value of item $e$. 
\end{definition}

\bbb{Proof of Unbiasedness (Theorem \ref{theo:bucket:unbias}):}
\begin{proof}
We begin by assuming $\mathbb{E}[Q(e, i-1)] = R(e, i-1)$ and seek to demonstrate that $\mathbb{E}[Q(e, i)] = R(e, i)$ follows.

When $op_i = \langle e_i, ``:=", v_i\rangle$ with $e_i = e$, we have $R(e, i) = v_i$. In the event of case 1 (refer to Section \ref{sec:algo:bucket}), $Q(e, i)$ is set to $v_i$. In cases 2 and 4, $e$ is assigned an entry, making $Q(e, i)$ equal to $v_i$. In case 3, merging $\langle e, v\rangle$ with $B[d]$ results in:
$$
\mathbb{E}[Q(e, i)] = \frac{|v_i|}{|v_i| + |B[d].V|} \cdot \mathtt{sgn}(v_i)(|v_i| + |B[d].V|) = v_i,
$$
establishing that $\mathbb{E}[Q(e, i)] = R(e, i)$.

For an increment update $op_i = \langle e_i, ``+=", v_i\rangle$ where $e_i = e$, $R(e, i)$ is updated to $R(e, i-1) + v_i$. In case 1, $Q(e, i)$ becomes $Q(e, i-1) + v_i$. In cases 2 and 4, where $Q(e, i-1) = 0$, $e$ receives an entry, and thus $Q(e, i) = v_i$. In case 3, where $Q(e, i-1) = 0$, merging is necessary, and so $\mathbb{E}[Q(e, i)] = v_i$. This confirms that $\mathbb{E}[Q(e, i)] = \mathbb{E}[Q(e, i-1)] + v_i = R(e, i)$.

For all other updates $op_i = \langle e_i, *, v_i\rangle$ with $e_i \neq e$, $R(e, i)$ remains $R(e, i-1)$. Cases 1 and 2 ensure $Q(e, i) = Q(e, i-1)$. In cases 3 and 4, if $e$ is present and must merge, then $\mathbb{E}[Q(e, i)] = Q(e, i-1)$; otherwise, $Q(e, i)$ remains $Q(e, i-1)$. Thus, we conclude that $\mathbb{E}[Q(e, i)] = \mathbb{E}[Q(e, i-1)] = R(e, i)$.

Since $\mathbb{E}[Q(e, 0)] = R(e, 0) = 0$, and the inductive hypothesis holds for all $i$, it follows that $\mathbb{E}[Q(e, n)] = R(e, n)$, hence $\mathbb{E}[Q(e)] = R(e)$.
\end{proof}

\bbb{Proof of Variance Bound (Theorem \ref{theo:bucket:var}):}
\begin{proof}
Consider the incremental variance $\Delta_i$ expressed by:
\begin{align*}
    &\Delta_i=
    \begin{aligned}
        &\mathbb{E}\left[\sum_{e\in U}\left(Q(e,i)-R(e,i)\right)^2\right]\\
        -&\mathbb{E}\left[\sum_{e\in U}\left(Q(e,i-1)-R(e,i-1)\right)^2\right]
    \end{aligned}\\
    =&\mathbb{E}\left[\sum_{e\in U}\left(Q(e,i)-R(e,i)\right)^2-\left(Q(e,i-1)-R(e,i-1)\right)^2\right],
\end{align*}
pertaining to items where either $Q(e,i) \neq Q(e,i-1)$ or $R(e,i) \neq R(e,i-1)$.

In the scenario where $op_i=\langle e_i, ``:=", v_i\rangle$, and specifically for case 1 and case 2, we have:
$$(Q(e_i, i)-R(e_i,i))^2-(Q(e_i, i-1)-R(e_i,i-1))^2\leqslant 0;$$
For case 3, the merging of entry $\langle e_i, v_i\rangle$ with $B[d]$ leads to:
\begin{align*}
&\mathbb{E}\left[
\begin{aligned}
    &(Q(e_i, i)-R(e_i, i))^2\\
    +&(Q(B[d].K,i)-R(B[d].K,i))^2\\
    -&(Q(e_i, i-1)-R(e_i, i-1))^2\\
    -&(Q(B[d].K,i-1)-R(B[d].K,i-1))^2
\end{aligned}\right]\\
=&(|v_i|+|B[d].V|)^2-|v_i|^2-|B[d].V|^2-R(e_i,i-1)^2\\
=&2|v_i||B[d].V|-R(e_i,i-1)^2\leqslant2|v_i||B[d].V|.
\end{align*}
When dealing with the insertion of entry $B[d-1]$ alongside $B[d]$ in case 4, the variance increases by:
\begin{align*}
&\mathbb{E}\left[
\begin{aligned}
    &(Q(B[d-1].K, i)-R(B[d-1].K, i))^2\\
    +&(Q(B[d].K,i)-R(B[d].K,i))^2\\
    -&(Q(B[d-1].K, i-1)-R(B[d-1].K, i-1))^2\\
    -&(Q(B[d].K,i-1)-R(B[d].K,i-1))^2
\end{aligned}\right]\\
=&(|B[d-1].V|+|B[d].V|)^2-|B[d-1].V|^2-|B[d].V|^2\\
=&2|B[d-1].V||B[d].V|\leqslant2|v_i||B[d].V|.
\end{align*}
Therefore, each $\Delta_i\leqslant 2|v_i||B[d].V|$.

Applying similar logic to increment updates $op_i=\langle e_i, ``+=", v_i\rangle$, we also deduce that $\Delta_i\leqslant 2|v_i||B[d].V|$.

Acknowledging that 
$$
|B[d].V|\leqslant\frac{\sum_{i=1}^{d} |B[i].V|}{d}\leqslant\frac{\lVert\mathcal{S}\rVert_1}{d},
$$
it follows that $\Delta_i\leqslant 2|v_i|\frac{\lVert\mathcal{S}\rVert_1}{d}$. Aggregating all $\Delta_i$ values, we obtain:
\begin{align*}
 &\mathbb{E}\left[\sum_{e\in U}\left(Q(e)-R(e)\right)^2\right]\leqslant\sum_{i=1}^{n}\Delta_i\\
 \leqslant &\left(\sum_{i=1}^{n}2|v_i|\right)\frac{\lVert\mathcal{S}\rVert_1}{d}=\frac{2\lVert\mathcal{S}\rVert_1^2}{d}.
\end{align*}

\end{proof}

\subsection{\textbf{Analysis of \overflow}}
\label{sec:math:search}

\begin{figure}[t]
    \centering
    \includegraphics[width=0.4\textwidth]{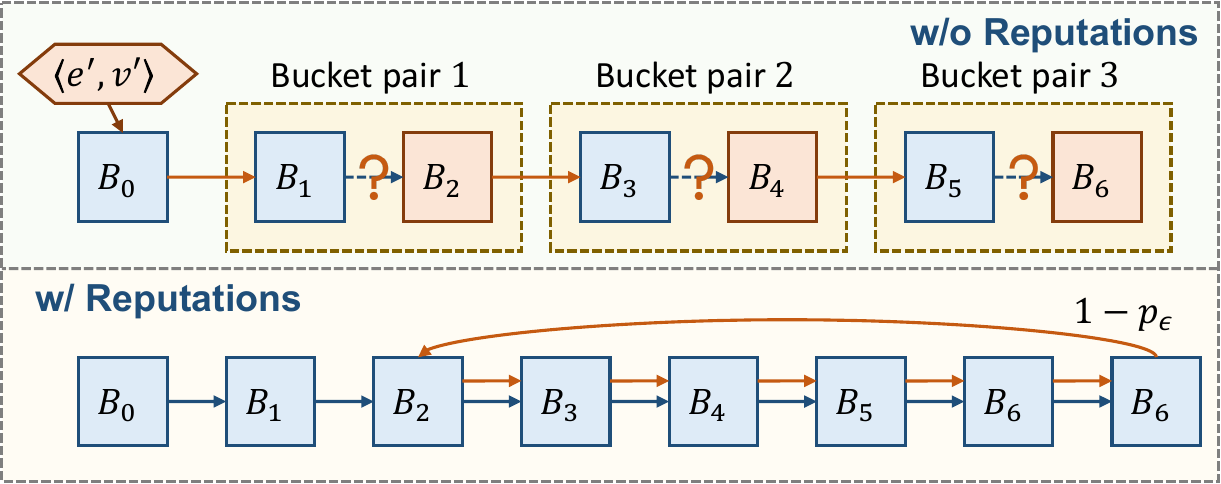}
    \caption{Illustration of Time Complexity.}
    \label{fig:math:time}
\end{figure}

\bbb{Proof of Time Complexity (Theorem \ref{theo:search}):}
\begin{proof}
Assume the searching stage begins at bucket $B_0$, with successive buckets denoted as $B_1, B_2, \cdots$. We initially address the case without bucket duplication along the path.
Illustrated in Figure \ref{fig:math:time}, we group buckets in consecutive pairs, thereby considering the local optimal merge cost $min_{local}$ within bucket $B_{2k}$ and $B_{2k+2}$ as independent random variables.
We simplify the stopping condition, only contemplating the cessation of the searching stage between buckets $B_{2k-1}$ and $B_{2k}$, and not between $B_{2k}$ and $B_{2k+1}$.

With this setup, the searching stage is segmented into two phases: the first seeks a sufficiently small $min_{local}$ as the near-global minimal merge cost $min_{global}$; the second estimates the expected number of steps to halt the process.
To find a bucket $B*$ with $min_{local}$ smaller than that of $\phi w$ buckets, the expected number of steps required is
\begin{align*}
&\mathbb{E}[\text{number of steps in phase 1}]\\
\leqslant&\sum_{k=1}^{w/2} 2k(1-\phi)^{k-1}\phi\leqslant\frac{2}{\phi}.
\end{align*}
Upon updating the near-global optimal merge cost $min_{global}$ to $min_{local}$ of $B*$, the stopping probability at each bucket $B_{2k}$ becomes 
$$
\Pr[\text{stop probability}]=(1-\phi)p_\epsilon.
$$
The expected step count to stop in phase two is then
\begin{align*}
&\mathbb{E}[\text{number of steps in phase 2}]\\
\leqslant&\sum_{k=1}^{w/2} 2k(1-(1-\phi)p_\epsilon)^{k-1}((1-\phi)p_\epsilon)\leqslant\frac{2}{(1-\phi)p_\epsilon}.
\end{align*}
For any $\phi\in(0, 1)$, the expected step count in the searching stage is
\begin{align*}
&\mathbb{E}[\text{number of steps in searching stage}]\\
\leqslant&\min_{\phi}\left(\frac{2}{\phi}+\frac{2}{(1-\phi)p_\epsilon}\right)=2\frac{(\sqrt{p_\epsilon}+1)^2}{p_\epsilon}.
\end{align*}
Substituting $p_\epsilon=\frac{\epsilon}{(1-\sqrt\epsilon)^2}$ yields
\begin{align*}
&\mathbb{E}[\text{number of steps in searching stage}]\\
\leqslant&2\frac{\left(\sqrt{\frac{\epsilon}{(1-\sqrt\epsilon)^2}}+1\right)^2}{\frac{\epsilon}{(1-\sqrt\epsilon)^2}}=\frac{2}{\epsilon}=O\left(\frac{1}{\epsilon}\right).
\end{align*}

Accounting for duplicate buckets on the search path, as depicted in Figure \ref{fig:math:time}, the search loops upon encountering the first repeat bucket. Given that there is no item displacement in this stage, and that $min_{global}$ ceases to update, the search is likely to conclude sooner than in the non-duplicate scenario, maintaining the expected step count at $O\left(\frac{1}{\epsilon}\right)$.
\end{proof}

\subsection{\textbf{Analysis of \sketch}}
\label{sec:math:subset}

The proof of unbiasedness for \sketch is similar to the proof of unbiasedness for \bucket, but it additionally considers the effects of \overflow.

\bbb{Proof of Unbiasedness (Theorem \ref{theo:subset}):}
\begin{proof}
Theorem \ref{theo:subset}'s proof closely follows the rationale of Theorem \ref{theo:bucket:unbias}, where we establish that $\mathbb{E}[Q(e, i-1)]=R(e, i-1)$ implies $\mathbb{E}[Q(e, i)]=R(e, i)$.

For a set update $op_i=\langle e_i, ":=", v_i\rangle$ with $e_i=e$, we have $R(e, i)=v_i$. In situations where item $e$ is recorded in either bucket $B_{h_1(e)}$ or $B_{h_2(e)}$, or when there are empty slots in these buckets, or if the searching stage identifies another bucket as $B_{opt}$, it follows that $Q(e, i)=v_i$. 
If the searching stage designates $B_{h_1(e)}$ or $B_{h_2(e)}$ as $B_{opt}$, the analysis in cases 3 and 4 of Theorem \ref{theo:bucket:unbias} ensures $\mathbb{E}[Q(e,i)]=v_i$. Hence, $\mathbb{E}[Q(e,i)]=R(e,i)$ holds.

For an increment update $op_i=\langle e_i, "+=", v_i\rangle$ with $e_i=e$, $R(e, i)=R(e,i-1)+v_i$. If item $e$ is recorded or if a vacancy exists in $B_{h_1(e)}$ or $B_{h_2(e)}$, or if $B_{opt}$ is determined during the searching stage, then $Q(e, i)=Q(e,i-1)+v_i$ is maintained. 
Following the analysis in Theorem \ref{theo:bucket:unbias} for cases 3 and 4, if $B_{h_1(e)}$ or $B_{h_2(e)}$ becomes $B_{opt}$, we also deduce $\mathbb{E}[Q(e,i)]=Q(e,i-1)+v_i$. Consequently, $\mathbb{E}[Q(e,i)]=\mathbb{E}[Q(e,i-1)]+v_i=R(e,i)$.

In instances where $op_i=\langle e_i, *, v_i\rangle$ and $e_i\neq e$, $R(e, i)$ stays consistent with $R(e, i-1)$. If item $e$ is absent from $B_{opt}$ in \overflow, then $Q(e,i)=Q(e,i-1)$ is valid. Otherwise, according to Theorem \ref{theo:bucket:unbias}, we infer $\mathbb{E}[Q(e,i)]=Q(e,i-1)$. Thus, $\mathbb{E}[Q(e,i)]=\mathbb{E}[Q(e,i-1)]=R(e,i)$.

Given that $\mathbb{E}[Q(e, 0)]=R(e, 0)=0$, and by induction, for $i=n$, we conclude $\mathbb{E}[Q(e, n)]=R(e, n)$, that is, $\mathbb{E}[Q(e)]=R(e)$. The linearity of expectation allows us to express
$$
\mathbb{E}[\sum_{e\in\mathcal{U}'}Q(e)]=\sum_{e\in\mathcal{U}'}\mathbb{E}[Q(e)]=\sum_{e\in\mathcal{U}'}R(e).
$$

\end{proof}

        \section{\textbf{Experimental Results}}
\label{sec:exp}

We evaluate \sketch's performance from the following key aspects.
\begin{itemize}[leftmargin=*]
    \item Accuracy of approximately recording the entire data stream (\S\ref{sec:exp:point}).
    \item Identification and approximation of Top-$K$ items in a compact memory space (\S\ref{sec:exp:topk}).
    \item Influence of \sketch's parameters and data stream traits on performance and complexity (\S\ref{sec:exp:param}).
    \item Efficiency of \sketch's shrinking algorithm versus full reconstruction (\S\ref{sec:exp:shrink}).
\end{itemize}
This concise exploration confirms \sketch's robustness and adaptability.


\begin{figure*}[t]
    \centering
    \subfigure[CAIDA dataset]{
        \includegraphics[height=3.2cm]{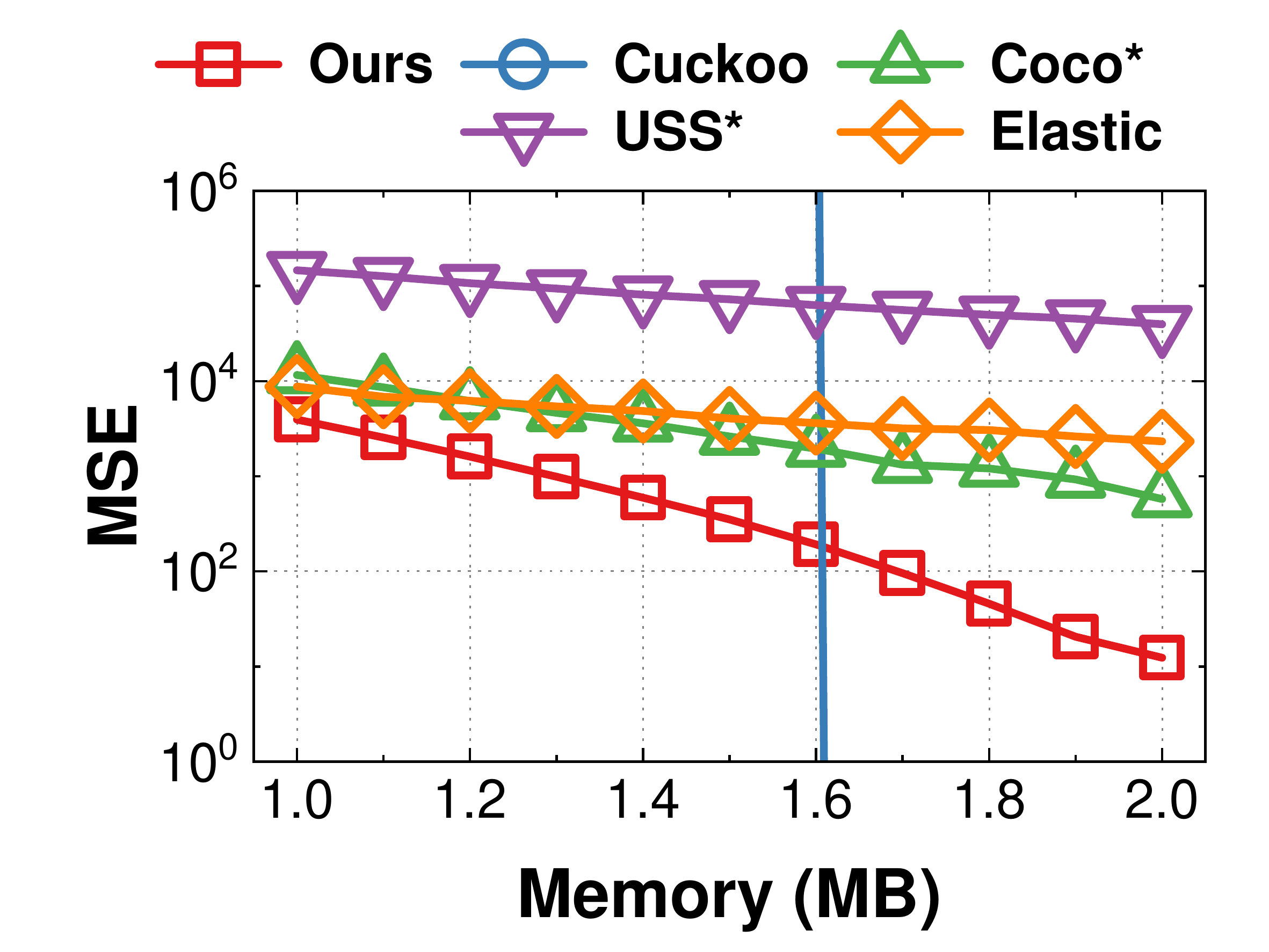}
        \label{fig:err:mse:caida}
    }
    \hspace{-0.4cm}
    \subfigure[Synthetic dataset]{
        \includegraphics[height=3.2cm]{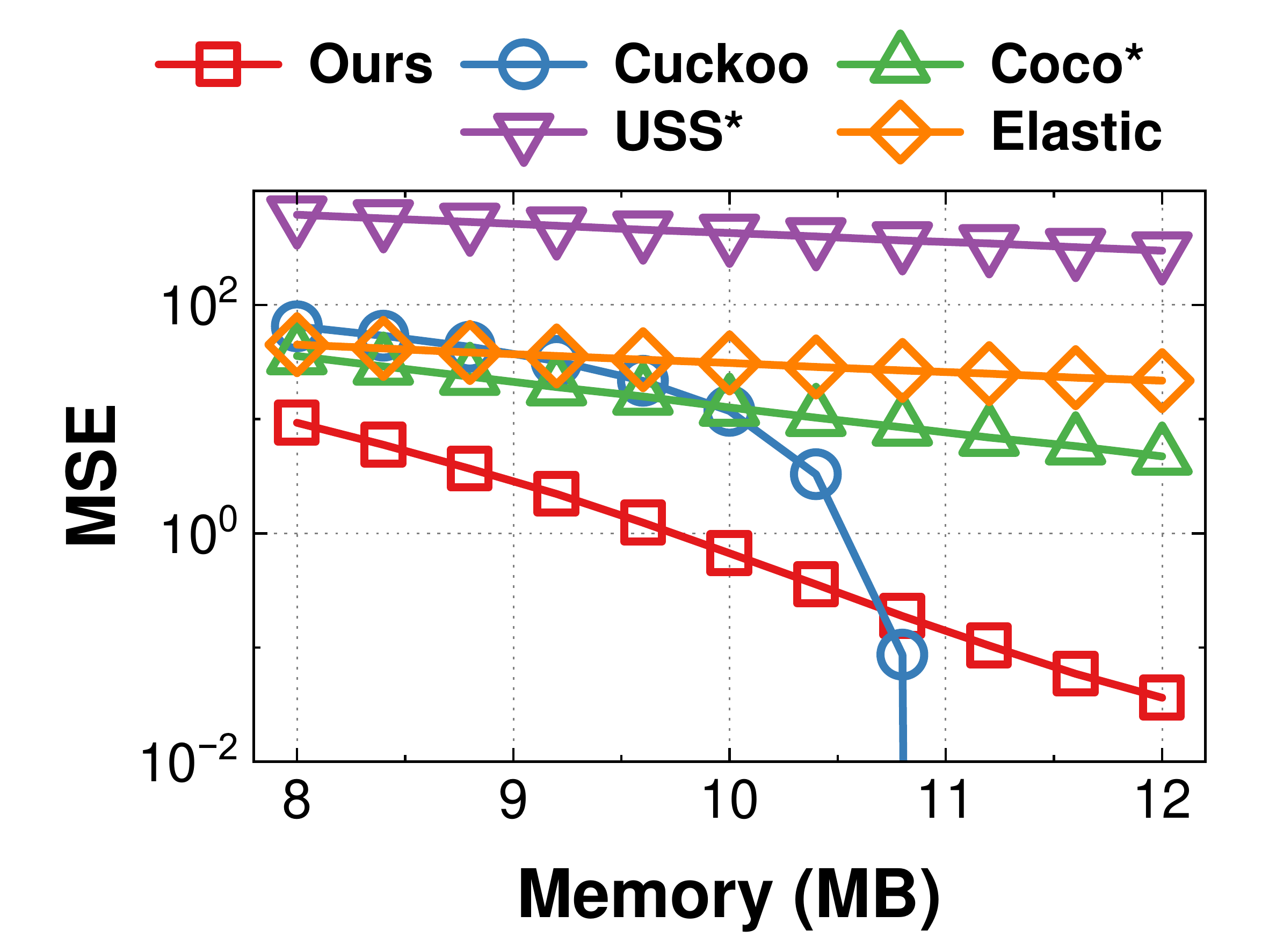}
        \label{fig:err:mse:zipf}
    }
    \hspace{-0.4cm}
    \subfigure[Webpage dataset]{
        \includegraphics[height=3.2cm]{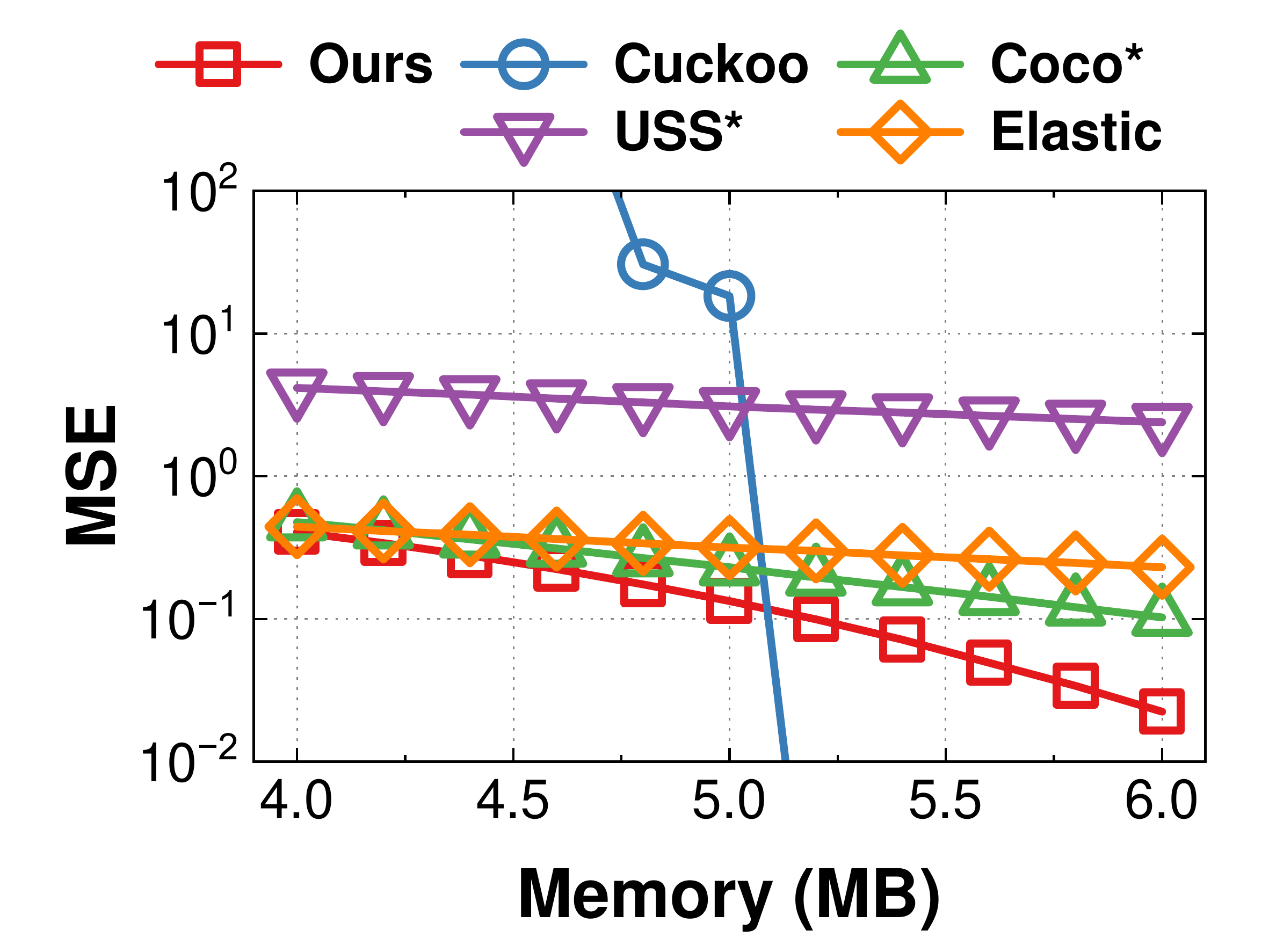}
        \label{fig:err:mse:web}
    }
    \hspace{-0.4cm}
    \subfigure[Criteo dataset]{
        \includegraphics[height=3.2cm]{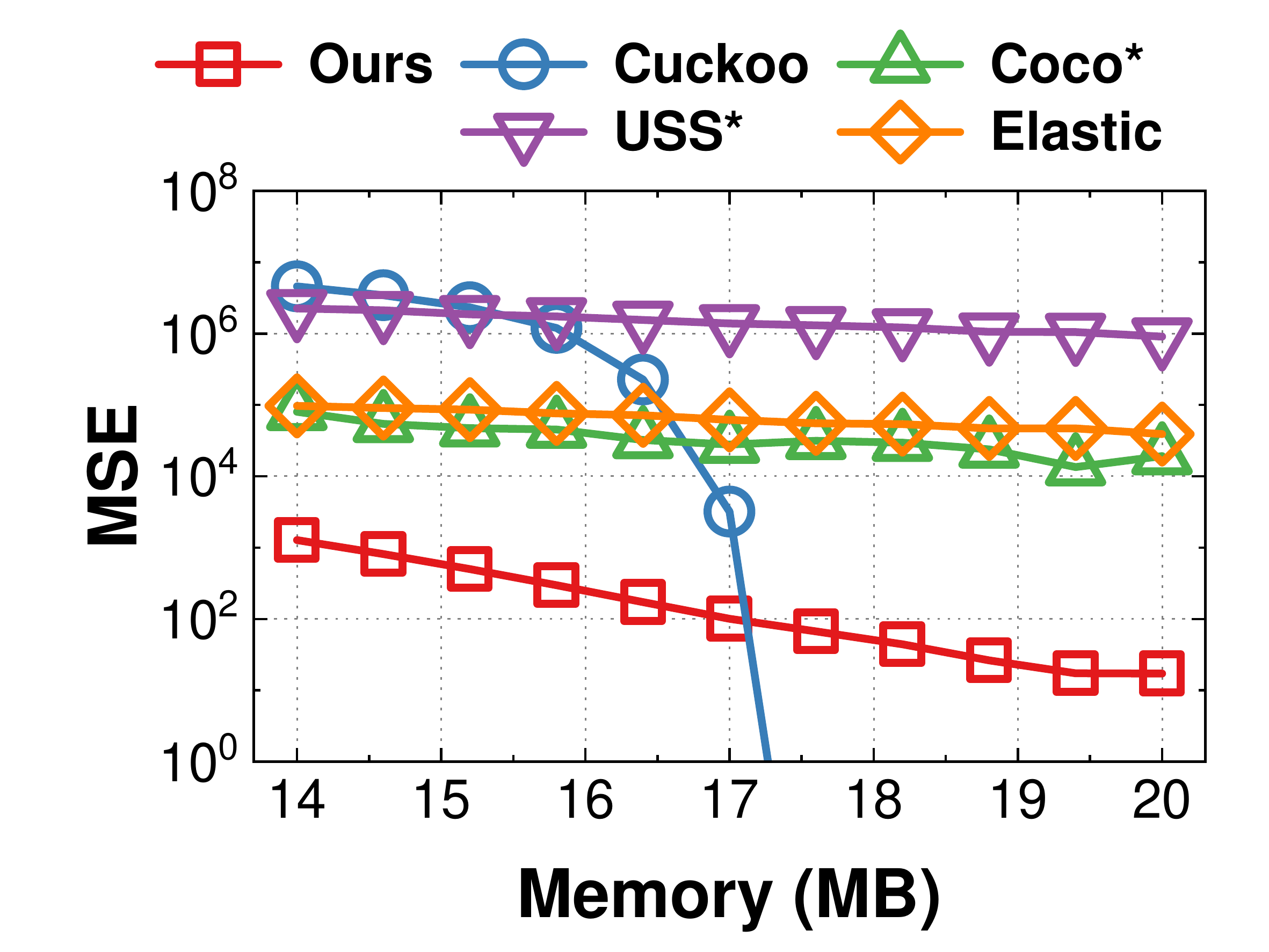}
        \label{fig:err:mse:criteo}
    }
    
    \caption{Point Query, MSE.}
    \label{fig:err:mse}
\end{figure*}

\begin{figure*}[t]
    \centering
    \subfigure[CAIDA dataset]{
        \includegraphics[height=3.2cm]{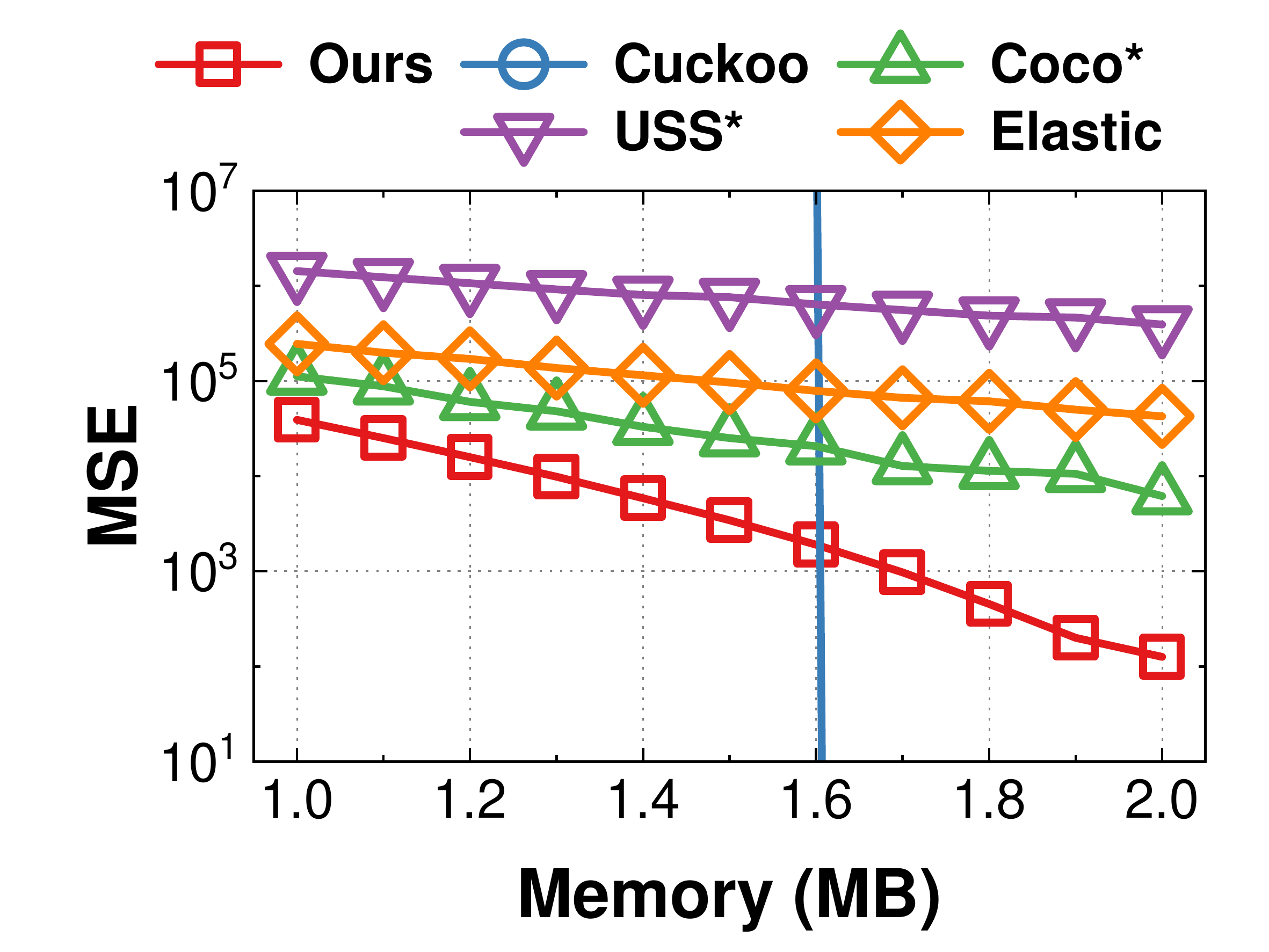}
        \label{fig:set:mse:caida}
    }
    \hspace{-0.4cm}
    \subfigure[Synthetic dataset]{
        \includegraphics[height=3.2cm]{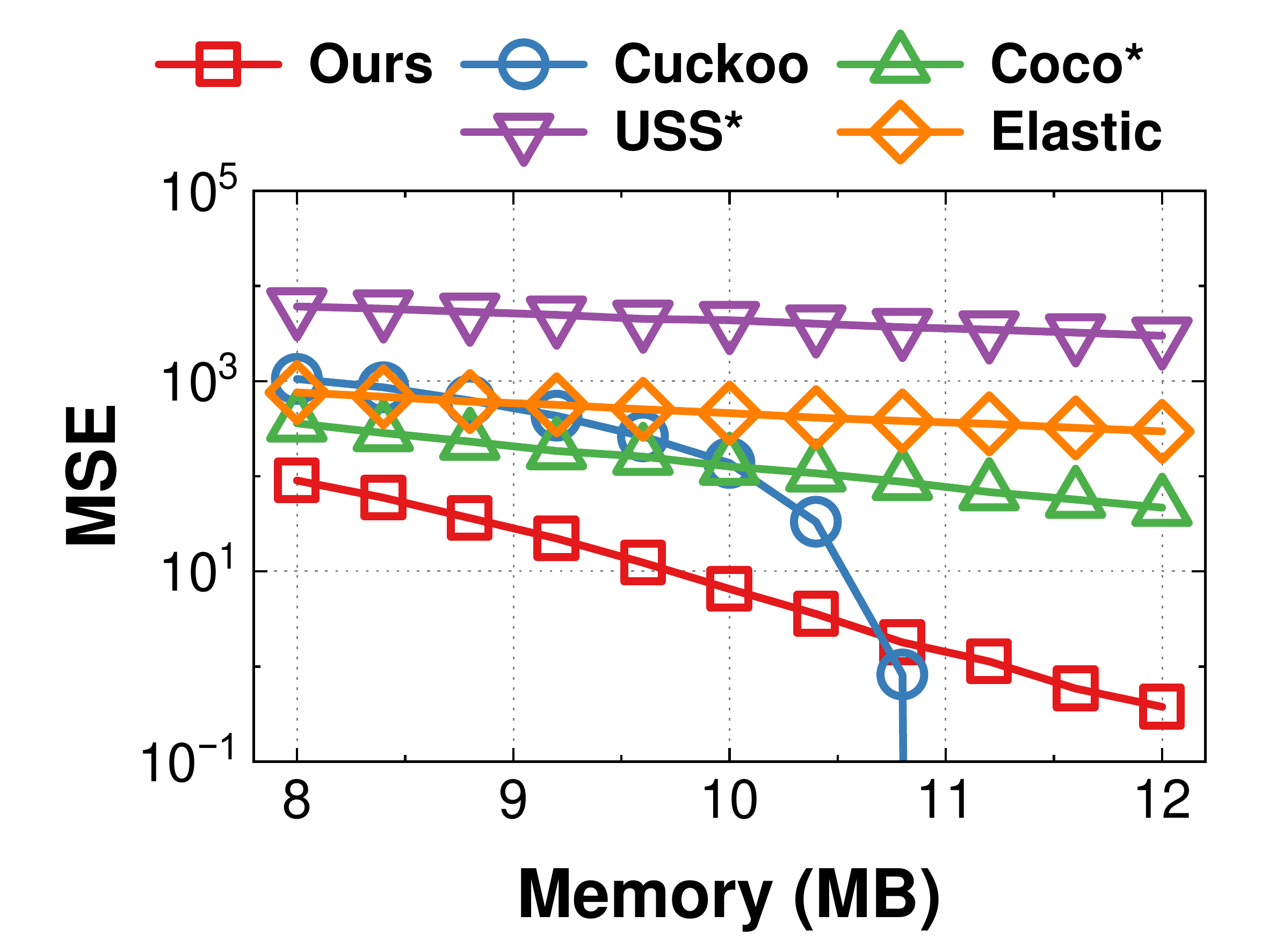}
        \label{fig:set:mse:zipf}
    }
    \hspace{-0.4cm}
    \subfigure[Webpage dataset]{
        \includegraphics[height=3.2cm]{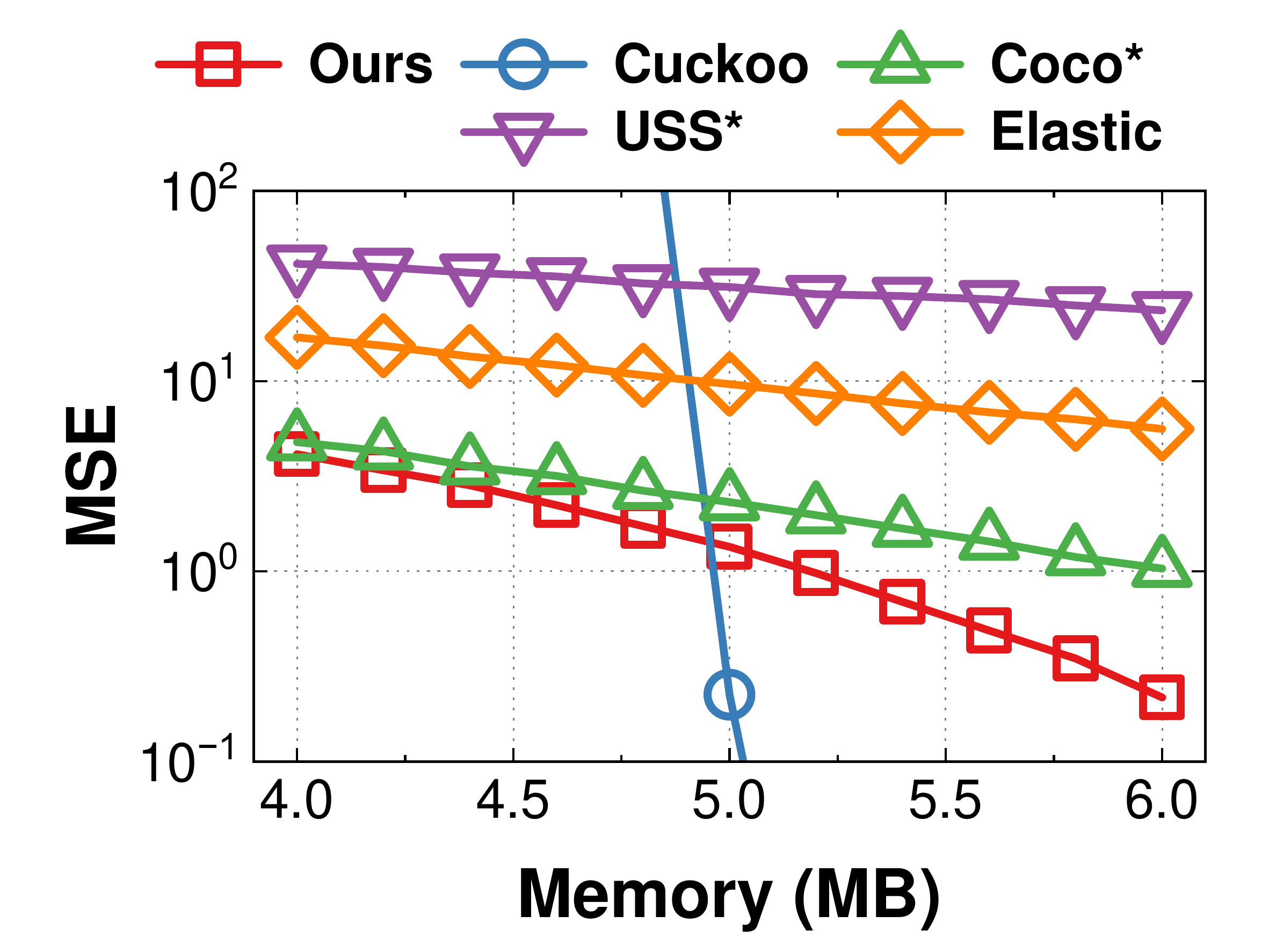}
        \label{fig:set:mse:web}
    }
    \hspace{-0.4cm}
    \subfigure[Criteo dataset]{
        \includegraphics[height=3.2cm]{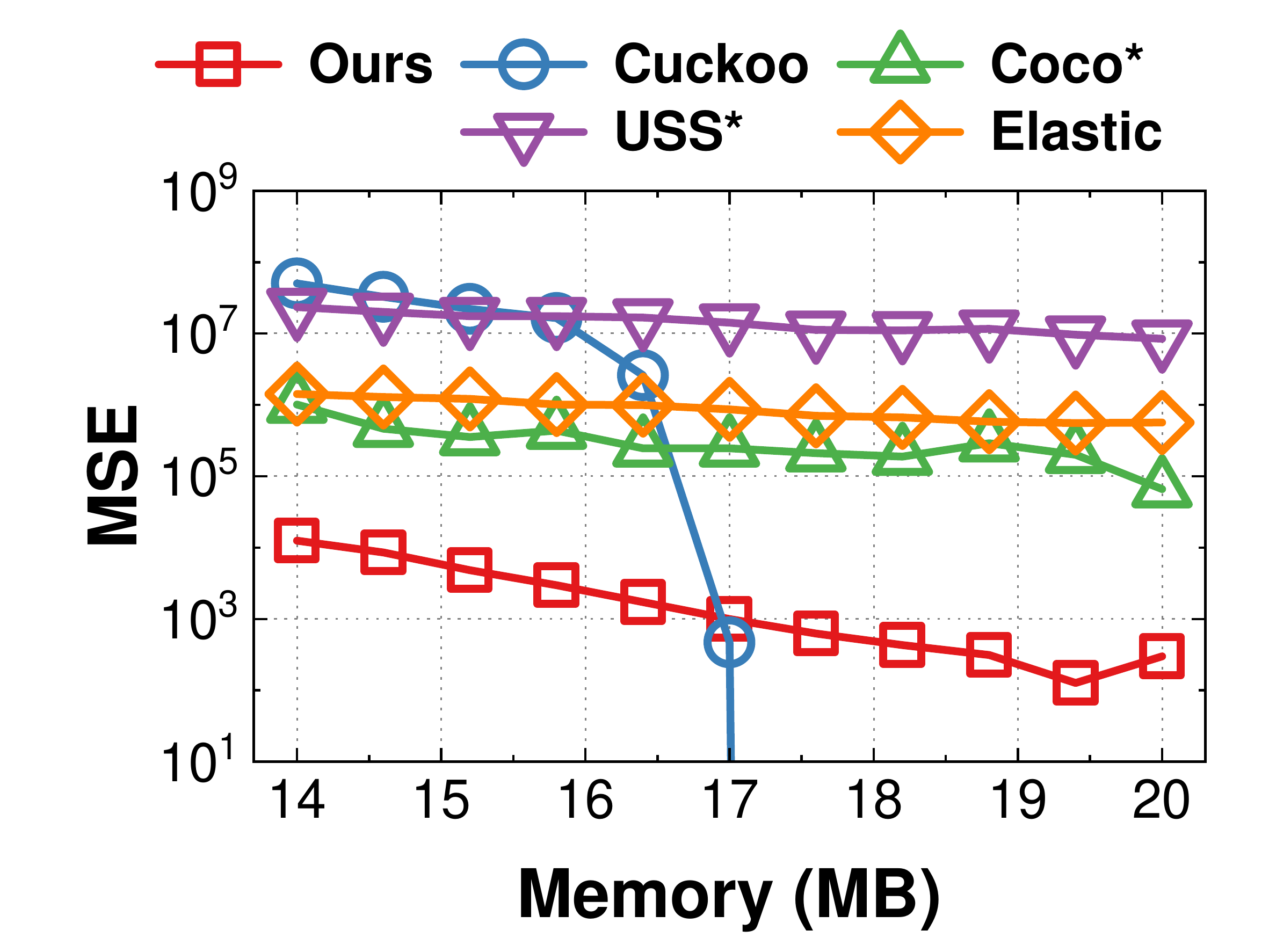}
        \label{fig:set:mse:criteo}
    }
    
    \caption{Subset Query, MSE.}
    \label{fig:set:mse}
\end{figure*}

\begin{figure}[t]
    \centering
    \subfigure[CAIDA dataset, Point Query]{
        \includegraphics[height=3.2cm]{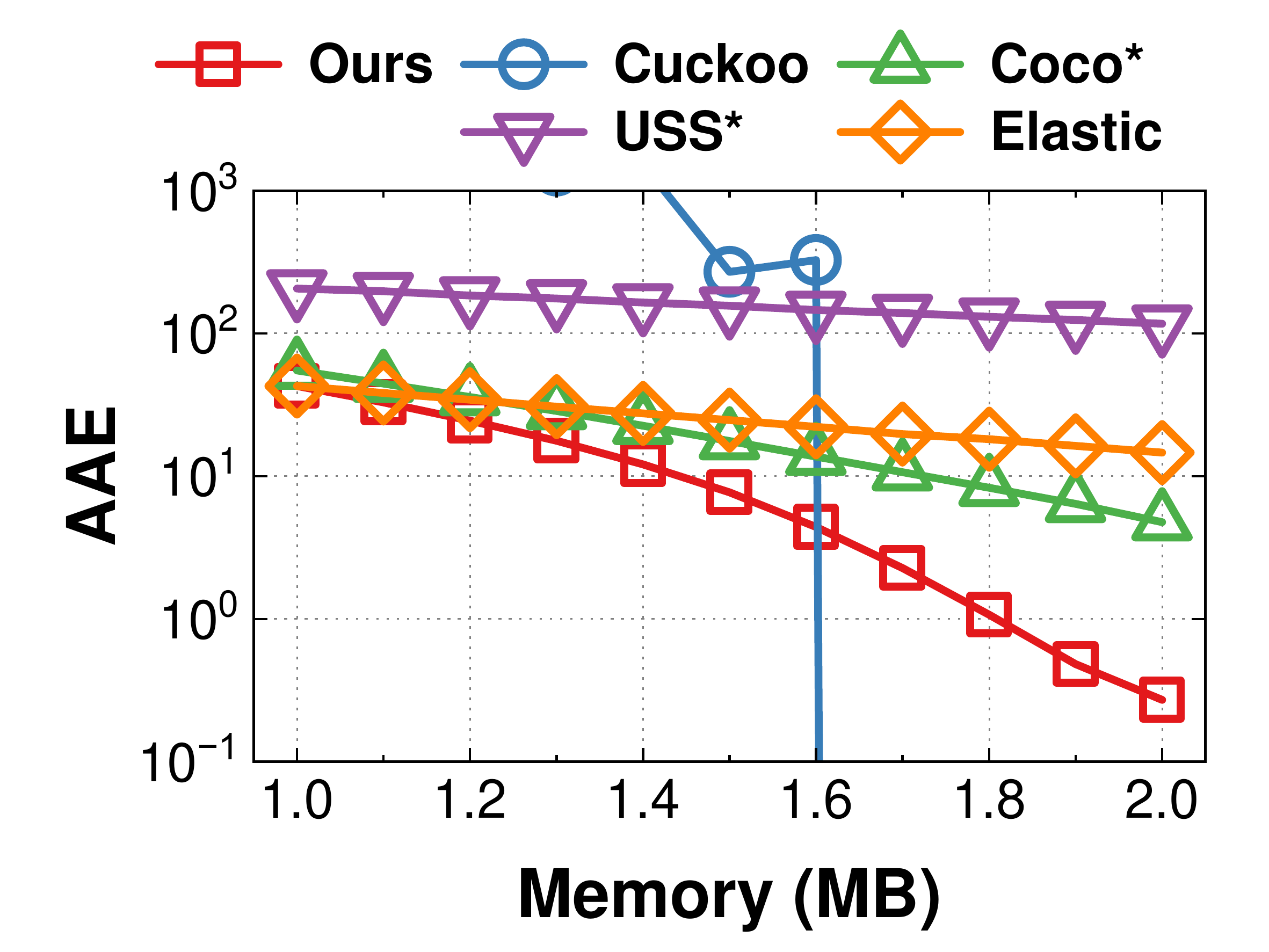}
        \label{fig:err:aae:caida}
    }
    \hspace{-0.6cm}
    \subfigure[CAIDA dataset, Subset Query]{
        \includegraphics[height=3.2cm]{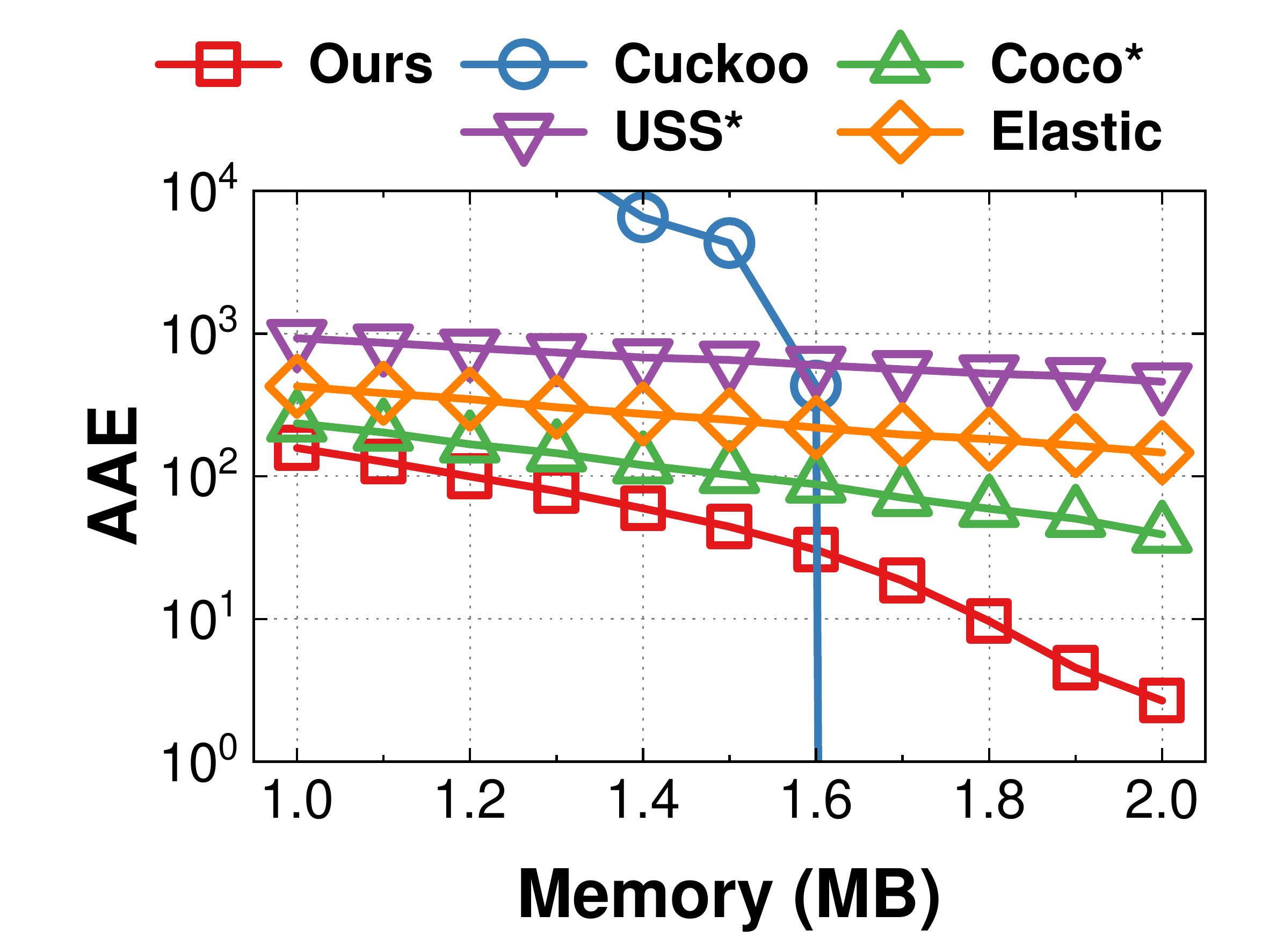}
        \label{fig:set:aae:caida}
    }
    
    \caption{Point Query \& Subset Query, AAE.}
    \label{fig:aae:caida}
\end{figure}

\begin{figure}[t]
    \centering
    \subfigure[Insertion]{
        \includegraphics[height=3.2cm]{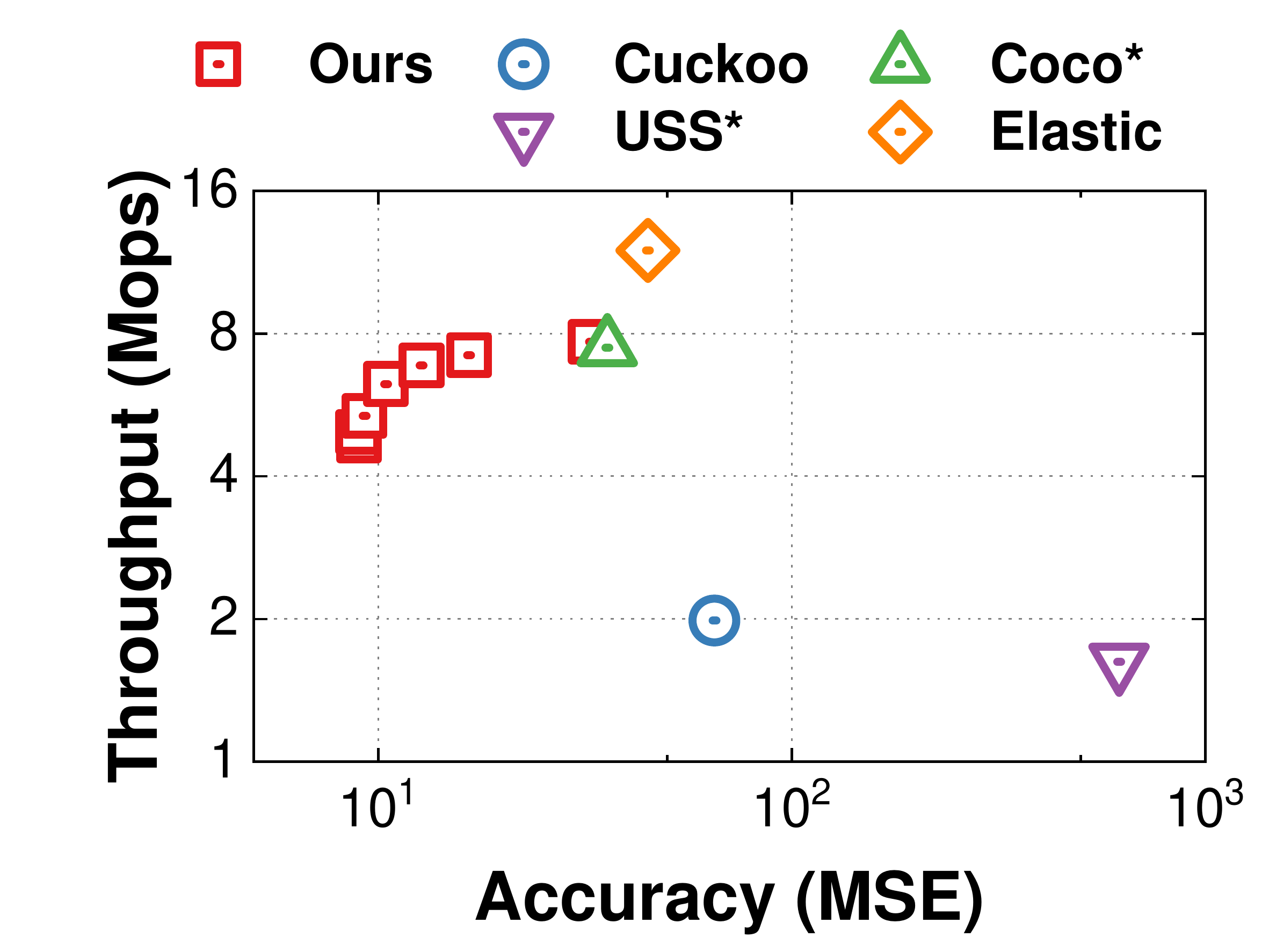}
        \label{fig:zipf:throughput:insert}
    }
    \hspace{-0.6cm}
    \subfigure[Query]{
        \includegraphics[height=3.2cm]{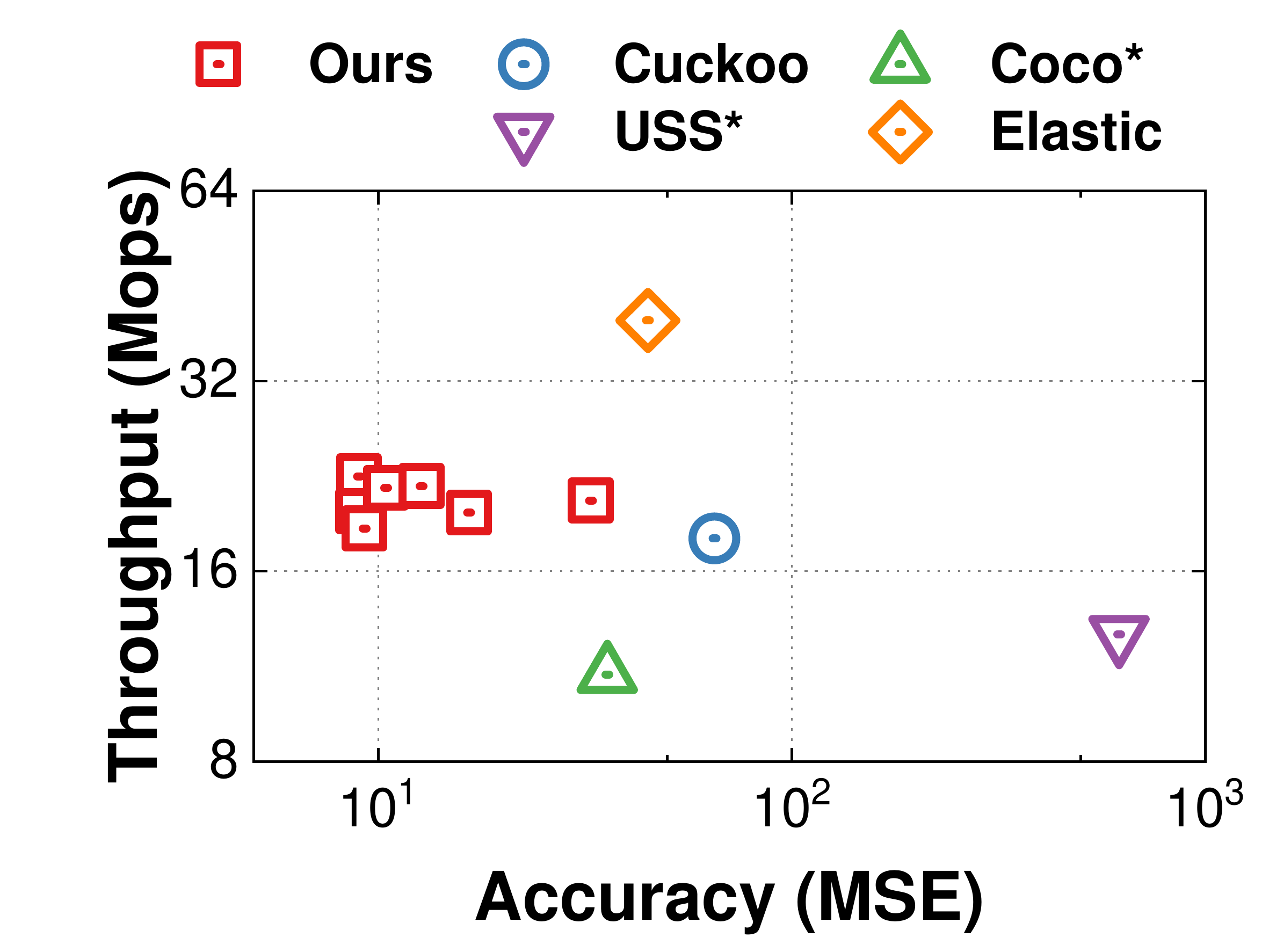}
        \label{fig:zipf:throughput:query}
    }
    
    \caption{Throughput, Synthetic dataset.}
    \label{fig:zipf:throughput}
\end{figure}

\subsection{\textbf{Experiment Setup}}

\textbf{Dataset}:
\begin{itemize}[leftmargin=*]
\item \textit{The CAIDA dataset} \cite{caida} includes anonymized network trace data over one hour. A 4-byte source IP serves as the key, coupled with a 2-byte packet size as the value. A new flowlet is distinguished if the time between packets exceeds $1ms$, indicated by a $:="$ operation; otherwise, a $+="$. For analysis, we focus on the first $10M$ items.
\item \textit{The Synthetic dataset} is generated with 4-byte keys according to a Zipfian distribution \cite{zipfian} with $\alpha=0.9$, replicating real-world long-tail distributions. Half of the dataset's items are associated with $:="$ operations with values derived from an $\mathtt{EXP}(10)$ distribution, and the other half with $+="$ operations with values from a $\mathtt{N}(0,10)$ distribution. For analysis, we focus on the first $10M$ items.
\item \textit{The Webpage dataset} \cite{webpage} records sequences of word occurrences in web documents, using a 4-byte word code as the key and frequency $1$ as the value. A new burst is recognized when words are separated by more than 1000 others, signified by a $:="$ operation; otherwise, a $+="$. For analysis, we focus on the first $10M$ items.
\item \textit{The Criteo dataset} \cite{criteo} details advertisement features and feedback spanning 24 days, utilizing a 4-byte categorical feature as the key and an average of 13 numerical features as the value. Operations are equally divided between $:="$ and $+="$. For analysis, we focus on the first $10M$ items.
\end{itemize}

\bbb{Metrics}:
The assessment of \sketch employs the following metrics:
\begin{itemize}[leftmargin=*]
    \item \textit{ARE} (Average Relative Error): Defined as the mean of the relative errors for each item key, it is calculated as 
    \(
    \frac{\sum_{e\in\mathcal{U}}\frac{|R(e)-Q(e)|}{|R(e)|}}{|\mathcal{U}|},
    \)
    where \( R(e) \) is the actual value of item \( e \), and \( Q(e) \) is the estimated value.
    
    \item \textit{AAE} (Average Absolute Error): This metric represents the average of the absolute differences between true and estimated values for all items, defined as
    \(
    \frac{\sum_{e\in\mathcal{U}}|R(e)-Q(e)|}{|\mathcal{U}|}.
    \)
    
    \item \textit{MSE} (Mean Squared Error): This measures the mean of the squared differences, formulated as
    \(
    \frac{\sum_{e\in\mathcal{U}}(R(e)-Q(e))^2}{|\mathcal{U}|}.
    \)
    
    \item \textit{Recall Rate}: Employed to evaluate the precision in identifying Top-$K$ items, the recall rate is calculated as the proportion of accurately identified Top-$K$ items in the estimated set \( \hat{\mathcal{T}} \) compared to the true set \( \mathcal{T} \), given by
    \(
    \frac{|\hat{\mathcal{T}} \cap \mathcal{T}|}{|\mathcal{T}|}.
    \)

    \item \textit{Throughput (Mops)}: A metric assessing the number of insertions or queries an algorithm can perform in a unit of time (1 second).
\end{itemize}
These metrics facilitate a thorough evaluation of \sketch's performance across different use cases and are crucial for substantiating its superiority over competing algorithms.

\bbb{Default Setting}:
We implement \sketch alongside comparison algorithms—Cuckoo hash \cite{cuckoo} (Cuckoo), Unbiased SpaceSaving \cite{uss} with \textsc{Set} (USS*), CocoSketch \cite{coco} with \textsc{Set} (Coco*), and Elastic Sketch \cite{elastic} (Elastic) in C++.
All associated code is openly available on Github \cite{github}. Our experiments are conducted on i9-10980XE CPUs under Linux, with a baseline frequency of 4GHz. For \sketch, Cuckoo, Coco*, and Elastic, the parameter $d$ is set to 4. In \sketch, $M$ is 10, and $p_\epsilon$ is 0.1. 
For \textsc{Set} updates, Coco*, USS*, and Elastic overwrite the existing value with a new value if the key is already recorded, or treat it as an \textsc{Increment} update if the key is not recorded.
Elastic only uses the heavy part. USS* is emulated with a heap and dictionary, adding four pointers per key-value pair to estimate memory overhead. For Cuckoo, we discard items that fail to insert to accommodate compact memory. We do not compare with counting sketches such as CM \cite{cmsketch}, Count \cite{csketch}, and SALSA \cite{salsa}, as they cannot be naively adapted to SIM updates.
For the synthetic dataset, our default memory setting is 8MB for point and subset queries, and 3MB for Top-$K$ queries.

\bbb{Statement}:
Our experimentation encompasses point, subset, and Top-$K$ queries across all datasets, evaluating ARE, AAE, MSE, and recall rates. While the paper presents select results, additional findings are accessible on Github \cite{github}, acknowledging space constraints.

\begin{figure*}[t]
    \centering
    \subfigure[CAIDA dataset]{
        \includegraphics[height=3.2cm]{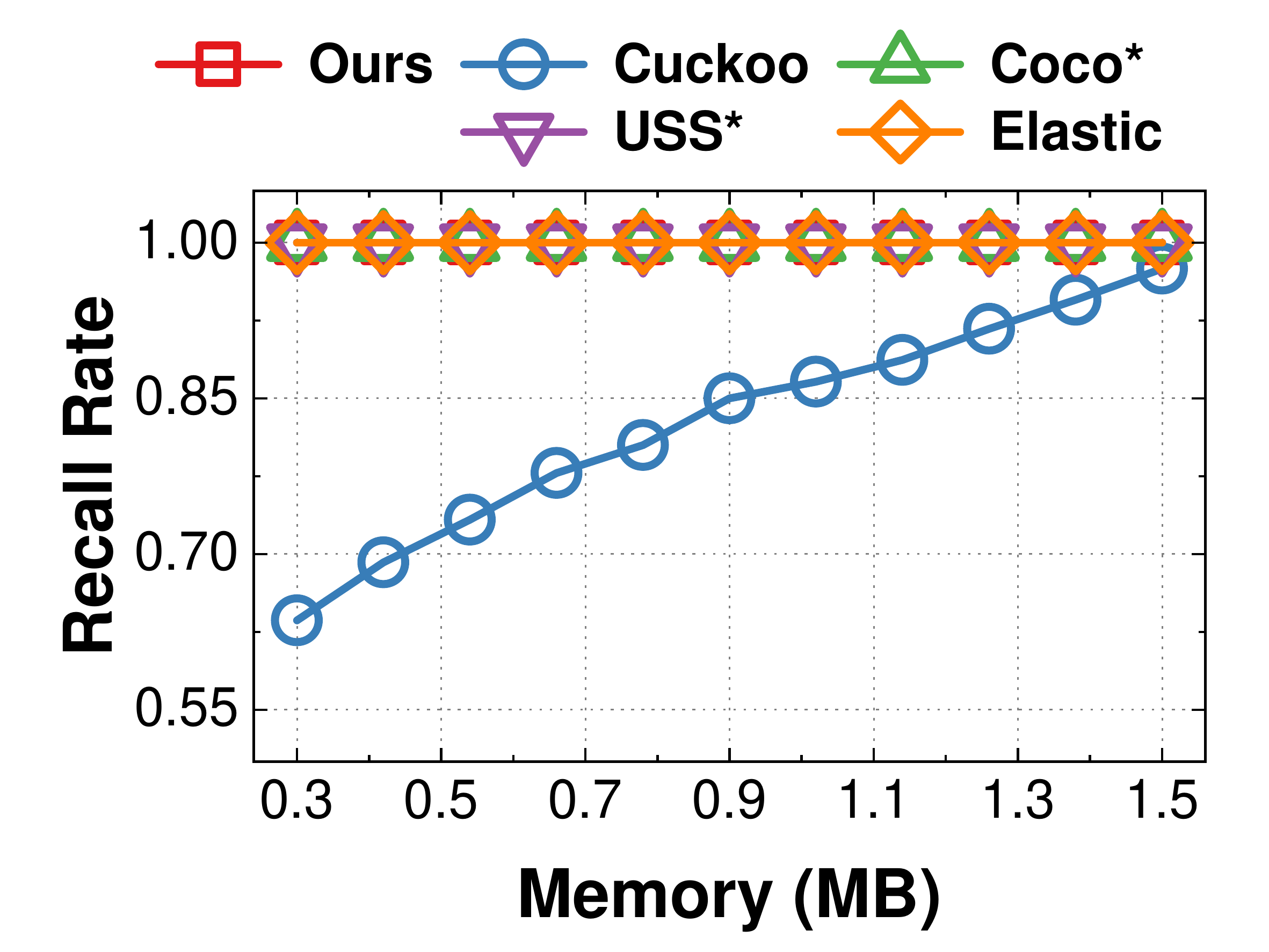}
        \label{fig:topk:recall:caida}
    }
    \hspace{-0.4cm}
    \subfigure[Synthetic dataset]{
        \includegraphics[height=3.2cm]{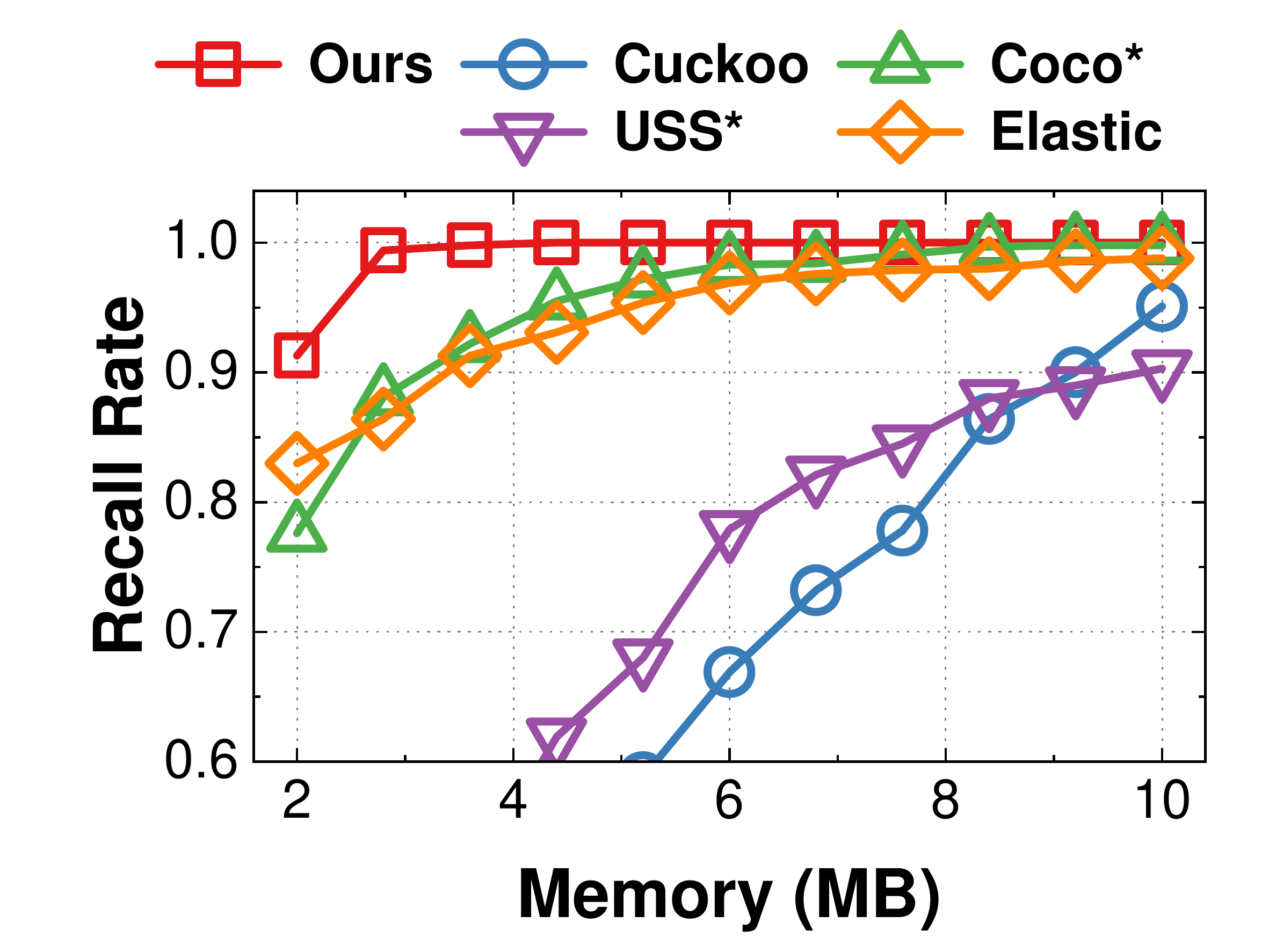}
        \label{fig:topk:recall:zipf}
    }
    \hspace{-0.4cm}
    \subfigure[Webpage dataset]{
        \includegraphics[height=3.2cm]{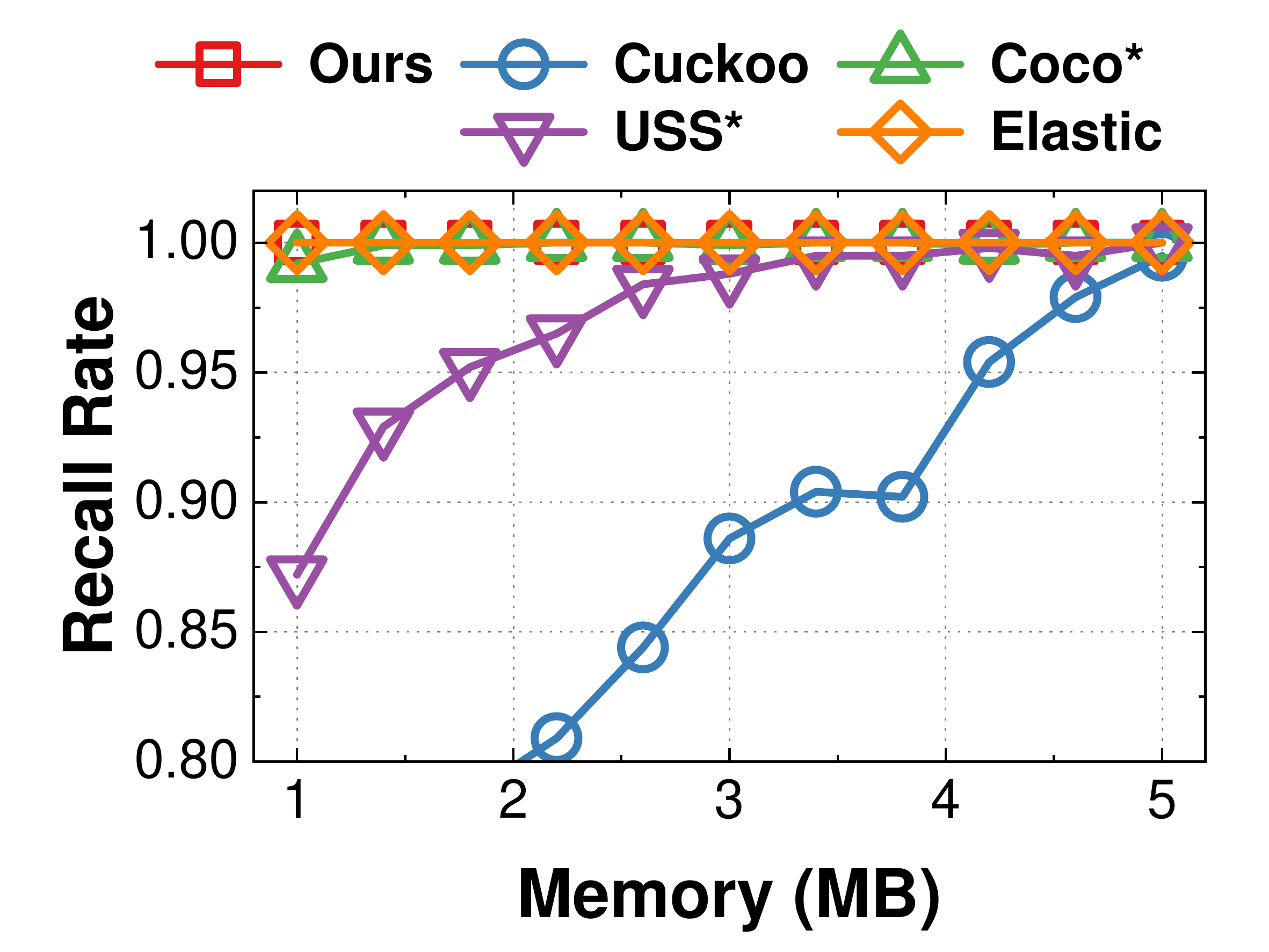}
        \label{fig:topk:recall:web}
    }
    \hspace{-0.4cm}
    \subfigure[Criteo dataset]{
        \includegraphics[height=3.2cm]{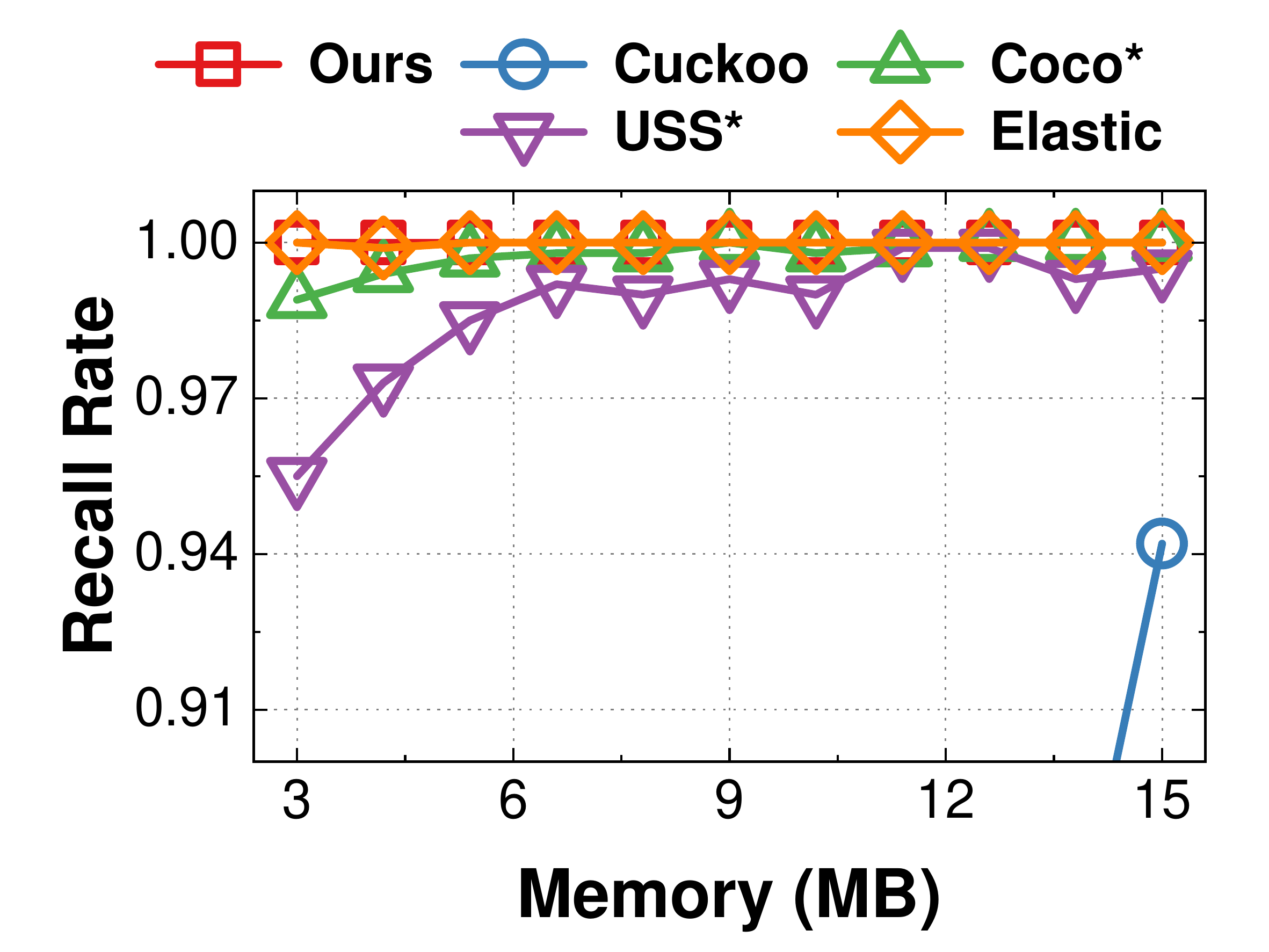}
        \label{fig:topk:recall:criteo}
    }
    
    \caption{Top-$K$ Query, Recall rate.}
    \label{fig:topk:recall}
\end{figure*}

\begin{figure*}[t]
    \centering
    \subfigure[CAIDA dataset]{
        \includegraphics[height=3.2cm]{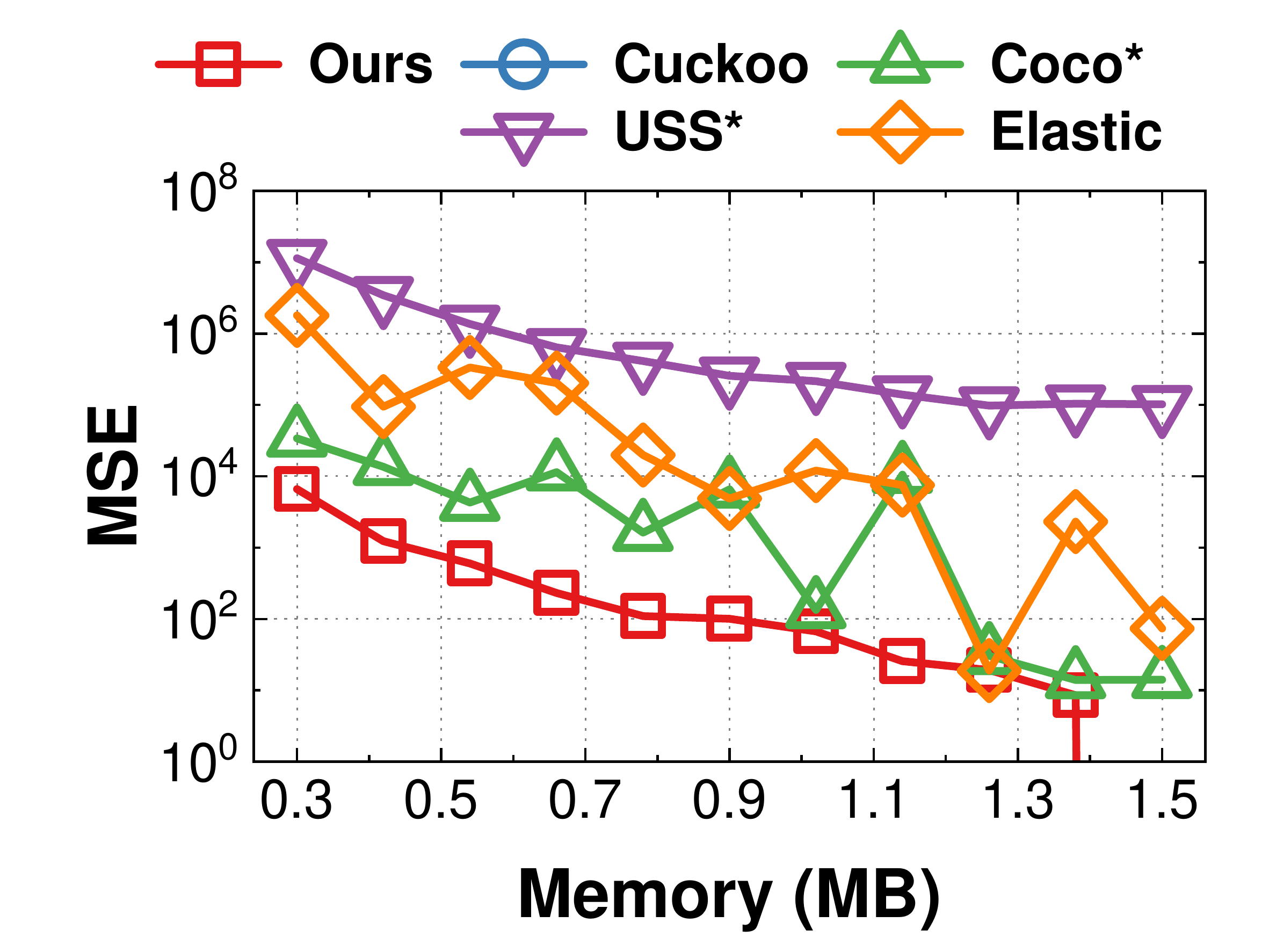}
        \label{fig:topk:mse:caida}
    }
    \hspace{-0.4cm}
    \subfigure[Synthetic dataset]{
        \includegraphics[height=3.2cm]{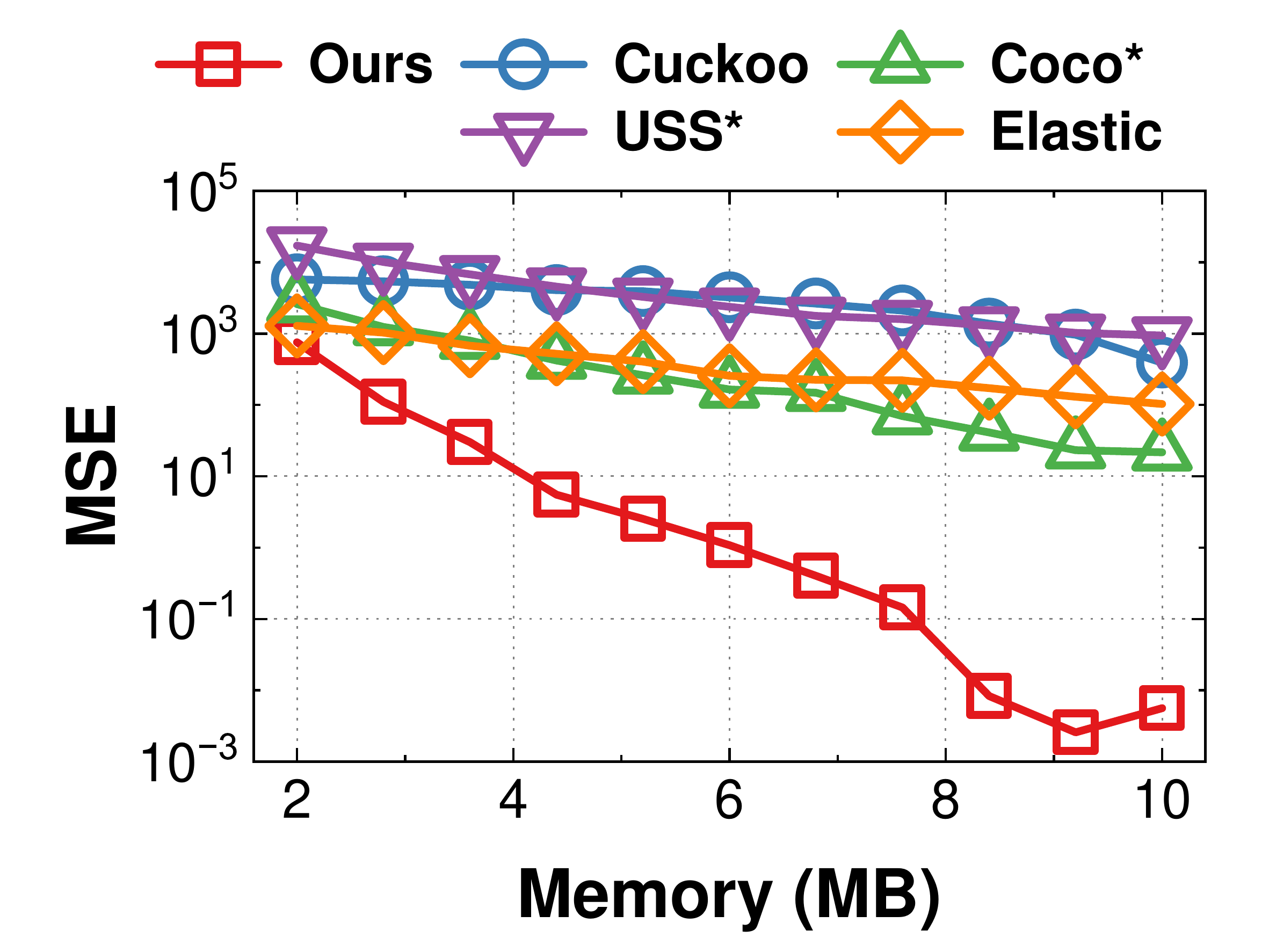}
        \label{fig:topk:mse:zipf}
    }
    \hspace{-0.4cm}
    \subfigure[Webpage dataset]{
        \includegraphics[height=3.2cm]{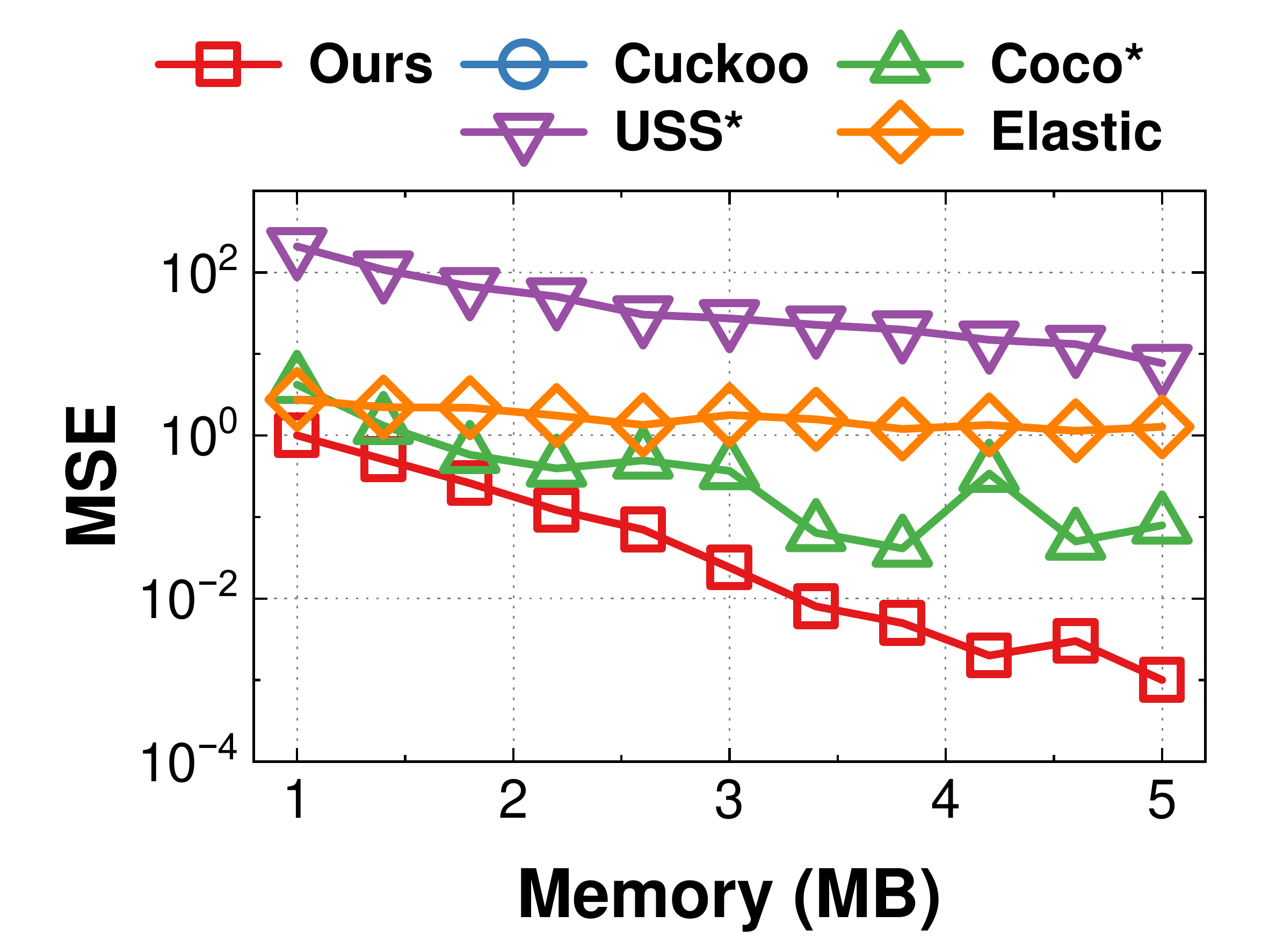}
        \label{fig:topk:mse:web}
    }
    \hspace{-0.4cm}
    \subfigure[Criteo dataset]{
        \includegraphics[height=3.2cm]{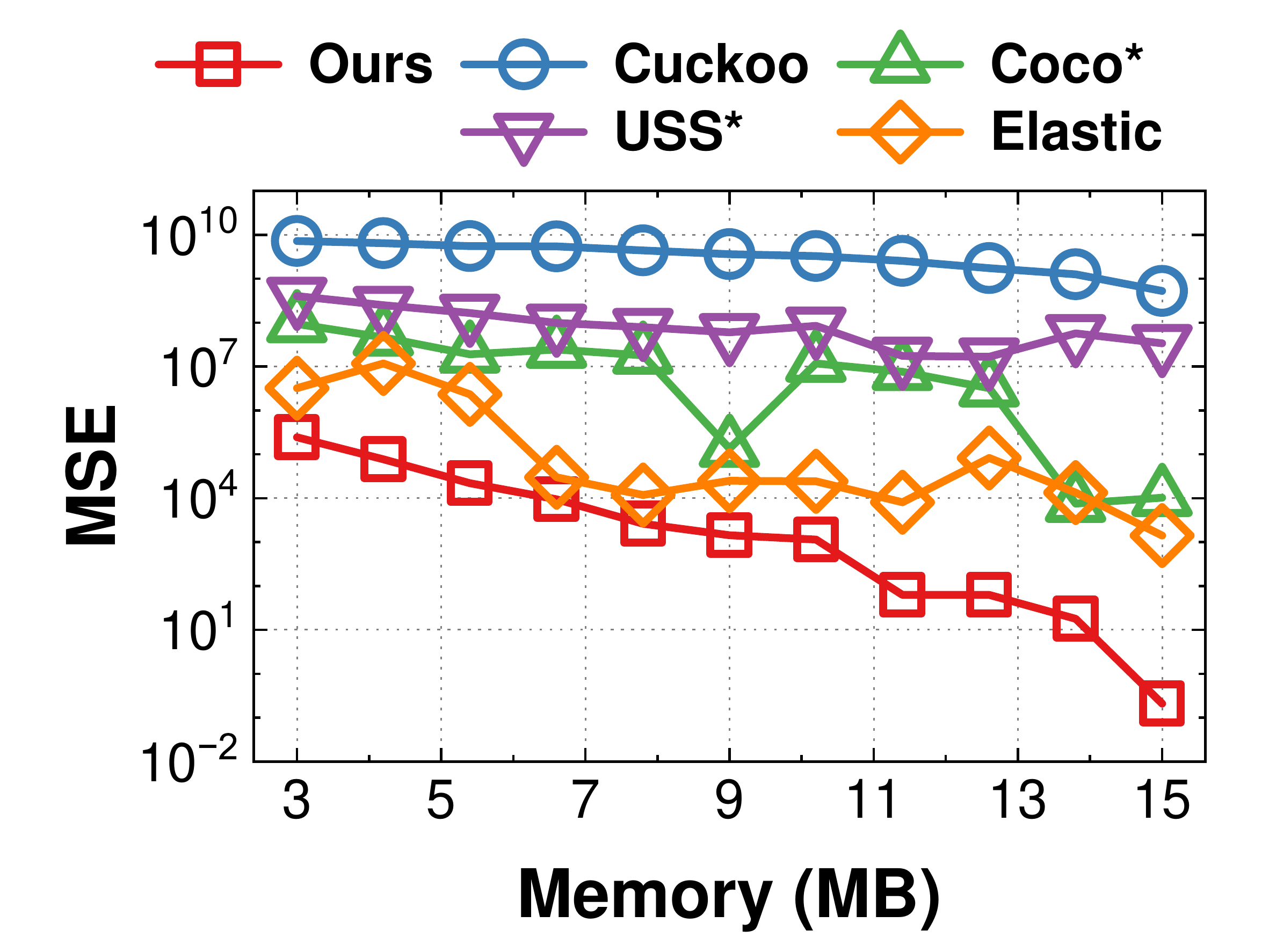}
        \label{fig:topk:mse:criteo}
    }
    
    \caption{Top-$K$ Query, MSE.}
    \label{fig:topk:mse}
\end{figure*}

\begin{figure}[t]
    \centering
    \subfigure[CAIDA dataset, ARE]{
        \includegraphics[height=3.2cm]{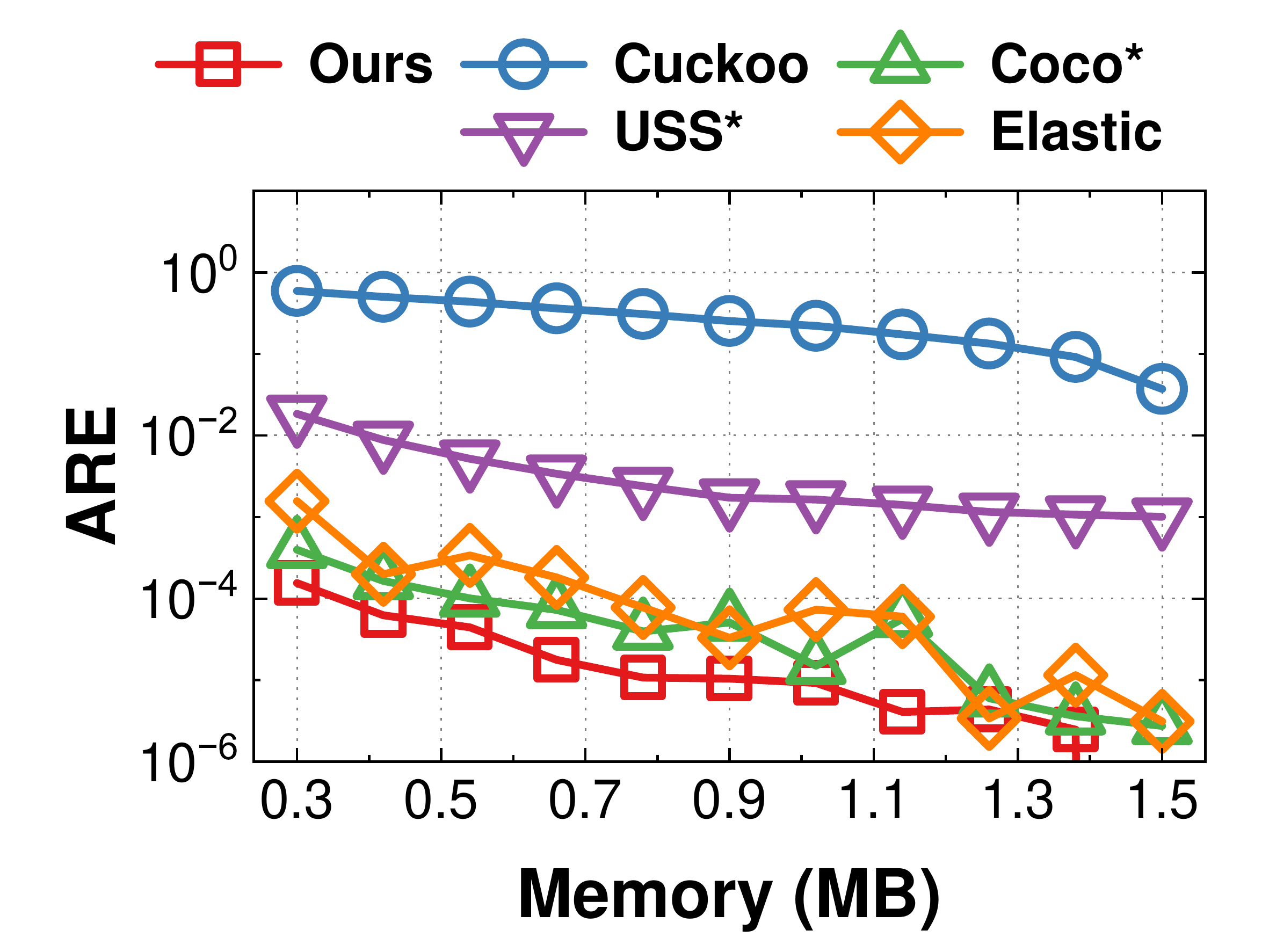}
        \label{fig:topk:are:caida}
    }
    \hspace{-0.6cm}
    \subfigure[CAIDA dataset, AAE]{
        \includegraphics[height=3.2cm]{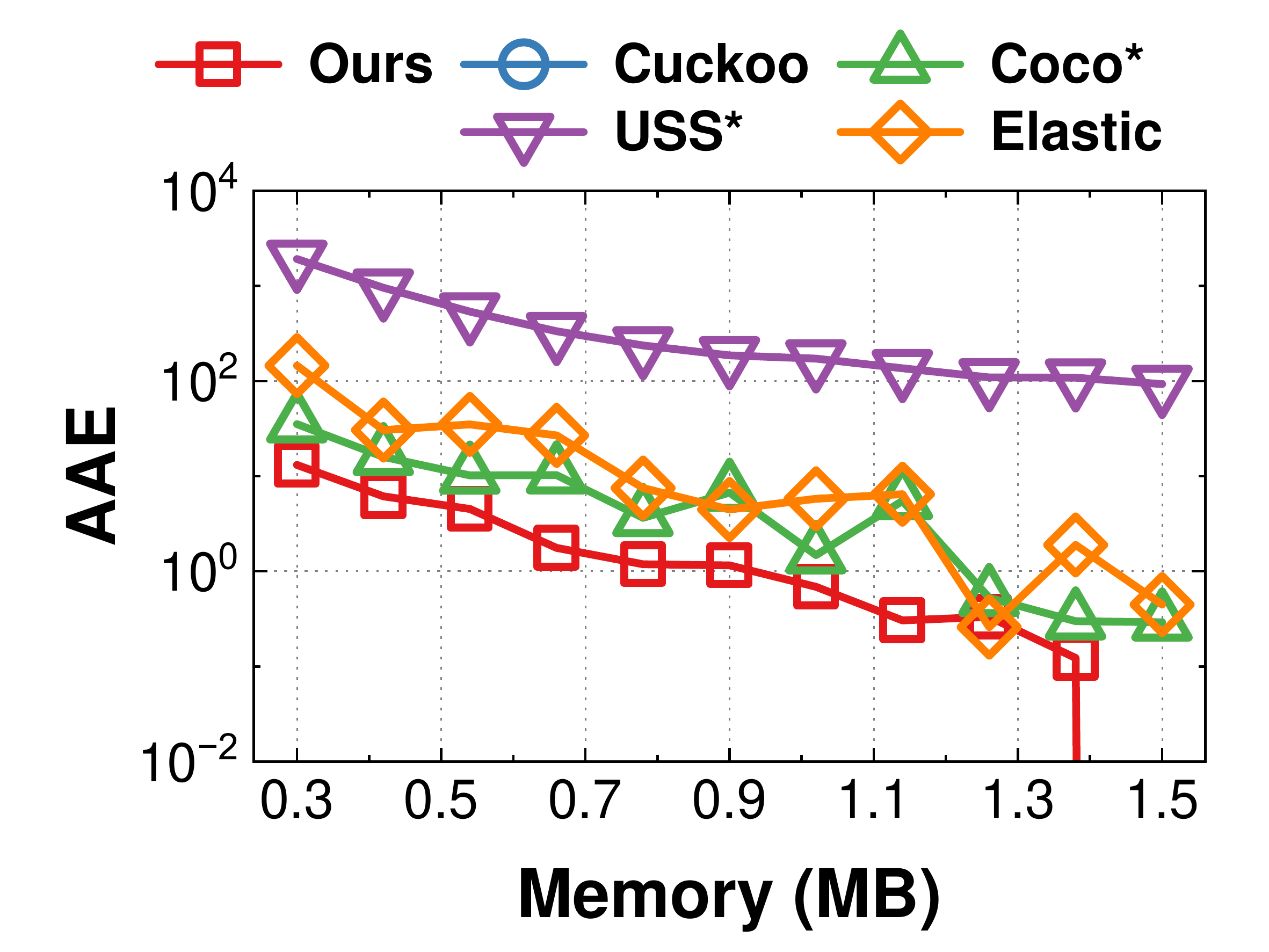}
        \label{fig:topk:aae:caida}
    }
    
    \caption{Top-$K$ Query, ARE \& AAE.}
    \label{fig:topk:caida}
\end{figure}

\subsection{\textbf{Experiments on Point and Subset Query}}
\label{sec:exp:point}

In this section, we vary the memory consumption of different algorithms, evaluating its performance in point and subset query tasks. We ensure that the memory consumption of the comparison algorithms is on par with \sketch. For different datasets, we allocate memory based on the number of distinct items they contain\footnote{Synthetic dataset: 898935, CAIDA dataset: 137796, Webpage dataset: 430831, Criteo dataset: 1403213.}.
For point queries, we interrogate all items that have appeared in the data stream; for subset queries, we randomly generate 10,000 sets, each containing 10 items, for our queries. Conclusion draw from other construction methods are similar.

\bbb{MSE for Point Queries (Figure \ref{fig:err:mse}):}
As depicted in Figures \ref{fig:err:mse:caida} to \ref{fig:err:mse:criteo}, within point query tasks, \sketch consistently exhibits lower MSE than Coco*, Elastic, and USS*, with reductions of up to $1.11\times10^3$-fold compared to Coco* ($1.22\times10^2$ on average), up to $2.69\times10^3$-fold compared to Elastic ($2.83\times10^2$ on average), and up to $6.06\times10^4$-fold compared to USS* ($6.08\times10^3$ on average).
Cuckoo's MSE remains at 0 when the load factor is at or below 94\%\footnote{A 94\% load factor corresponds different memory sizes across various datasets, specifically, Synthetic dataset: $10.9$MB, CAIDA dataset: $1.7$MB, Webpage dataset: $5.3$MB, Criteo dataset: $17.1$MB}, due to its ability to handle a load factor of slightly above 94\%. However, as the load factor surpasses 94\%, Cuckoo's MSE escalates dramatically, with \sketch demonstrating an MSE up to $2.08\times 10^8$ times lower.
%

\bbb{AAE for Point Queries (Figure \ref{fig:aae:caida}):}
Figure \ref{fig:err:aae:caida} shows that in point query evaluations, \sketch's AAE remains below that of Coco*, Elastic, and USS*, potentially offering up to $17.6$ times reduction compared to Coco* ($5.11$ on average), up to $59.1$ times to Elastic ($11.8$ on average), and up to $4.33\times10^2$ times to USS* ($88.1$ on average).
Once the load factor exceeds 94\%, Cuckoo's AAE rises steeply, whereas \sketch's AAE is at most $1.63\times10^2$ times lower.
%

\bbb{Throughput for Point Queries (Figure \ref{fig:zipf:throughput}):}
We set the parameter $M$ of \sketch to 0, 1, 2, 4, 10, 20, and 30, respectively, and compare their insertion and query throughput with those of the comparison algorithms. \sketch's insertion throughput is higher than that of Cuckoo ($1.98$Mops) and USS* ($1.47$Mops), slightly lower than Coco* ($7.46$Mops) and Elastic ($12.0$Mops), and is affected by the parameter $M$.
Specifically, with $M=30$, \sketch will have the highest accuracy and the lowest throughput ($4.75$Mops); with $M=0$, the cascading overflow technique will not be used, and \sketch will have the lowest accuracy and highest throughput ($7.68$Mops). 
With $M=0$, \sketch and Coco* are very similar in terms of accuracy and throughput, because the operations of \sketch with $M=0$ and Coco* are very similar. Given that the accuracy at $M=10$,$20$, and $30$ is similar, we recommend using $M=10$ by default.
\sketch's query throughput ($20.7$Mops on average) is higher than Coco* ($11.0$Mops) and USS* ($12.7$Mops), similar to Cuckoo ($18.0$Mops), lower than Elastic ($39.9$Mops), and is not affected by the parameter $M$\footnote{Although there are some differences in query throughput when $M$ varies, we attribute these irregular differences to the randomness of CPU and cache performance.}. The query operations of \sketch are consistent with those of Cuckoo, thus their throughput is essentially identical.

\bbb{MSE for Subset Queries (Figure \ref{fig:set:mse}):}
Illustrated in Figures \ref{fig:err:mse:caida} to \ref{fig:err:mse:criteo}, \sketch's MSE in subset query tasks always undercuts those of Coco*, Elastic, and USS*, with a diminution of up to $1.57\times10^3$-fold compared to Coco* ($1.10\times10^2$ on average), up to $4.39\times10^3$-fold compared to Elastic ($3.71\times10^2$ on average), and up to $7.55\times10^4$-fold compared to USS* ($5.73\times10^3$ on average).
Beyond a 94\% load factor, Cuckoo's MSE surges sharply, while \sketch's MSE remains up to $2.35\times10^8$ times lower.
%

\bbb{AAE for Subset Queries (Figure \ref{fig:aae:caida}):}
As shown in Figure \ref{fig:set:aae:caida}, for subset query tasks, \sketch consistently achieves a lower AAE than Coco*, Elastic, and USS*, with up to $171$ times smaller values compared to them ($19.5$ on average).
Cuckoo's AAE increases significantly after the load factor crosses 94\%, whereas \sketch manages an AAE up to $3.80\times10^2$ times smaller.

\subsection{\textbf{Experiments on Top-K Query}}
\label{sec:exp:topk}

In this section, we adjust the memory consumption of different algorithms, assessing its accuracy in Top-$K$ query tasks. The memory usage of comparison algorithms is aligned with that of \sketch. By default, we set $K=1000$.

\bbb{Recall for Top-$K$ Queries (Figure \ref{fig:topk:recall}):}
Figures \ref{fig:topk:recall:caida} to \ref{fig:topk:recall:criteo} illustrate that in Top-$K$ query tasks, \sketch, Coco*, and Elastic maintain the highest recall rates. Conversely, the recall rates for Cuckoo and USS* diminish significantly with increasing load factors. Specifically, on the Synthetic dataset, notable disparities in recall rates among \sketch, Coco*, and Elastic are observed: \sketch attains a recall rate up to $13.7$\% higher than Coco* ($4.07$\% on average) and up to $13$\% more than Elastic ($4.86$\% on average). Overall, the recall rate of \sketch consistently outperforms and remains near 100\%.

\bbb{MSE for Top-$K$ Queries (Figure \ref{fig:topk:mse}):}
As depicted in Figures \ref{fig:topk:mse:caida} to \ref{fig:topk:mse:criteo}, within Top-$K$ query tasks, \sketch's MSE invariably falls below that of Coco*, Elastic, USS*, and Cuckoo, with reductions of up to $1.22\times10^5$-fold compared to Coco* ($6.15\times10^3$ on average), up to $5.05\times10^4$-fold compared to Elastic ($2.49\times10^3$ on average), up to $1.62\times10^8$-fold compared to USS* ($3.89\times10^3$ on average), and up to $8.58\times10^{11}$-fold compared to Cuckoo ($6.06\times10^{10}$ on average).
The MSE values for Cuckoo are excessively high and are therefore not presented in the Figure \ref{fig:topk:mse:caida} and \ref{fig:topk:mse:web}.

\bbb{AAE \& ARE for Top-$K$ Queries (Figure \ref{fig:topk:caida}):}
Figures \ref{fig:topk:are:caida} and \ref{fig:topk:aae:caida} show that in Top-$K$ query tasks, the ARE and AAE for \sketch are invariably lower than those for Coco*, Elastic, USS*, and Cuckoo, with \sketch's ARE and AAE being up to $18.0$ times ($4.25$ on average), up to $21.2$ times ($8.24$ on average), up to $8.84\times10^2$ times  ($2.53\times10^2$ on average), and up to $8.94\times10^5$ times ($2.52\times10^5$ on average) smaller than those of Coco*, Elastic, USS*, and Cuckoo, respectively.

\begin{figure}[t]
    \centering
    \subfigure[Parameter $M$]{
    \includegraphics[height=3.2cm]{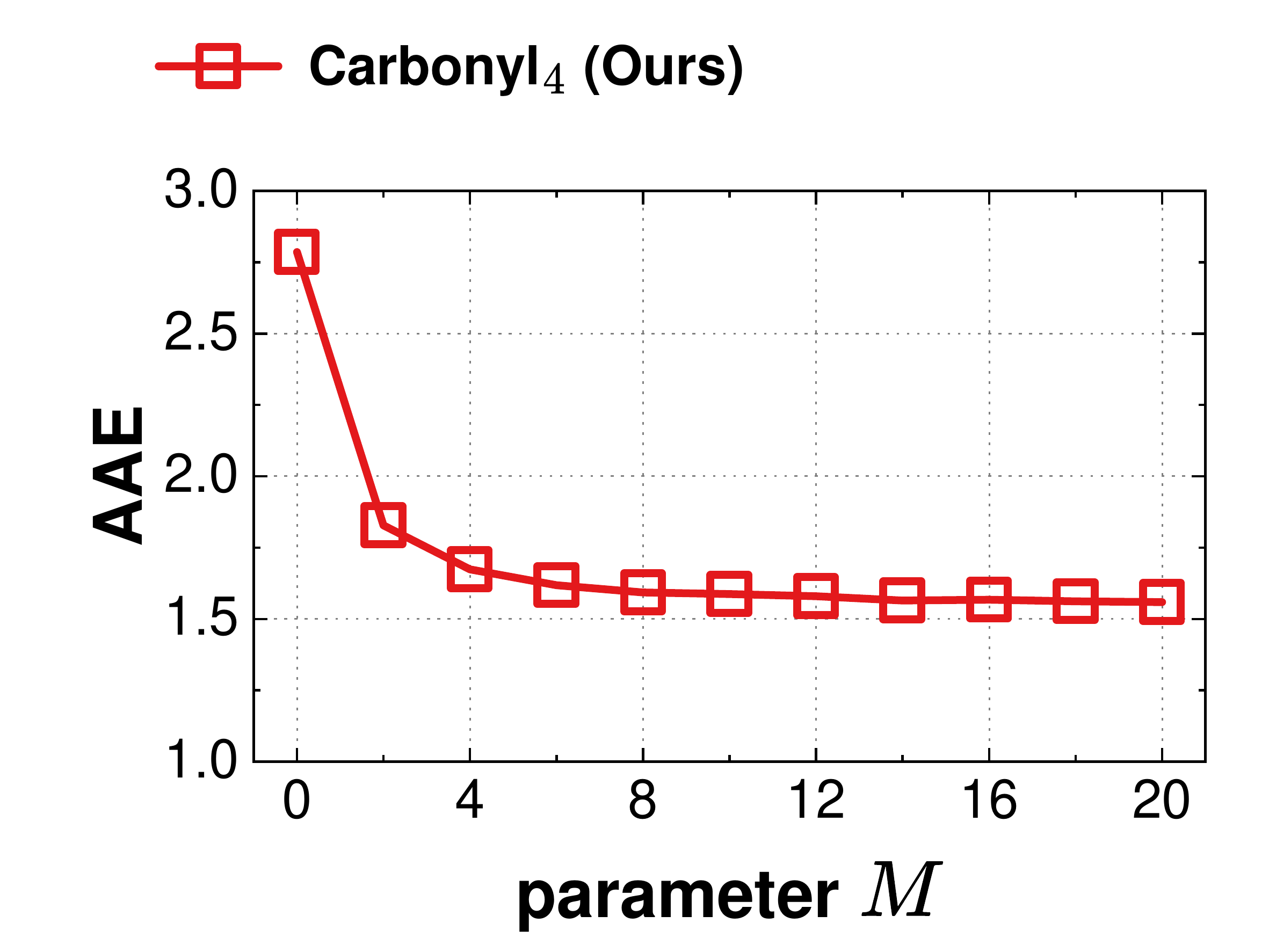}
        \label{fig:zipf:param_m:p_aae}
    }
    \hspace{-0.6cm}
    \subfigure[Parameter $M$]{
    \includegraphics[height=3.2cm]{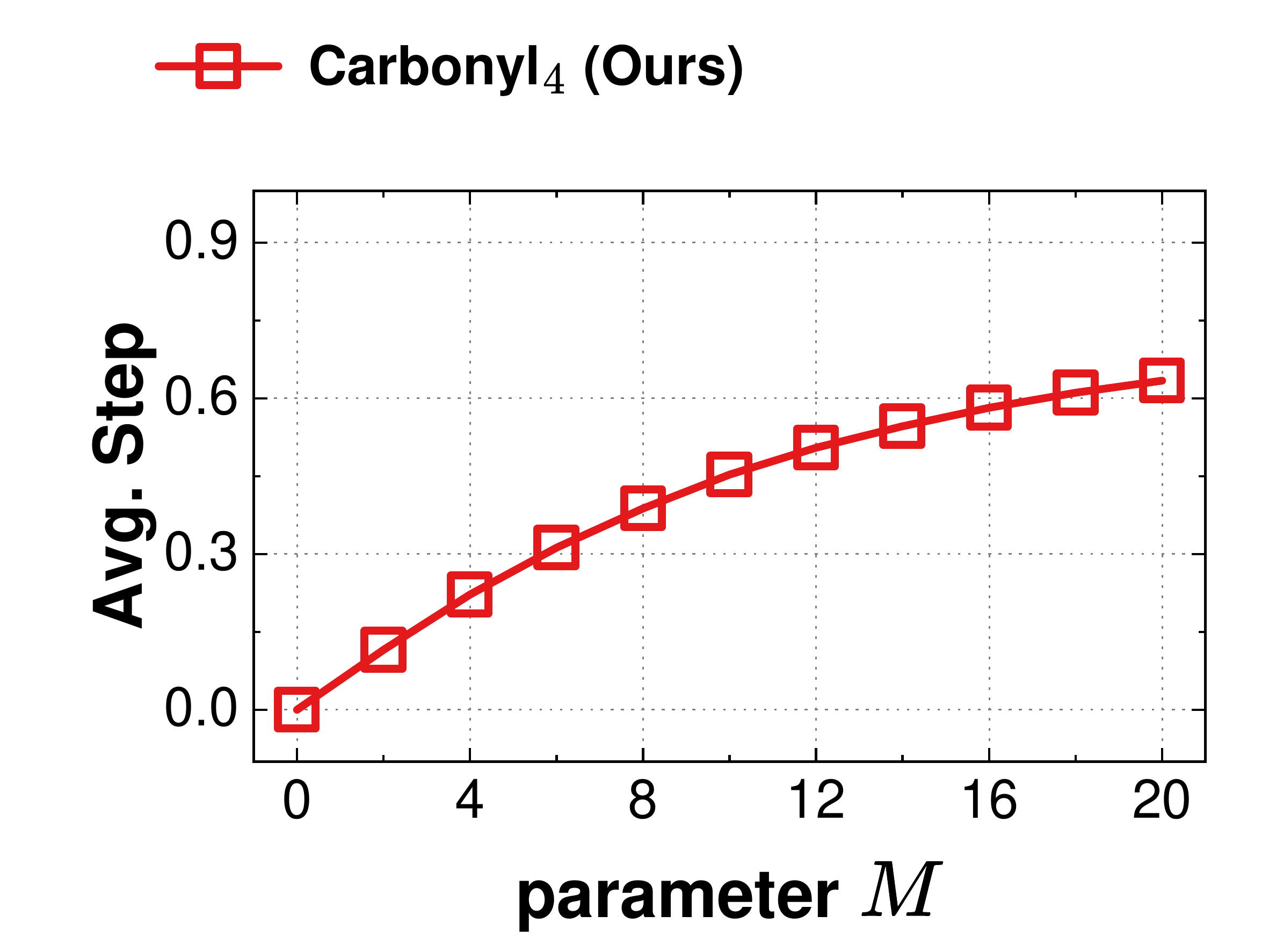}
        \label{fig:zipf:param_m:p_kick}
    }
    
    \vspace{-0.3cm}
    
    \subfigure[Parameter $p_\epsilon$]{
    \includegraphics[height=3.2cm]{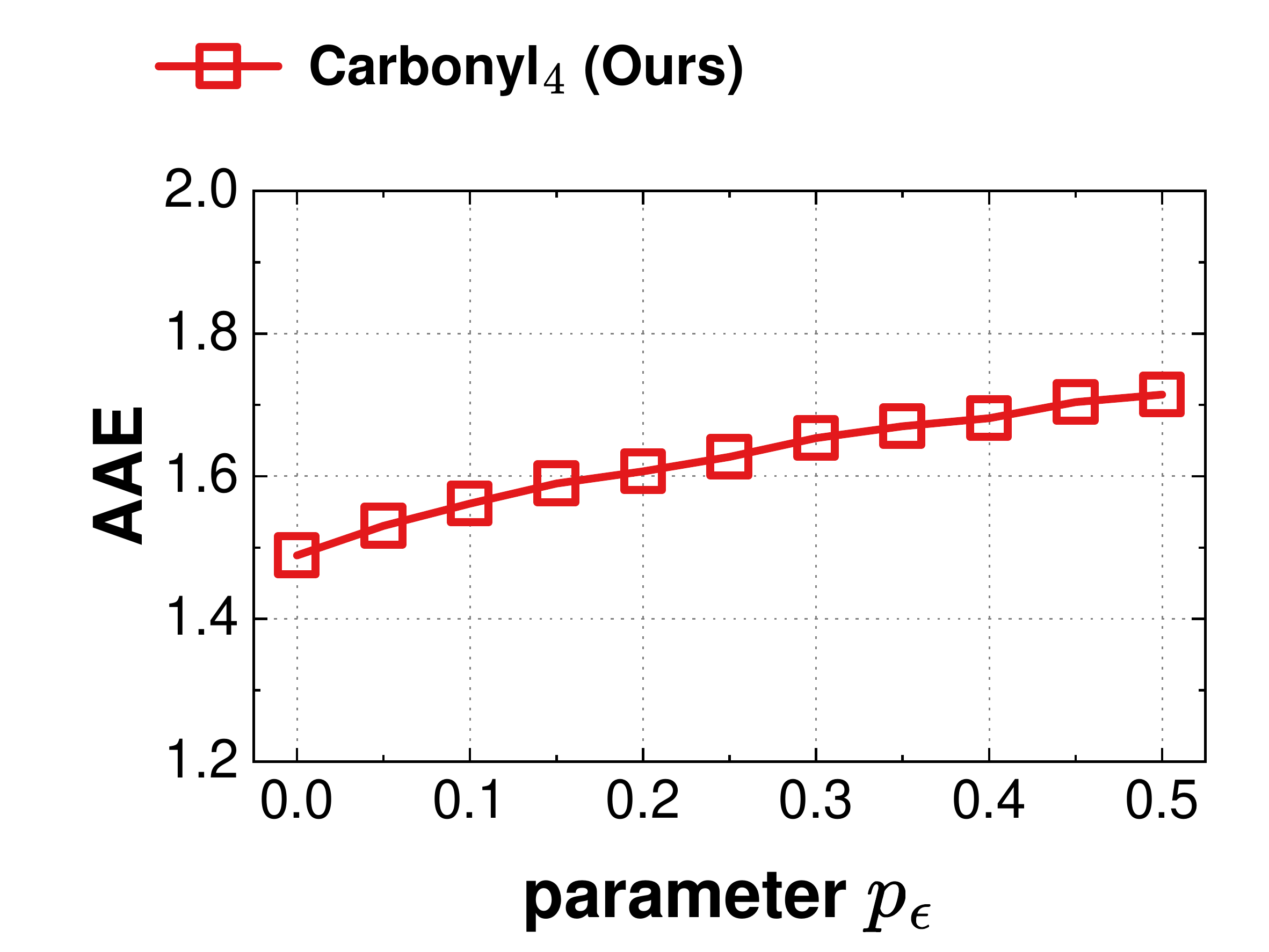}
        \label{fig:zipf:param_e:p_aae}
    }
    \hspace{-0.6cm}
    \subfigure[Parameter $p_\epsilon$]{
    \includegraphics[height=3.2cm]{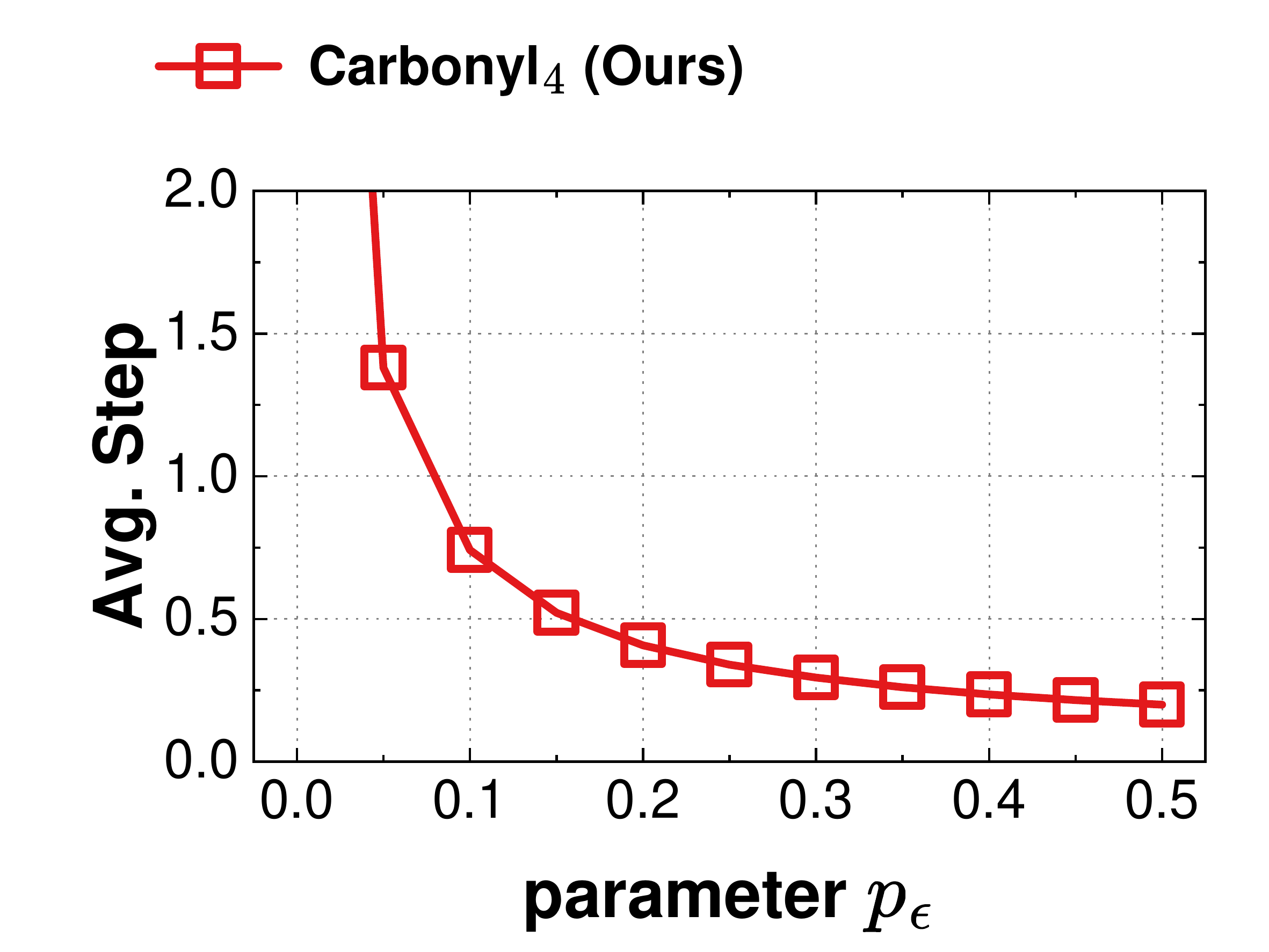}
        \label{fig:zipf:param_e:p_kick}
    }
    \caption{Parameter $M$ \& $p_\epsilon$, Synthetic dataset.}
    \label{fig:zipf:param}
\end{figure}

\subsection{\textbf{Experiments on Parameters}}
\label{sec:exp:param}

This section explores the variation in parameters of \sketch, including maximum search steps $M$, stop probability $p_\epsilon$, and the number of entries per \bucket $d$. We evaluate the accuracy and time complexity of \sketch for point and Top-$K$ queries and examine the effects of set and increment updates proportions in Synthetic datasets, as well as the skewness of item keys on accuracy.

\bbb{Accuracy v.s. $M$ and $p_\epsilon$ (Figure \ref{fig:zipf:param}):}
Figures \ref{fig:zipf:param_m:p_aae} and \ref{fig:zipf:param_e:p_aae} show that in point queries, \sketch's AAE declines as $M$ increases, stabilizing when $M \geqslant 10$. Conversely, AAE rises with increasing $p_\epsilon$, displaying a proportional linear trend.
For Figures \ref{fig:zipf:param_m:p_aae} and \ref{fig:zipf:param_m:p_kick}, we set $p_\epsilon=0.1$; For Figures \ref{fig:zipf:param_e:p_aae} and \ref{fig:zipf:param_e:p_kick}, we set $M=100$.

\bbb{Average Search Steps v.s. $M$ and $p_\epsilon$ (Figure \ref{fig:zipf:param}):}
Figures \ref{fig:zipf:param_m:p_kick} and \ref{fig:zipf:param_e:p_kick} illustrate that in point queries, \sketch's average search steps grow with larger $M$, indicating a direct linear relationship. As $p_\epsilon$ increases, the average search steps decrease inversely, aligning with Theorem \ref{theo:search}.

\begin{figure}[t]
    \centering
    \subfigure[Point Query]{
        \includegraphics[height=3.2cm]{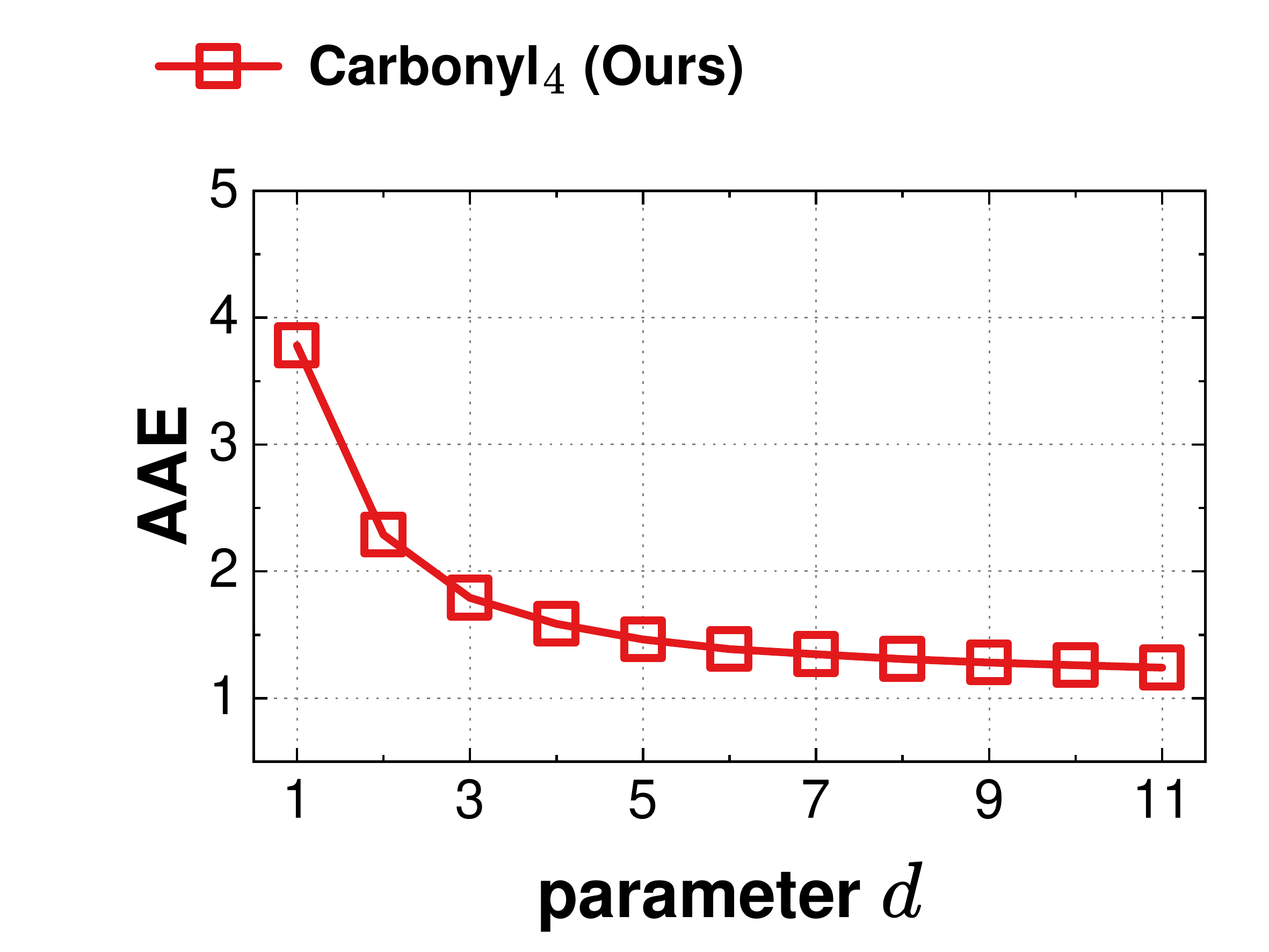}
        \label{fig:zipf:param_d:p_aae}
    }
    \hspace{-0.6cm}
    \subfigure[TopK Query]{
        \includegraphics[height=3.2cm]{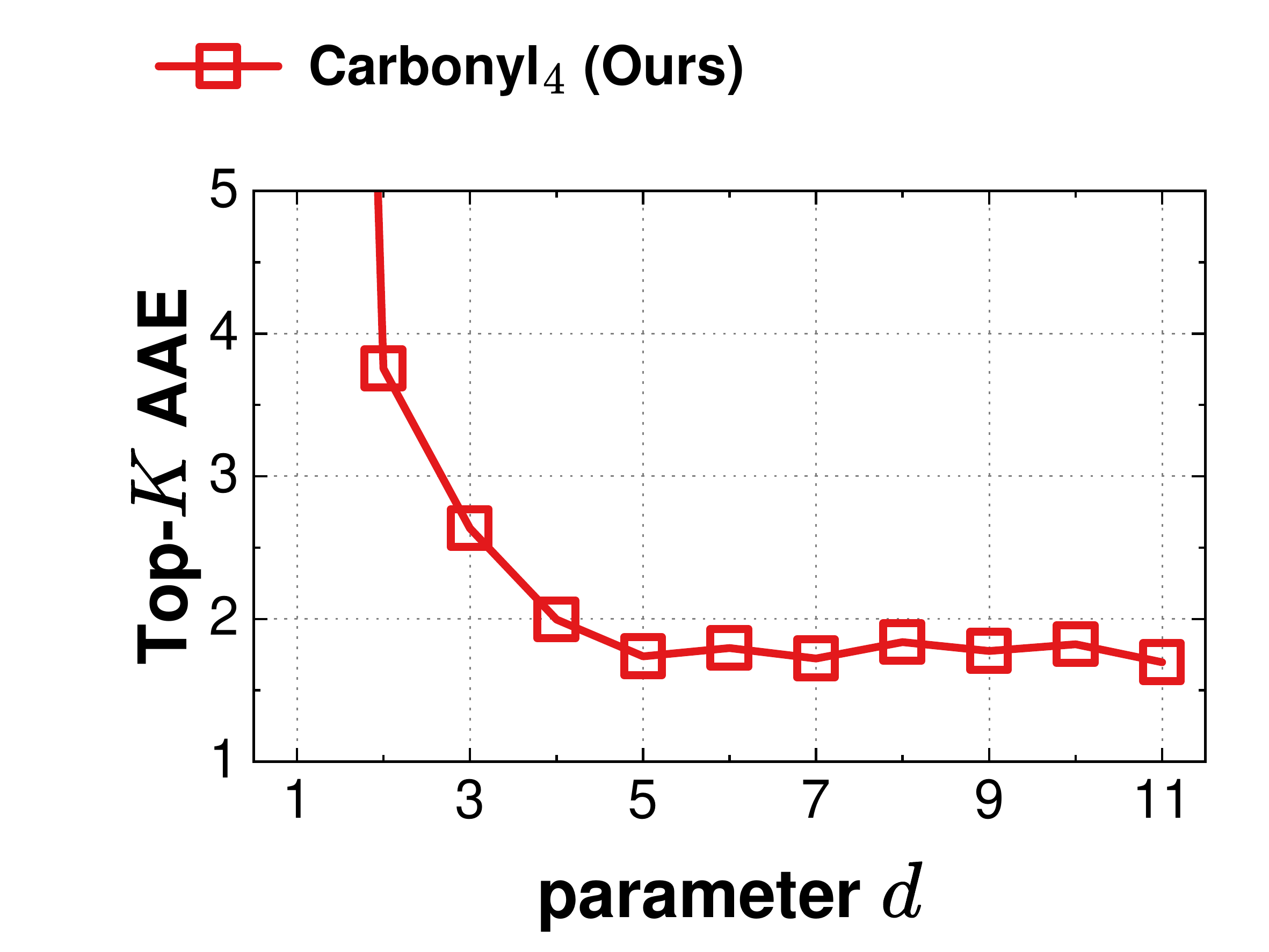}
        \label{fig:zipf:param_d:t_aae}
    }

    \caption{Parameter $d$, Synthetic dataset.}
    \label{fig:zipf:param_d}
\end{figure}

\begin{figure}[t]
    \centering
    \subfigure[AAE]{
        \includegraphics[height=3.2cm]{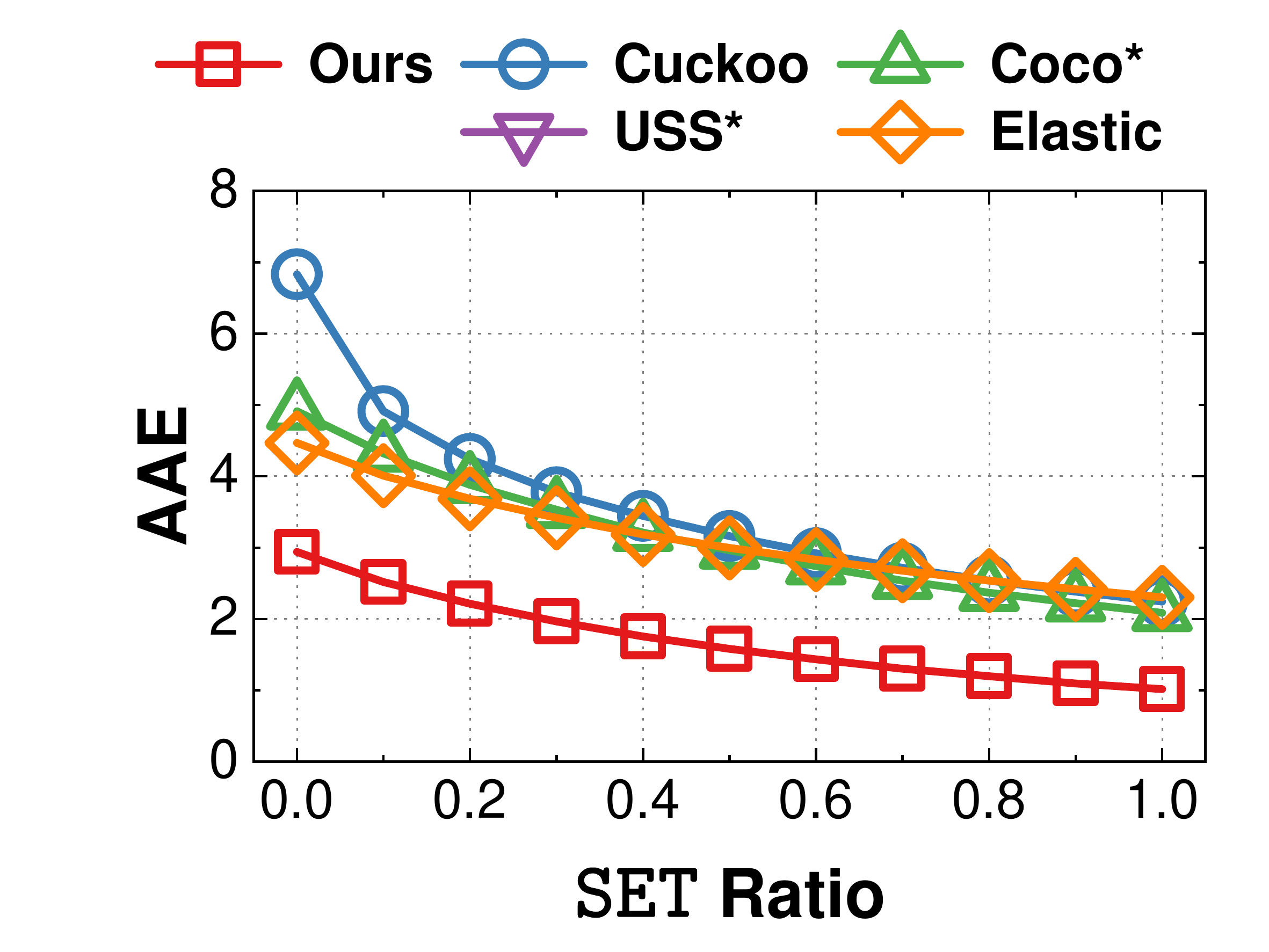}
        \label{fig:zipf:ratio:aae}
    }
    \hspace{-0.6cm}
    \subfigure[MSE]{
        \includegraphics[height=3.2cm]{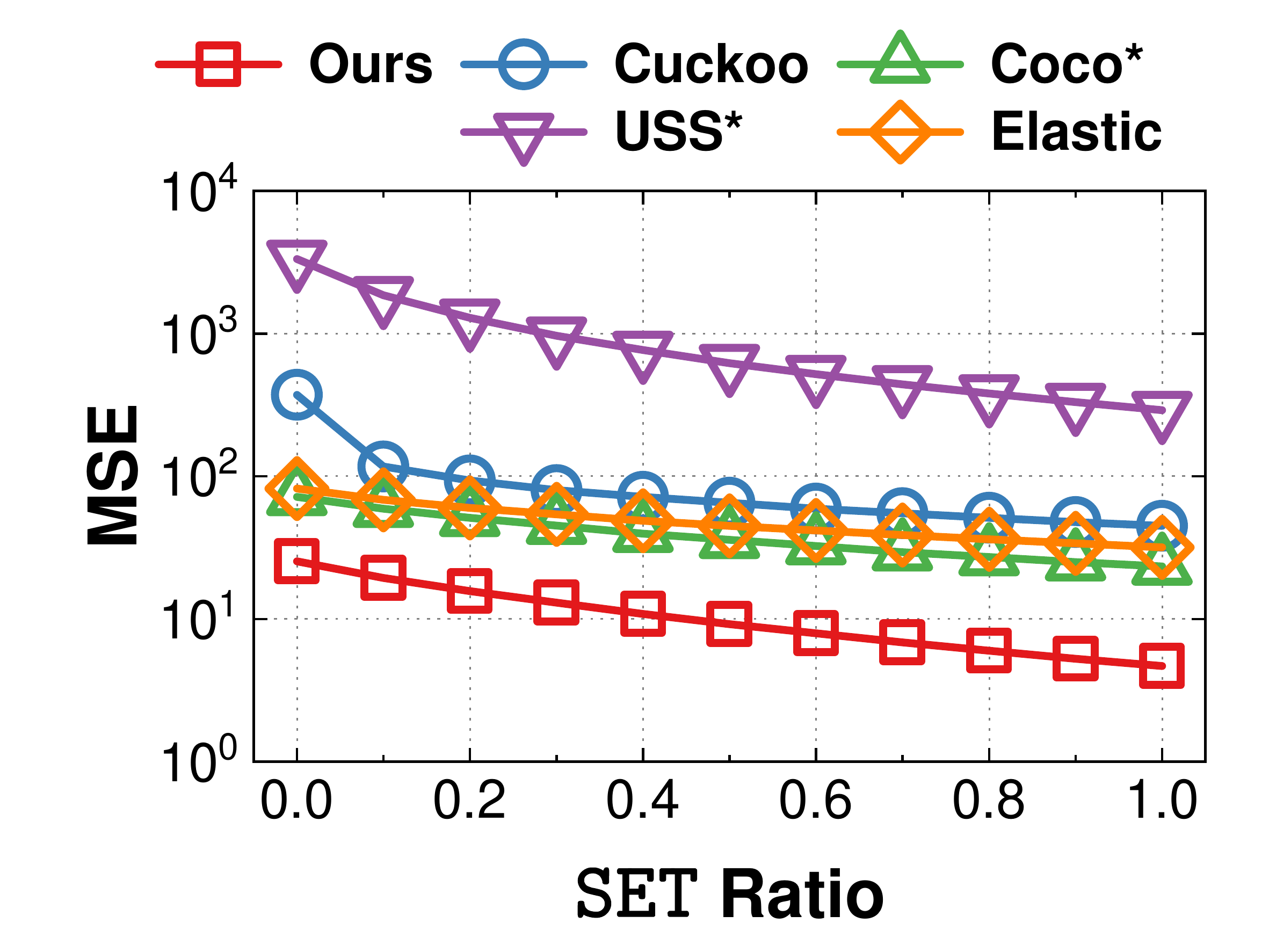}
        \label{fig:zipf:ratio:mse}
    }
    
    \caption{\textsc{Set} Ratio, Synthetic dataset.}
    \label{fig:zipf:ratio}
\end{figure}

\begin{figure}[t]
    \centering
    \subfigure[AAE]{
        \includegraphics[height=3.2cm]{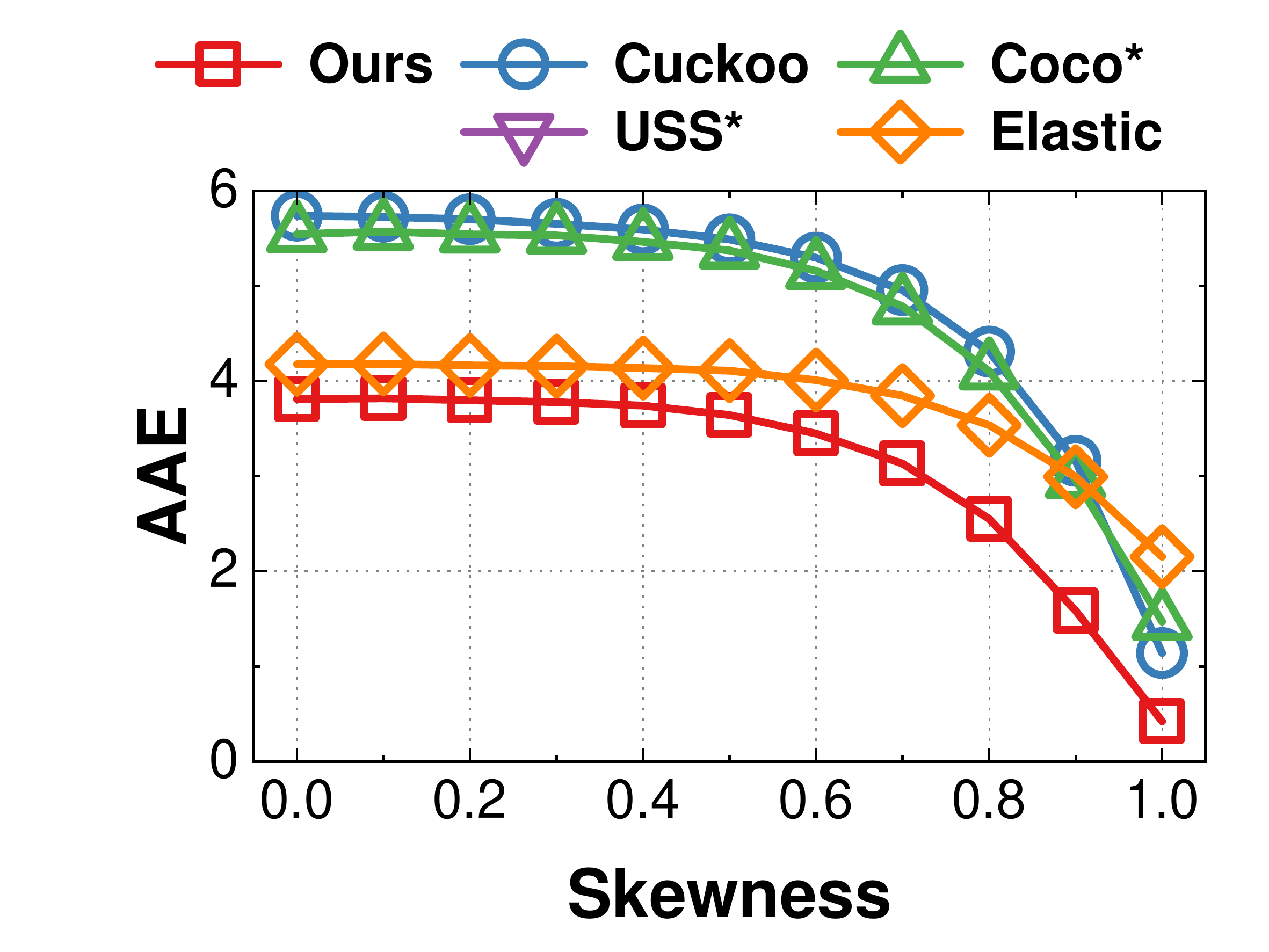}
        \label{fig:zipf:skew:aae}
    }
    \hspace{-0.6cm}
    \subfigure[MSE]{
        \includegraphics[height=3.2cm]{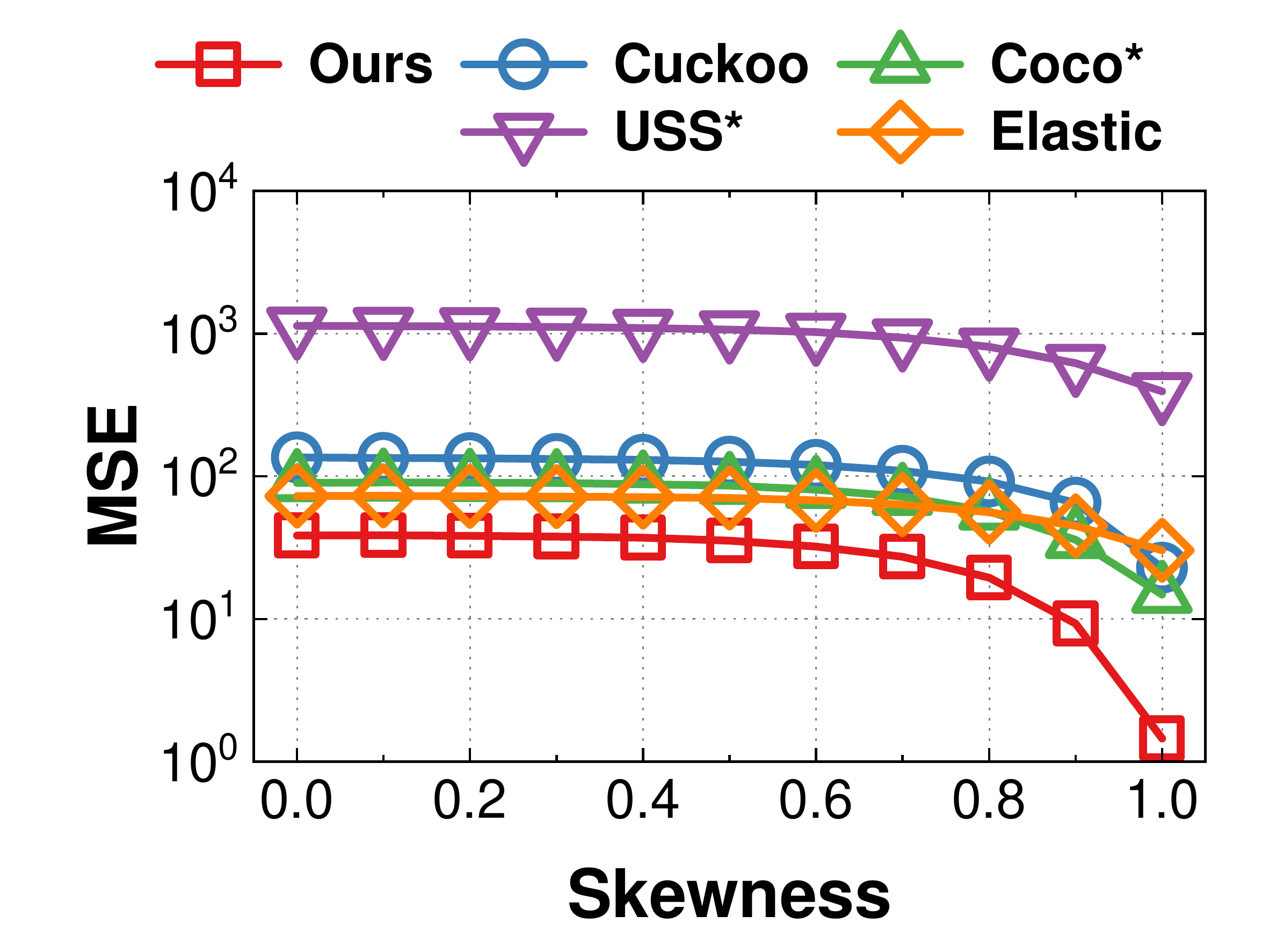}
        \label{fig:zipf:skew:mse}
    }
    
    \caption{Skewness, Synthetic dataset.}
    \label{fig:zipf:skew}
\end{figure}

\begin{figure}[t]
    \centering
    \subfigure[$\Delta$ MSE]{
        \includegraphics[height=3.2cm]{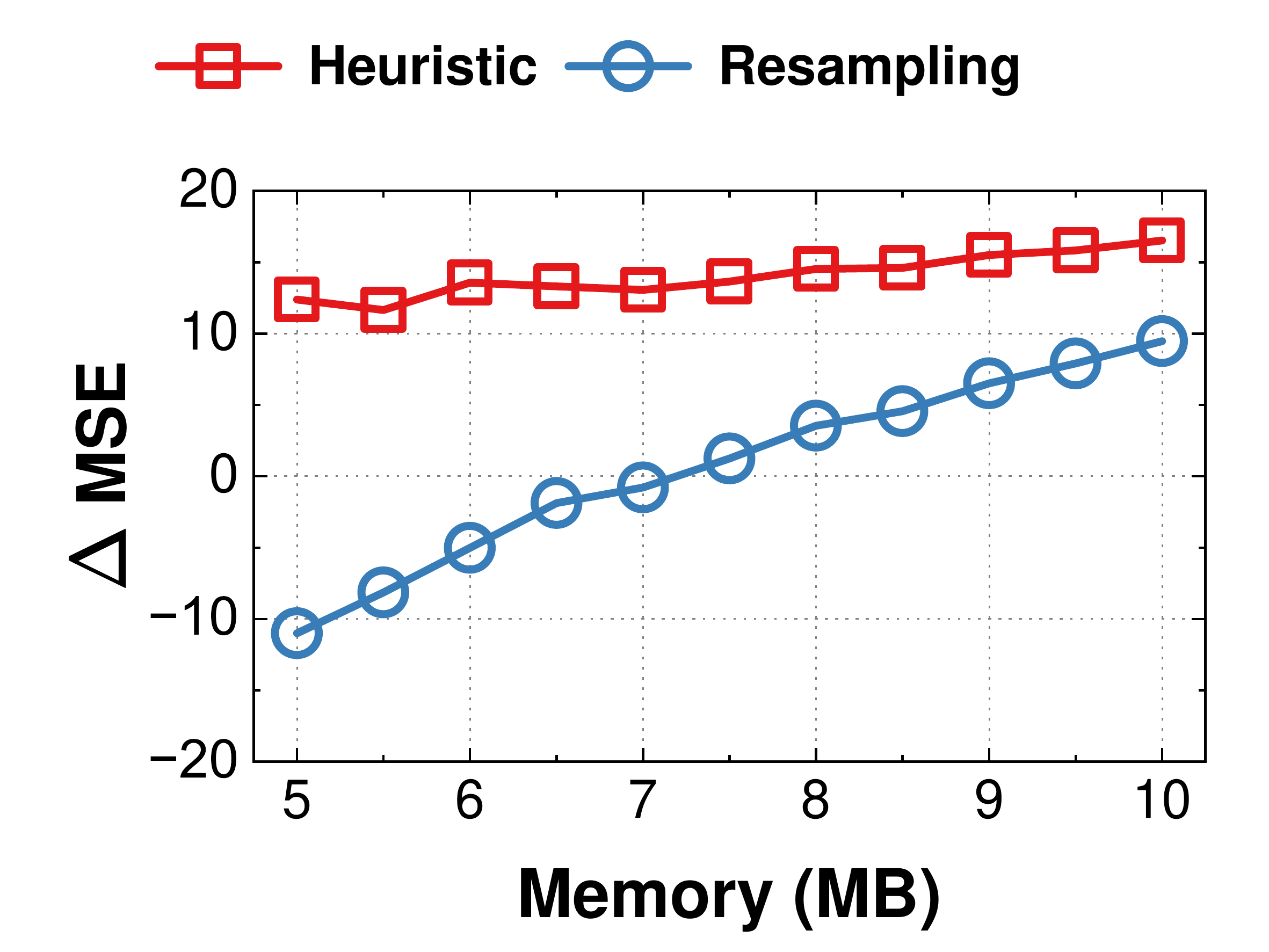}
        \label{fig:shrink:point:mse}
    }
    \hspace{-0.6cm}
    \subfigure[Speedup]{
        \includegraphics[height=3.2cm]{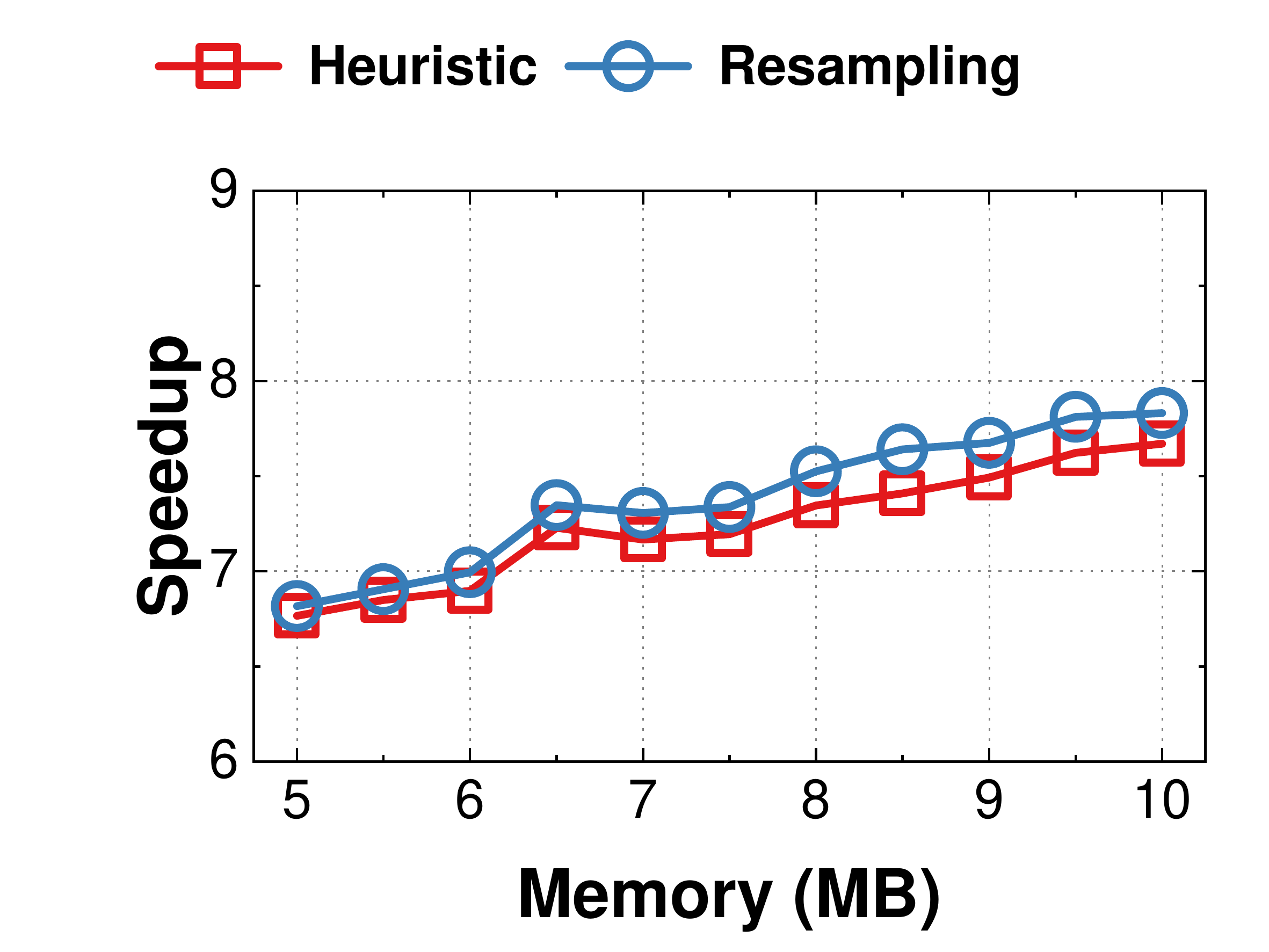}
        \label{fig:shrink:point:speedup}
    }
    
    \vspace{-0.3cm}
    
    \centering
    \subfigure[$\Delta$ Recall]{
        \includegraphics[height=3.2cm]{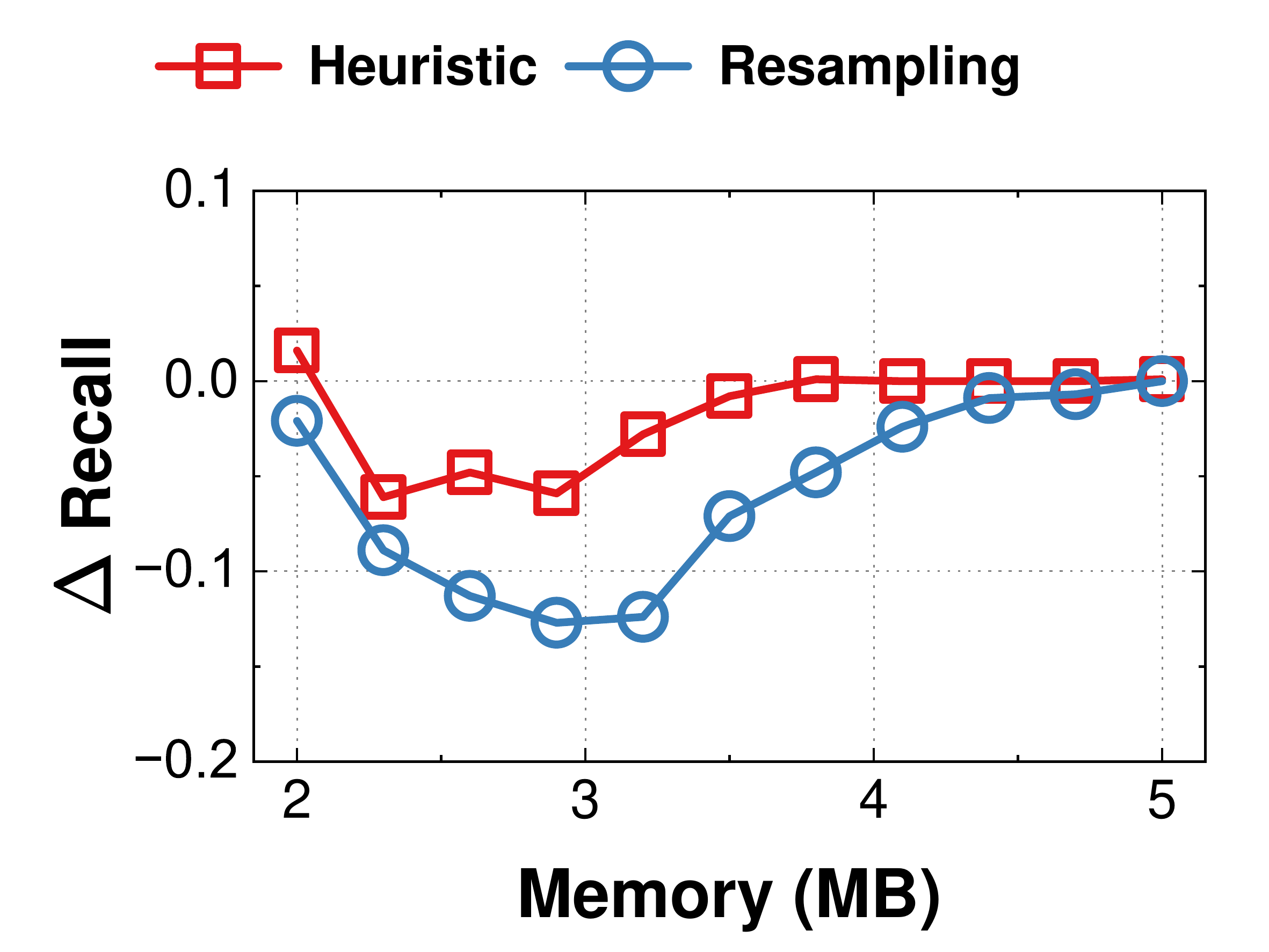}
        \label{fig:shrink:topk:recall}
    }
    \hspace{-0.6cm}
    \subfigure[Speedup]{
        \includegraphics[height=3.2cm]{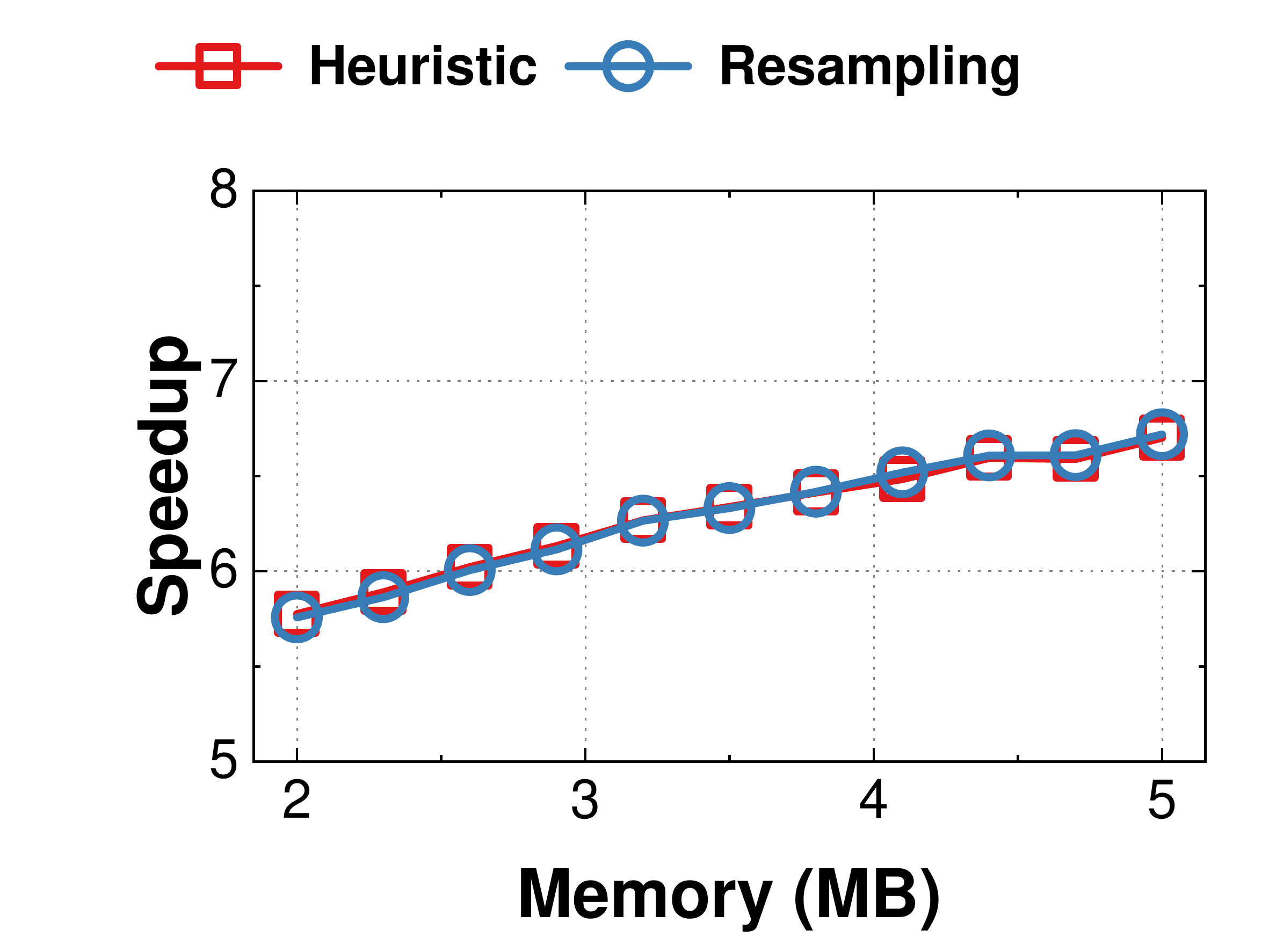}
        \label{fig:shrink:topk:speedup}
    }
    
    \caption{Performance of Shrinking, Synthetic dataset.}
    \label{fig:shrink}
\end{figure}

\bbb{Accuracy v.s. $d$ (Figure \ref{fig:zipf:param_d}):}
Figures \ref{fig:zipf:param_d:p_aae} and \ref{fig:zipf:param_d:t_aae} reveal that \sketch's AAE diminishes with higher $d$, reaching stability for $M \geqslant 6$ in point queries and for $M \geqslant 4$ in Top-$K$ queries. We recommend $d=4$ for optimal results.

\bbb{Accuracy v.s. \textsc{Set} Ratio (Figure \ref{fig:zipf:ratio}):}
Figures \ref{fig:zipf:ratio:aae} and \ref{fig:zipf:ratio:mse} indicate that on the Synthetic dataset for point queries, all algorithms' AAE and MSE decrease with an increased \textsc{Set} update ratio. \sketch shows $2.05$, $2.27$, $12.0$, and $2.80$ times lower AAE than Coco*, Elastic, USS*, and Cuckoo respectively; MSE is also $4.98$, $6.79$, $1.32\times10^2$, and $14.7$ times lower respectively. This trend confirms that \sketch's design is highly effective for mixed set-increment data streams.
Taking Elastic as an example, when the \textsc{Set} ratio is 0, \sketch's ARE and MSE are respectively $1.52$ times and $2.27$ times lower than those of Elastic; when the \textsc{Set} ratio is 1, \sketch's ARE and MSE are respectively $3.24$ times and $6.79$ times lower than those of Elastic. This indicates that the cascading overflow technique used by \sketch is very effective for \textsc{Set} updates.

\bbb{Accuracy v.s. Skewness (Figure \ref{fig:zipf:skew}):}
Figures \ref{fig:zipf:ratio:aae} and \ref{fig:zipf:ratio:mse} demonstrate that for point queries on the Synthetic dataset, the AAE and MSE for all algorithms decrease with increased data stream skewness. \sketch consistently maintains the highest accuracy among all compared algorithms.

\subsection{\textbf{Experiments on Shrinking Algorithm}}
\label{sec:exp:shrink}

This section delves into the performance of the re-sampling and heuristic-based shrinking algorithms regarding both accuracy and speed. We display the difference ($\Delta$) in MSE and recall rate between our proposed two fast shrinking algorithms and the MSE and recall of the shrinking algorithm based on re-building.

\bbb{Performance of Re-Sampling (Figure \ref{fig:shrink}):}
Figure \ref{fig:shrink:point:mse} elucidates the MSE disparities between re-sampling based shrinking and re-building. The MSE of re-sampling is higher than that of re-building when the original \sketch's memory consumption exceeds $7$MB, while its MSE is lower than re-building when blow this memory consumption, attesting to re-sampling's optimality.
Figure \ref{fig:shrink:topk:recall} contrasts recall rates, and Figures \ref{fig:shrink:point:speedup} to \ref{fig:shrink:topk:speedup} highlight the re-sampling algorithm's velocity, showcasing a $5.76$ to $7.83$ times acceleration over re-building.

\bbb{Performance of Heuristic Shrinking (Figure \ref{fig:shrink}):}
Figure \ref{fig:shrink:point:mse} compares MSE outcomes between heuristic shrinking and re-building, noting a consistent increase in MSE for the heuristic approach.
Figure \ref{fig:shrink:topk:recall} differentiates recall rates, where heuristic shrinking prevails over re-sampling, validating its efficacy in optimizing Top-$K$ queries.
Figures \ref{fig:shrink:point:speedup} and \ref{fig:shrink:topk:speedup} document the speedup afforded by heuristic shrinking, which achieves a $5.78$ to $7.67$ hastening compared to re-building.

        \section{\textbf{Conclusion}}
In summary, \sketch emerges as a robust solution for SIM data streams, bridging the gap where current sketches falter. The algorithm's novel \bucket and \overflow techniques pave the way for high accuracy and variance optimization, while its memory efficiency is exemplified through innovative shrinking methods. Through rigorous experimentation, \sketch has proven to outperform existing algorithms, bolstering its potential as a new standard for data stream analysis. The source codes of \sketch are available at GitHub \cite{github}.



	{
	\balance
        \bibliographystyle{unsrt}
	\bibliography{reference}	
	}

\end{document}